\documentclass[useAMS]{mn2e} 
 
\usepackage{graphicx} 
\usepackage{txfonts} 
\newcommand\aj{AJ} 
\newcommand\apj{ApJ} 
\newcommand\apjs{ApJS}       
 
\newcommand\aap{A\&A} 
\newcommand\mnras{MNRAS} 
\newcommand\apjl{ApJ} 
\newcommand\pasp{PASP} 
\newcommand\nat{Nature} 
\newcommand\aaps{A\&AS} 

\newcommand\aapr{ARA\&A}
 
\title[]{Multiple stellar populations in Magellanic Cloud clusters. VI. A survey of multiple sequences and Be stars in young clusters}
\author[A.\,P.\,Milone et al.] 
        {A.\,P.\,Milone$^{1,2}$,
         A.\,F.\,Marino$^{2}$,
         M.\,Di Criscienzo$^{3}$,         
         F.\,D'Antona$^{3}$,
         L.\,R.\,Bedin$^{4}$,\newauthor
         G.\,Da Costa$^{2}$,
         G.\,Piotto$^{1,4}$,
         M.\,Tailo$^{5}$,
         A.\,Dotter$^{6}$,
         R.\,Angeloni$^{7, 8}$,
         J.\,Anderson$^{9}$, \newauthor                  
         H.\,Jerjen$^{2}$, 
         C.\,Li$^{10}$, 
         A.\,Dupree$^{6}$,
         V.\,Granata$^{1,4}$,
         E.\,P.\,Lagioia$^{1,11,12}$,         
         A.\,D.\,Mackey$^{2}$, \newauthor                  
         D.\,Nardiello$^{1,4}$,
         E.\,Vesperini$^{13}$
\\ 
$^{1}$Dipartimento di Fisica e Astronomia ``Galileo Galilei'', Univ. di Padova, Vicolo dell'Osservatorio 3, Padova, IT-35122\\
$^{2}$Research School of Astronomy \& Astrophysics, Australian National University, Canberra, ACT 2611, Australia \\
$^{3}$Istituto Nazionale di Astrofisica - Osservatorio Astronomico di Roma, Via Frascati 33, I-00040 Monteporzio Catone, Roma, Italy\\
$^{4}$Istituto Nazionale di Astrofisica - Osservatorio Astronomico di Padova, Vicolo dell'Osservatorio 5, Padova, IT-35122\\
$^{5}$ Dipartimento di Fisica, Universita' degli Studi di Cagliari, SP Monserrato-Sestu km 0.7, 09042 Monserrato, Italy\\
$^{6}$ Harvard-Smithsonian Center for Astrophysics, Cambridge, MA, USA \\
$^{7}$ Departamento de Fisica y Astronomia, Universidad de La Serena, Av. Juan Cisternas 1200 N, La Serena, Chile\\
$^{8}$ Instituto de Investigacion Multidisciplinar en Ciencia y Tecnologia, Universidad de La Serena, Raul Bitran 1305, La Serena, Chile\\ 
$^{9}$Space Telescope Science Institute, 3800 San Martin Drive, Baltimore,  MD 21218, USA\\
$^{10}$ Department of Physics and Astronomy, Macquarie University, Sydney, NSW 2109, Australia\\
$^{11}$Instituto de Astrof\`isica de Canarias, E-38200 La Laguna, Tenerife, Canary Islands, Spain\\
$^{12}$Department of Astrophysics, University of La Laguna, E-38200 La Laguna, Tenerife, Canary Islands, Spain\\
$^{13}$Department of Astronomy, Indiana University, Bloomington, IN 47405, USA\\
       } 
\begin{document} 
\date{Accepted 2018 February 27. Received 2018 February 26; in original form 2017 December 21} 
 
\maketitle 
\label{firstpage} 
 
\begin{abstract}
The split main sequences (MSs) and extended MS turnoffs (eMSTOs) detected in a few young clusters have demonstrated that these stellar systems host multiple populations differing in a number of properties such as rotation and, possibly, age. 
   
We analyze Hubble Space Telescope photometry for thirteen clusters with ages between $\sim$40 and $\sim$1000 Myrs and  of different masses. Our goal is to investigate for the first time the occurrence of multiple populations  in a large sample of young clusters.
  
 We find that all the clusters exhibit the eMSTO phenomenon and that MS stars more massive than $\sim$1.6 $\mathcal M_{\rm \odot}$ define a blue and red MS, with the latter hosting the majority of MS stars.  The comparison between the observations and isochrones suggests that the blue MSs are made of  slow-rotating stars, while the red MSs host stars with rotational velocities close to the breakup value. 

About half of the bright MS stars in the youngest clusters are H-alpha emitters. These Be stars populate the red MS and the reddest part of the eMSTO thus supporting the idea that the red MS is made of fast rotators. 

We conclude that the split MS and the eMSTO are a common feature of young clusters in both Magellanic Clouds.
 The phenomena of a split MS and an eMSTO occur for stars that are more massive than a specific threshold which is independent of the host-cluster mass.

As a by-product, we report the serendipitous discovery of a young SMC cluster, GALFOR1.
\end{abstract} 
 
\begin{keywords} 
techniques: photometric --- binaries: visual --- stars: rotation --- globular clusters: individual: KRON\,34, NGC\,294, NGC\,330, NGC\,1755, NGC\,1801, NGC\,1805, NGC\,1818, NGC\,1844, NGC\,1850, NGC\,1856, NGC\,1866, NGC\,1868, NGC\,1953, NGC\,2164 --- Magellanic Clouds.
\end{keywords} 
 
\section{Introduction}\label{sec:intro} 
Recent studies, based on {\it Hubble Space Telescope} ({\it HST}\,) photometry, have revealed that the color-magnitude diagram (CMD) of some
 $\sim$30-400-Myr old stellar clusters in the Large and Small Magellanic Cloud (LMC, SMC) exhibit either a broadened or split main sequence
(MS, Milone et al.\,2013, 2015, 2016, 2017a; Correnti et al.\,2017; Bastian et al.\,2017; Li et al.\,2017a).
Moreover, most of these clusters show an extended MS turn off (eMSTO), in close analogy with what is
observed in nearly all the intermediate-age ($\sim$1-2-Gyr old) clusters of both Magellanic Clouds (e.g.\,Bertelli et al.\,2003; Mackey \& Broby Nelsen\,2007; Glatt et al.\,2008; Milone et al.\,2009).
These findings have challenged the traditional picture that the CMDs of young and intermediate-age star clusters are best described by simple isochrones and have triggered a to huge efforts to understand the physical reasons responsible for the double MSs and the eMSTOs.

The comparison between the observed CMDs of young clusters and stellar models shows that the split MS is consistent with two stellar populations with different rotation rates: a red MS, which hosts stars with rotational velocities close to the breakup value, and a blue MS, which is made of slow-rotating or non-rotating stars (D'Antona et al.\,2015; Milone et al.\,2016; Bastian et al.\,2017).
It is well known that B-type stars with H-$\alpha$ emission (Be stars) are fast rotators (see review by Rivinius et al.\,2013). As a consequence, the existence of fast-rotating stars in young clusters is further supported by the presence of a large number of Be stars (e.g.\,Keller et al.\,1998; Correnti et al.\,2017; Bastian et al.\,2017) and by direct spectroscopic measurements of rotational velocities in MS stars (Dupree et al.\,2017).

While the interpretation of the split MS is widely accepted, the physical reasons responsible for the eMSTO are still controversial.
The most straightforward interpretation for the eMSTO is that young and intermediate-age clusters have experienced a prolonged star formation and host multiple generations with different age (Mackey et al.\,2008; Goudfrooij et al.\,2011, 2014; Correnti et al.\,2014; Piatti \& Cole 2017). 
 The evidence the color-magnitude diagrams (CMDs) of both young Magellanic-Clouds clusters and old globular clusters (GCs) are not consistent with a single isochrone has suggested that the clusters with eMSTO could be the young counterparts of old GCs with multiple populations (e.g.\,Keller et al.\,2011; Conroy et al.\,2011).

As an alternative, it has been suggested that the eMSTO is another consequence of the presence of populations with different rotation rates (Bastian \& de Mink\,2009; Yang et al.\,2011, 2013; Li et al.\,2014; D'Antona et al.\,2015; Niederhofer et al.\,2015).
This would be consistent with the inferred picture for young clusters, but it seems that stellar models with different rotation rates alone are not able to entirely reproduce the observations and some age spread is still required (e.g.\,Correnti et al.\,2017; Milone et al.\,2017a; Goudfrooij et al.\,2017).   
D'Antona et al.\,(2017) have noticed that the blue MS of the studied young clusters host a sub-population of stars, which seems younger than the majority of blue-MS stars.
They suggested that stars caught in the stage of braking from a rapidly-rotating configuration are responsible for this feature of the CMD, and that these stars mimic a $\sim$30\% younger stellar population now. 
 
In this work we derive high-precision {\it HST} photometry to investigate  homogeneously the multiple populations and H-$\alpha$-emitting stars in thirteen clusters of both Magellanic Clouds younger than $\sim$1\,Gyr.
 Our sample comprises both clusters that have not been investigated in the
context of multiple populations, namely KRON\,34, NGC\,294, NGC\,1801 and NGC\,1953 and clusters with previous evidence of multiple sequences in the CMD. The latters include NGC\,330, NGC\,1755, NGC\,1805, NGC\,1818, NGC\,1850, NGC\,1856, NGC\,1866, NGC\,1868, and NGC\,2164 (Li et al.\,2014, 2017; Milone et al.\,2015, 2016, 2017; Correnti et al.\,2015, 2017). 
The paper is organized as follows. In Sect.~\ref{sec:data} we describe the data and the data reduction, and present the CMDs in Sect.~\ref{sec:cmd}. The MS and the eMSTO are first investigated in Sect.~\ref{sec:MS}, while Sect.~\ref{sec:iso} compares the observed CMDs with stellar models. In Sect.~\ref{sec:pratio} we derive the fraction of stars in the distinct MSs of each cluster while Sect.~\ref{sec:Be} is dedicated to stars with H-$\alpha$ emission. Finally, summary and discussion are provided in Sect.~\ref{sec:conclusion}.

 \section{Data and data analysis} \label{sec:data}
The dataset of this paper is mostly collected as part of our programs GO\,13379, GO\,14204, and GO\,14710. Multi-band images of three SMC clusters and ten LMC clusters were obtained through the Ultraviolet and Visual channel of the Wide-Field Camera 3 (UVIS/WFC3) on board of {\it HST}.
We used images in the F336W and F814W bands for all the analyzed clusters whereas F656N images are available for all the targets but NGC\,1755 and NGC\,1866.
 In addition, we exploited F225W data of NGC\,330, NGC\,1805, NGC\,1818, and NGC\,2164 and F275W images of NGC\,1850. Details on the data are provided in Table~\ref{tab:data}.

\begin{centering} 
\begin{figure} 
 \includegraphics[width=8.7cm]{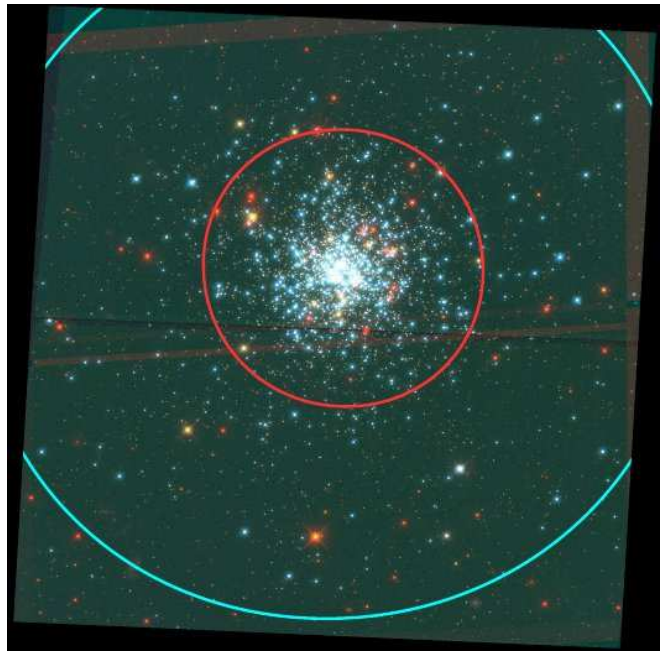}
 \caption{Three-colour image of NGC\,2164. The regions within the red circle and outside the azure circle are indicated as cluster field and reference field, respectively.} 
 \label{fig:NGC2164picture} 
\end{figure} 
\end{centering} 

 Stellar astrometry and photometry have been carried out by using the flt images and the procedure and the computer programs developed by one of us (J.\,A.\,).
  In a nutshell, we first removed the effect of the poor charge-transfer efficiency (CTE) of UVIS/WFC3 by using the empirical pixel-based method by Anderson \& Bedin (2010)  and derived for each image a 5$\times$5 array of perturbation point-spread functions (PSFs). To do this, we used a library array of empirical PSFs plus a spatially-varying array of PSF residuals that we obtained from unsaturated bright and isolated stars.

  We followed two distinct approaches to measure stars with different luminosities.
  All the bright stars have been measured in each image separately, by using the derived PSFs and the program {\it img2xym\_wfc3uv}, which is similar to the Wide-Field Camera package {\it img2xym\_WFC} by Anderson \& King (2006) but for UVIS/WFC3. This software routine also identifies saturated stars and estimates their fluxes and positions. In particular, it determines the amount of flux from the bled-into pixels thus allowing accurate flux determination for each saturated star (see Gilliland\,2004; Anderson et al.\,2008;  Gilliland et al.\,2010).  Faint stars have been measured by using a software package from Anderson et al.\,(in preparation) that analyzes all the images together to derive the stellar fluxes and luminosities. It is an improved version of the program by Anderson et al.\,(2008) and we refer to papers by Sabbi et al.\,(2016) and Bellini et al.\,(2017) for details on the method.

  Stellar positions have been corrected for the effect of geometric distortion by using the solution from Bellini et al.\,(2011) and have been transformed into the Gaia DR1 reference frame (Gaia Collaboration et al.\,2016). 
   Only relatively-isolated stars that are well fitted by the PSF and have small rms scatter in the position measurements have been included in the following analysis.
   The sample of stars with high-quality photometry has been selected by using the procedure described in Milone et al.\,(2009) and the various diagnostics of the astrometric and photometric quality provided by the computer programs. 

As an example of a typical cluster, we show in Fig.~\ref{fig:NGC2164picture} the trichromatic image of the analyzed field of NGC\,2164\footnote{The stacked images of all the clusters are available at http://galfor.astro.unipd.it }. The inner circle has been derived by eye with the criteria of selecting a region that is mostly populated by cluster members. It will be called the cluster field, hereafter. 
The analysis of each cluster 
 is based on stars in the cluster field, while in forthcoming papers of this series we will investigate the radial distribution of multiple populations in close analogy with what we have done in Milone et al.\,(2016) and Li et al.\,(2017b). Nevertheless, we verified that the main conclusions of this work are not affected by the choice of the cluster field.
 The region outside the azure circle plotted in  Fig.~\ref{fig:NGC2164picture} has the same area as the cluster field   but is mostly populated by field stars and defines the reference field. It will be used to statistically estimate the contamination of field stars in the cluster field.

The cluster-field radii, $R_{\rm cf}$, that we adopted for each cluster are provided in Table~\ref{tab:parameters}, where we also indicate the host Magellanic Cloud, the coordinates of the cluster center, the mass, the age, the distance modulus, and the reddening. The masses for the majority of clusters have been derived by  McLaughlin \& van der Marel\,(2005) by assuming a Wilson\,(1975) profile. In the cases of NGC\,1755, NGC\,1801, and NGC\,1953, which are not included in the sample analyzed by McLaughlin \& van der Marel, we used the masses from Popescu et al.\,(2012). Unfortunately, no mass estimates are available for Kron\,34 and NGC\,294.
 All the remaining parameters listed in in Table~\ref{tab:parameters} are derived in this paper. The cluster centers have been estimated as in Li et al.\,(2017a) by using the contours of the number density of stars brighter than $m_{\rm F814W}=22.0$. Ages, distances, and reddenings have been derived by comparing isochrones from the Geneva database with the observed CMDs (see Sect.~\ref{sec:iso} for details).  

\begin{table*}
 \caption{Description of the data used in this paper.}
 \begin{tabular}{ccccccc}
   \hline
   \hline
Cluster        & date      &  camera        & Filter & N$\times$ exposure time            & GO    &  PI \\
\hline
KRON\,34   & 2017, Jan, 13    & UVIS/WFC3 & F814W  & 90s$+$680s                         & 14710 & A.\,P.\,Milone \\
           & 2017, Jan, 13    & UVIS/WFC3 & F656N  & 2$\times$720s                      & 14710 & A.\,P.\,Milone \\
           & 2017, Jan, 13    & UVIS/WFC3 & F336W  & 3$\times$839s                      & 14710 & A.\,P.\,Milone \\
NGC\,294   & 2017, Feb, 08    & UVIS/WFC3 & F814W  & 90s$+$680s                         & 14710 & A.\,P.\,Milone \\
           & 2017, Feb, 08    & UVIS/WFC3 & F656N  & 2$\times$720s                      & 14710 & A.\,P.\,Milone \\
           & 2017, Feb, 08    & UVIS/WFC3 & F336W  & 3$\times$839s                      & 14710 & A.\,P.\,Milone \\
NGC\,330   & 2017, Jul, 03    & UVIS/WFC3 & F814W  & 90s$+$680s                         & 14710 & A.\,P.\,Milone \\
           & 2017, Jul, 03    & UVIS/WFC3 & F656N  & 2$\times$720s                      & 14710 & A.\,P.\,Milone \\
           & 2015, Jul, 26    & UVIS/WFC3 & F336W  & 10s$+$100s$+$805s$+$3$\times$960s  & 13727 & J.\,Kalirai \\
           & 2015, Jul, 26    & UVIS/WFC3 & F225W  & 10s$+$100s$+$5$\times$950s         & 13727 & J.\,Kalirai \\
NGC\,1755  & 2015, Oct, 05 - Dec, 28 - 2016, Mar, 26 - Jun 10   & UVIS/WFC3 &  F814W  & 4$\times$90s$+$4$\times$678s & 14204 & A.\,P.\,Milone \\
           & 2015, Oct, 05 - Dec, 28 - 2016, Mar, 26 - Jun 10   & UVIS/WFC3 &  F336W  & 8$\times$711s                & 14204 & A.\,P.\,Milone \\
NGC\,1801  & 2017, Feb, 26    & UVIS/WFC3 &  F814W  & 90s$+$666s                         & 14710 & A.\,P.\,Milone \\
           & 2017, Feb, 26    & UVIS/WFC3 &  F656N  & 2$\times$720s                      & 14710 & A.\,P.\,Milone \\
           & 2017, Feb, 26    & UVIS/WFC3 &  F336W  & 2$\times$834s$+$835                & 14710 & A.\,P.\,Milone \\
NGC\,1805  & 2017, Oct, 06    & UVIS/WFC3 &  F814W  & 90s$+$666s                         & 14710 & A.\,P.\,Milone \\
           & 2017, Oct, 06    & UVIS/WFC3 &  F656N  & 2$\times$720s                      & 14710 & A.\,P.\,Milone \\
           & 2015, Oct, 28    & UVIS/WFC3 &  F336W  & 10s$+$100s$+$790s$+$3$\times$947s  & 13727 & J.\,Kalirai \\
           & 2015, Oct, 28    & UVIS/WFC3 &  F225W  & 10s$+$100s$+$5$\times$938s         & 13727 & J.\,Kalirai \\
NGC\,1818  & 2017, Feb, 03    & UVIS/WFC3 &  F814W  & 90s$+$666s                         & 14710 & A.\,P.\,Milone \\
           & 2017, Feb, 03    & UVIS/WFC3 &  F656N  & 2$\times$720s                      & 14710 & A.\,P.\,Milone \\
           & 2015, Oct, 29    & UVIS/WFC3 &  F336W  & 10s$+$100s$+$790s$+$3$\times$947s  & 13727 & J.\,Kalirai \\
           & 2015, Oct, 29    & UVIS/WFC3 &  F225W  & 10s$+$100s$+$5$\times$938s         & 13727 & J.\,Kalirai \\
NGC\,1850  & 2015, Oct, 22    & UVIS/WFC3 &  F814W  & 7s$+$350s$+$440s                   & 14174 & P.\,Goudfrooij \\
           & 2015, Oct, 22    & UVIS/WFC3 &  F656N  & 900s$+$925s$+$950s                 & 14174 & P.\,Goudfrooij \\
           & 2015, Dec, 19    & UVIS/WFC3 &  F336W  & 260s$+$370s$+$600s$+$650s$+$675s   & 14069 & N.\,Bastian \\
           & 2015, Oct, 22    & UVIS/WFC3 &  F275W  & 150s$+$780s$+$790s                 & 14174 & P.\,Goudfrooij \\
NGC\,1856  & 2013, Nov, 12    & UVIS/WFC3 &  F814W  & 51s$+$360s$+$450s               & 13011 & P.\,Goudfrooij \\
           & 2014, Feb, 09 - Mar, 24 - May, 18 - Jun 11   & UVIS/WFC3 & F814W  & 4$\times$90s$+$4$\times$704s & 13379 & A.\,P.\,Milone \\
           & 2013, Nov, 12    & UVIS/WFC3 &  F656N  & 735s$+$2$\times$940s               & 13011 & P.\,Goudfrooij \\
           & 2014, Feb, 09 - Mar, 24 - May, 18 - Jun 11   & UVIS/WFC3 & F336W  & 8$\times$711s                & 13379 & A.\,P.\,Milone \\
NGC\,1866  & 2016, Mar, 01 - May, 31 - Jun, 01 - Aug 18   & UVIS/WFC3 & F814W  & 4$\times$90s$+$4$\times$678s & 14204 & A.\,P.\,Milone \\
           & 2016, Mar, 01 - May, 31 - Jun, 01 - Aug 18   & UVIS/WFC3 & F336W  & 8$\times$711s                & 14204 & A.\,P.\,Milone \\
NGC\,1868  & 2016, Dec, 16    & UVIS/WFC3 &  F814W  & 90s$+$666s                         & 14710 & A.\,P.\,Milone \\
           & 2016, Dec, 16    & UVIS/WFC3 &  F656N  & 2$\times$720s                      & 14710 & A.\,P.\,Milone \\
           & 2016, Dec, 16    & UVIS/WFC3 &  F336W  & 830s$+$2$\times$831s               & 14710 & A.\,P.\,Milone \\
NGC\,1953  & 2017, Jul, 18    & UVIS/WFC3 &  F814W  & 90s$+$666s                         & 14710 & A.\,P.\,Milone \\
           & 2017, Jul, 18    & UVIS/WFC3 &  F656N  & 2$\times$720s                      & 14710 & A.\,P.\,Milone \\
           & 2017, Jul, 18-19 & UVIS/WFC3 &  F336W  & 2$\times$834s$+$835s               & 14710 & A.\,P.\,Milone \\
NGC\,2164  & 2017, Feb, 02    & UVIS/WFC3 &  F814W  & 90s$+$758s                         & 14710 & A.\,P.\,Milone \\
           & 2017, Feb, 02    & UVIS/WFC3 &  F656N  & 2$\times$720s                      & 14710 & A.\,P.\,Milone \\
           & 2015, Sep, 05-06 & UVIS/WFC3 &  F336W  & 10s$+$100s$+$790s$+$3$\times$947s  & 13727 & J.\,Kalirai \\
           & 2015, Sep, 05    & UVIS/WFC3 &  F225W  & 10s$+$100s$+$5$\times$938s         & 13727 & J.\,Kalirai \\
 \hline\hline
 \end{tabular}\\
 \label{tab:data}
 \end{table*}

\begin{table*}
 \caption{Parameters of the studied clusters. Masses of NGC\,1755, NGC\,1801 and NGC\,1953 are from Popescu et al.\,(2012), while the masses of the other clusters are taken from McLaughlin \& van der Marel\,(2005) by assuming a Wilson\,(1975) profile. The remaining cluster parameters are derived in this paper.}
 \begin{tabular}{ccccccccc}
   \hline
   \hline
Cluster   & galaxy &RA (J2000) & DEC (J2000) & log($\mathcal{M}/\mathcal{M}_{\odot}$) & Age [Myr] & (m$-$M)$_{0}$ & E(B$-$V) & R$_{\rm cf}$[arcsec] \\
\hline
KRON\,34  & SMC & 00 55 33.50 & $-$72 49 56.0 & ---  &630 & 18.80 & 0.08 & 20 \\
NGC\,294  & SMC & 00 53 04.94 & $-$73 22 45.7 & ---  &500 & 18.80 & 0.15 & 22 \\
NGC\,330  & SMC & 00 56 18.59 & $-$72 27 48.1 & 4.61 & 40 & 18.80 & 0.11 & 24 \\
NGC\,1755 & LMC & 04 55 15.53 & $-$68 12 16.8 & 3.60 & 80 & 18.35 & 0.13 & 32 \\
NGC\,1801 & LMC & 05 00 34.57 & $-$69 36 46.2 & 4.26 &315 & 18.40 & 0.16 & 28 \\
NGC\,1805 & LMC & 05 02 21.73 & $-$66 06 42.6 & 3.70 & 50 & 18.40 & 0.10 & 20 \\
NGC\,1818 & LMC & 05 04 13.42 & $-$66 26 03.1 & 4.41 & 40 & 18.40 & 0.10 & 48 \\
NGC\,1850 & LMC & 05 08 45.20 & $-$68 45 38.7 & 5.04 & 80 & 18.35 & 0.15 & 40 \\
NGC\,1856 & LMC & 05 09 30.21 & $-$69 07 44.4 & 5.07 &250 & 18.30 & 0.19 & 20 \\
NGC\,1866 & LMC & 05 13 38.77 & $-$65 27 51.4 & 4.91 &200 & 18.30 & 0.12 & 32 \\
NGC\,1868 & LMC & 05 14 35.80 & $-$63 57 14.0 & 4.34 &1000& 18.45 & 0.09 & 40 \\
NGC\,1953 & LMC & 05 25 28.07 & $-$68 50 16.5 & 4.20 &250 & 18.40 & 0.18 & 24 \\ 
NGC\,2164 & LMC & 05 58 55.89 & $-$68 30 57.5 & 4.18 &100 & 18.35 & 0.13 & 36 \\
 \hline\hline
 \end{tabular}\\
 \label{tab:parameters}
 \end{table*}

\subsection{Differential reddening}\label{subsec:reddening}
 A significant spatial variation of the reddening across the field of view results in a broadening of the stellar colors and the magnitudes in the CMD and we may need to correct the photometry for the effect of differential reddening in order to properly investigate multiple populations. 

 To estimate the differential reddening suffered by each star, we applied the method of Milone et al.\,(2012) to all the analyzed GCs.
 Briefly, we first derived the absorption coefficient in the filters available in this paper as in Milone et al.\,(2016). To do this, we simulated a synthetic spectrum, for a star with metallicity, Z=0.006, effective temperature $T_{\rm eff}=13,285$K, and gravity, $\log{g}=4.29$, by using the ATLAS12 and SYNTHE codes (Kurucz\,2005; Sbordone et al.\,2007). According to the isochrones from Georgy et al.\,(2014), the adopted atmospheric parameters correspond to a MS star of NGC\,1755 with $m_{\rm F814W}=19.1$ (see Milone et al.\,2016 for details).  The simulated spectrum, which includes the wavelength interval between 1,900 and 10,000 \AA, has been convolved over the transmission curves of the filters used in this paper by assuming the reddening law by Cardelli et al.\,(1989) to derive the following absorption coefficients: $A_{\rm F225W}$=8.25 $E(B-V)$, $A_{\rm F275W}$=6.38 $E(B-V)$, $A_{\rm F336W}$=5.16 $E(B-V)$, $A_{\rm F656N}$=2.165 $E(B-V)$, $A_{\rm F814W}$=2.04 $E(B-V)$. Moreover, we obtained $A_{\rm B}=$ 4.27 $E(B-V)$ and $A_{\rm V}=$ 3.27 $E(B-V)$.

  To estimate the amount of differential reddening affecting each star, we first determined the direction of reddening vector in the $m_{\rm F814W}$ vs.\,$m_{\rm F336W}-m_{\rm F814W}$ plane by using the values of the absorption coefficients above, and rotated the CMD into a reference frame where the abscissa is parallel to that direction.
 We have then derived the ridge line of the red MS and selected a sample of reference red-MS stars that have high-quality photometry and astrometry according to the criteria of selection described in Sect.~\ref{sec:data}. Finally, we calculated for each star the color residuals of the 50 nearest reference stars with respect to the red-MS ridge line. We converted the median of the color residual to the best estimate for the differential reddening, $\Delta E(B-V)$, while the corresponding error has been derived from the ratio between the rms of the 50 color-residual measurements and the square root of 49.

 The procedure for the determination of the differential reddening across the field of view of each cluster is based on the $m_{\rm F814W}$ vs.\,$m_{\rm F336W}-m_{\rm F814W}$ CMD only in most clusters. As exceptions, in NGC\,330, NGC\,1805, NGC\,1818, and NGC\,2164 we used the average $\Delta E(B-V)$ values inferred from the $m_{\rm F814W}$ vs.\,$m_{\rm F336W}-m_{\rm F814W}$ and $m_{\rm F814W}$ vs.\,$m_{\rm F225W}-m_{\rm F814W}$ CMDs. In the case of NGC\,1850 we combined information from the $m_{\rm F814W}$ vs.\,$m_{\rm F336W}-m_{\rm F814W}$ and $m_{\rm F814W}$ vs.\,$m_{\rm F275W}-m_{\rm F814W}$ CMDs.

 The inferred values of $\Delta E(B-V)$ associated to the stars of most clusters, namely NGC\,294, NGC\,330, NGC\,1755, NGC\,1801, NGC\,1805, NGC\,1818, NGC\,1866, NGC\,1868, NGC\,1953, and NGC\,2164 are typically $\sim$0.003 mag and are comparable to the measurement errors which are around 0.003 mag. This fact indicates that the reddening is almost constant across the field of view and that any reddening variation is either comparable or smaller than our photometric uncertainties. In these cases the CMD corrected for differential reddening is almost indistinguishable from the original CMD and we will use the latter to analyze the multiple stellar populations.
 KRON\,34, NGC\,1850 and NGC\,1856 are the only clusters with significant variations of reddening across the field of view (see also Correnti et al.\,2015, 2017; Milone et al.\,2015 for NGC\,1850 and NGC\,1856).  In the following, we will analyze the differential-reddening corrected CMDs of these  three clusters. 

 \subsection{Artificial stars}
 Artificial-star (AS) tests have been used to estimate the photometric errors and the completeness level of our sample and have been run by using  the procedure described by Anderson et al.\,(2008).
  Briefly, we first produced for each cluster a list of 300,000 stars with instrumental magnitudes from $-4.0$, which is below the detection limit of the data, to the saturation limit in the F814W band and calculated the corresponding F336W, F656N magnitudes by using the fiducial line of the MS and the MSTO derived from the observed CMD. In the cases of NGC\,330, NGC\,1805, NGC\,1818, and NGC\,2164 we included in the AS list only the F225W magnitude and in the case of NGC\,1850 we included the stellar magnitudes in the F275W band.
   At this stage we used the instrumental magnitudes that are defined as $-2.5 log_{10}{\rm (flux)}$, where the flux is provided in photo-electrons and is recorded to the reference exposure.  
 
   ASs have been generated with a flat luminosity function in the F814W band. We included in the list the stellar coordinates in the reference frame and assumed for ASs a spatial distribution that resembles the observed spatial distribution of stars brighter than $m_{\rm F814W}=20.0$. We emphasize here that, since ASs are used to estimate completeness and photometric errors the adopted luminosity function and spatial distribution do not affect the results.
   The AS test have been performed for one star at time and entirely in the software so that the ASs will never interfere with each other.
   For each AS in the input list, we added the star to each exposure with the appropriate flux and position and measured it by following the same method used for real stars.
   We considered a star as recovered if the input and the output position differ by less than 0.5 pixel and the fluxes by less than 0.75 mag.
   Moreover, we included in our analysis only ASs that pass the criteria of selection that we adopted for real stars.

   Photometric errors depend on stellar luminosity and on crowding.
   To infer the typical uncertainty associated to the magnitude of a star, we selected all the ASs with the same luminosity, within 0.1 mag, and with the same radial distance from the cluster center, within 200 WFC3/UVES pixels. The error is estimated as the 68.27$^{\rm th}$ percentile of the distribution of the absolute values of the difference between the measured magnitude and the input magnitude of the selected stars.

To calculate the completeness value associated to each star, we divided the WFC3/UVIS field of view into 5 concentric annuli centered on the cluster centre and, within each of them, we analyzed the ASs in 15 magnitude intervals, from $-4$ to the saturation level. We derived the average completeness corresponding to each of these grid points by calculating the fraction of the recovered to input artificial stars within that bin. The completeness value associated to each star has been derived by linearly  interpolating among the grid points.

\section{The color-magnitude diagrams}
\label{sec:cmd}
\begin{centering} 
\begin{figure*} 
 \includegraphics[width=12.1cm]{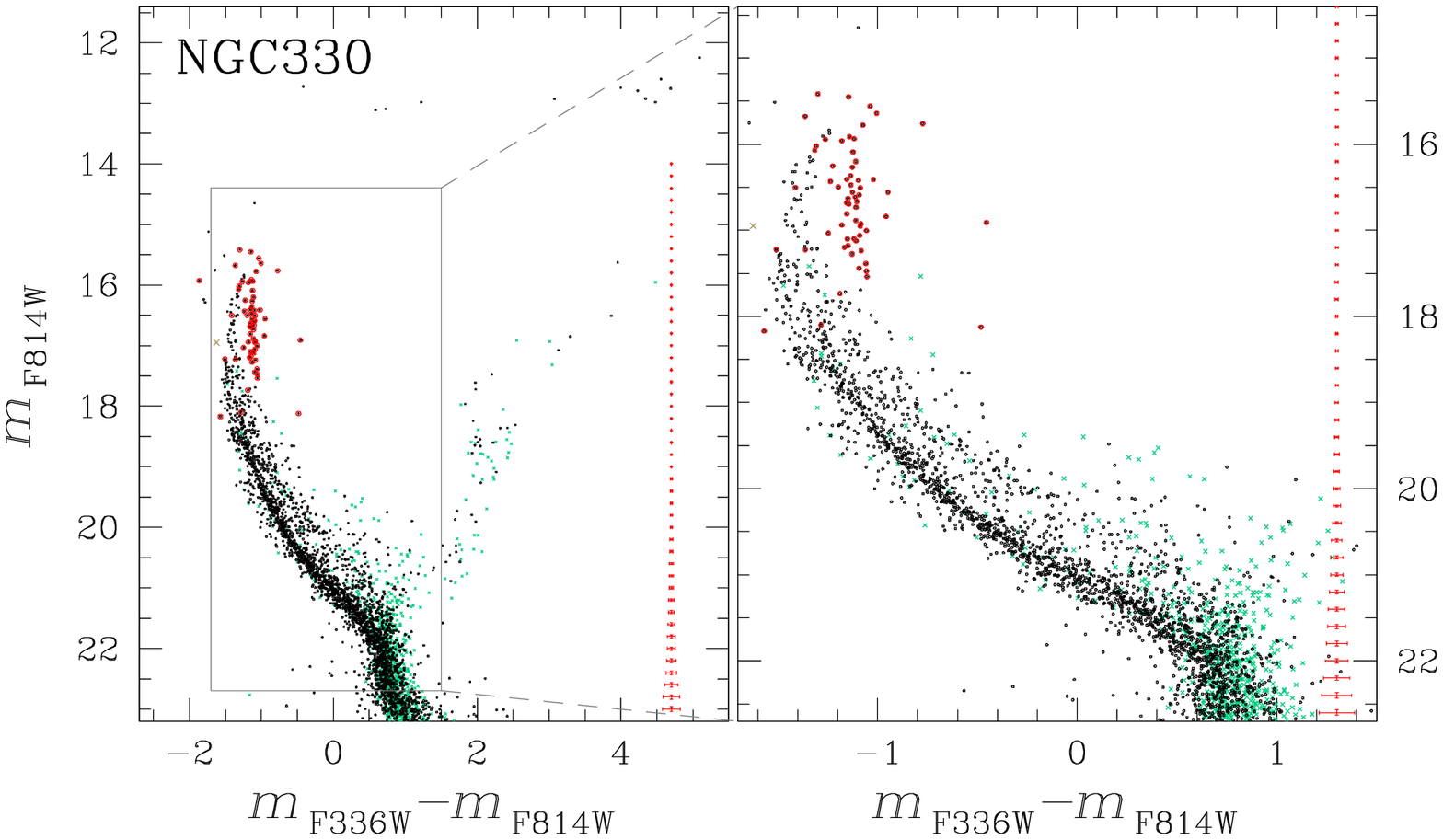}
 \includegraphics[width=5.418cm]{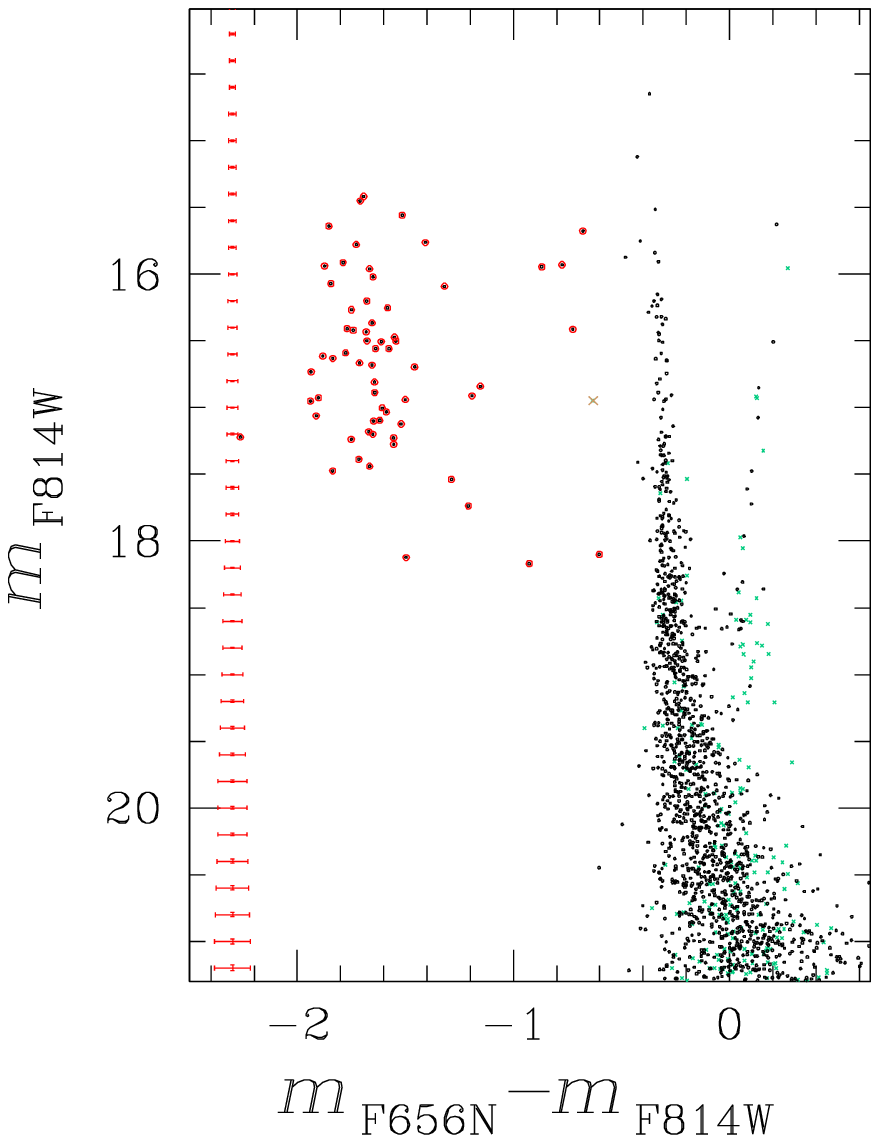}
  \includegraphics[width=12.1cm]{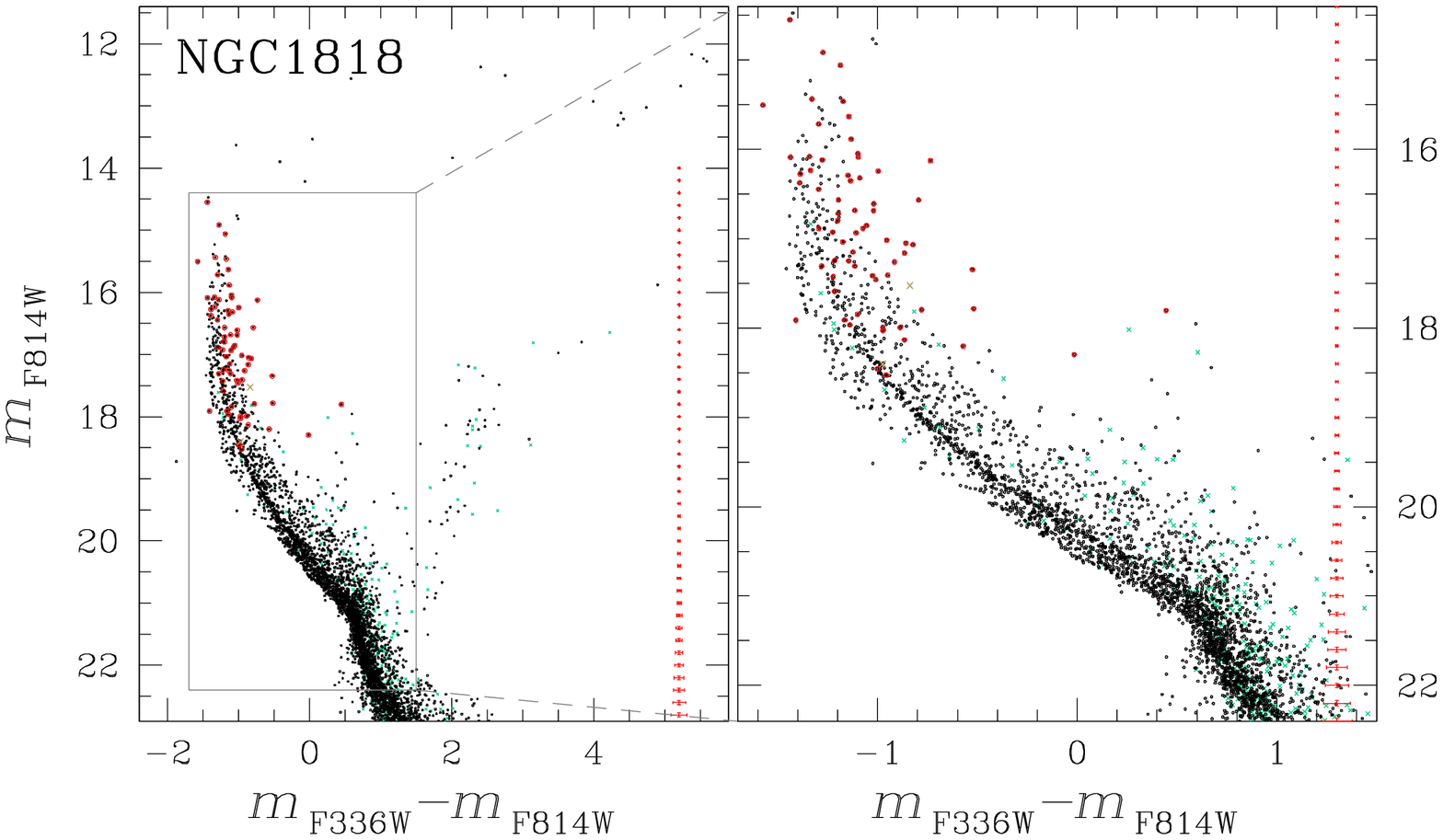}
 \includegraphics[width=5.418cm]{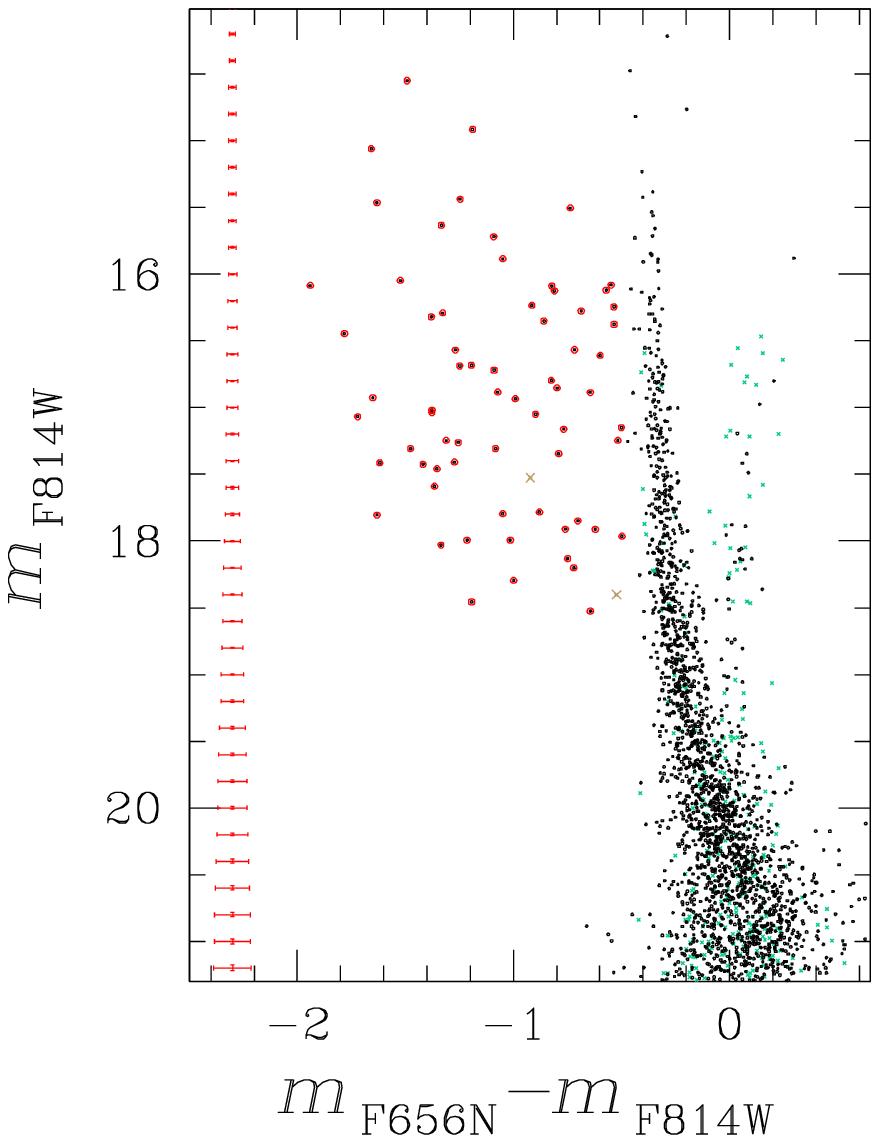}
  \includegraphics[width=12.1cm]{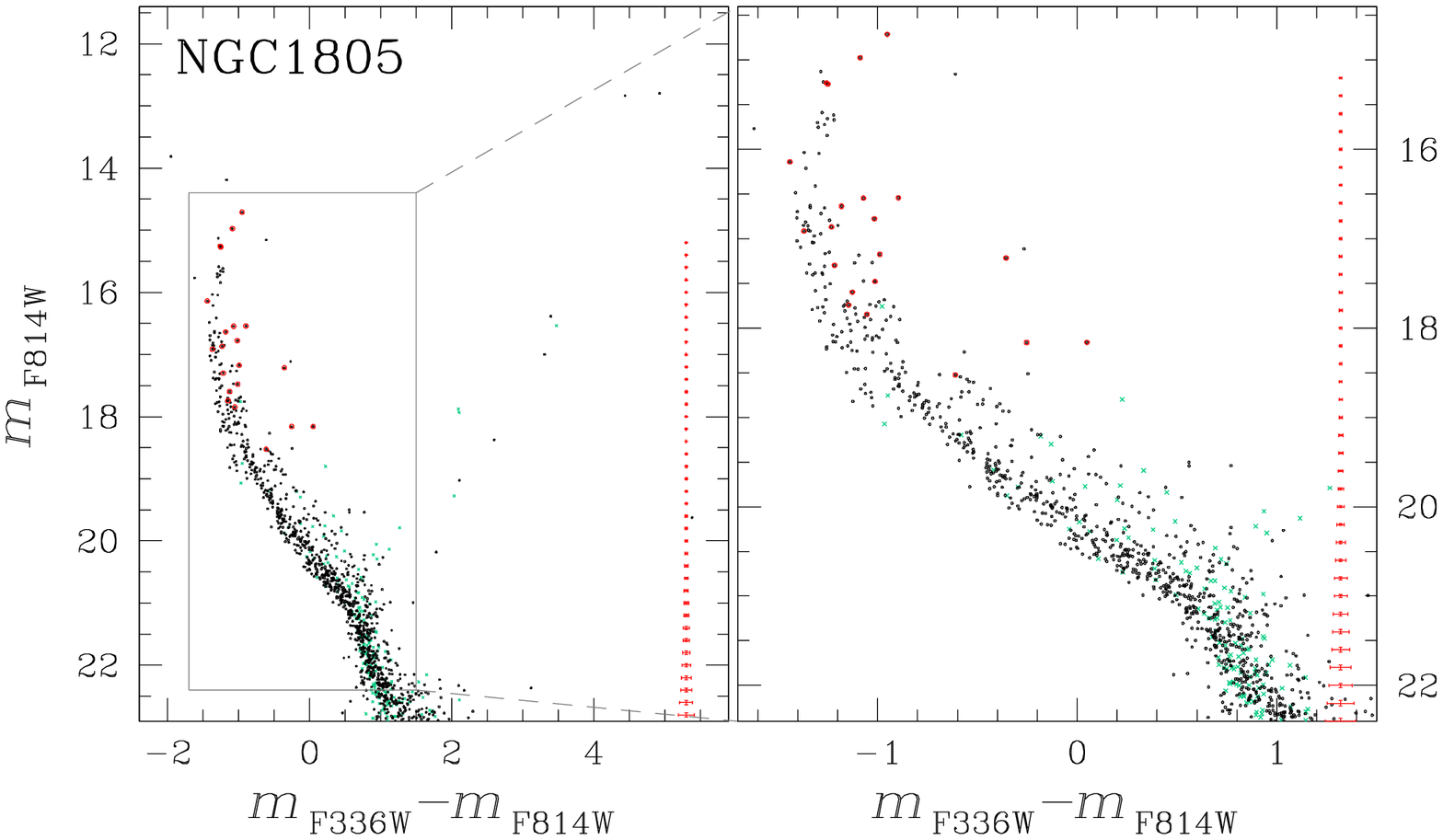}
 \includegraphics[width=5.418cm]{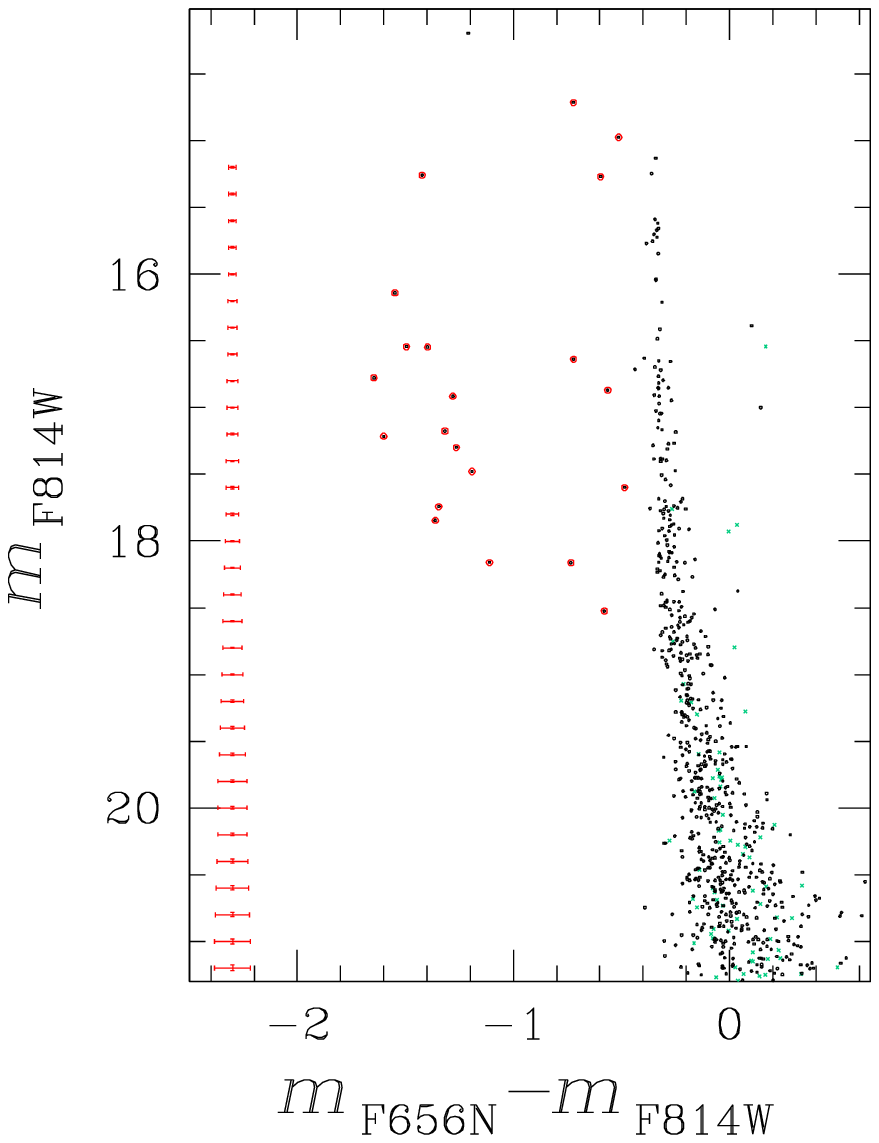}
 \caption{Collection of CMDs for stars in the field of view of NGC\,330, NGC\,1818, and NGC\,1805. Left panels show the $m_{\rm F814W}$ vs.\,$m_{\rm F336W}-m_{\rm F814W}$ CMD for stars in the cluster field (black points) and in the reference field (aqua crosses). Middle panels are a zoom of the left-panel CMDs around the upper MS and the MSTO.  Right panels plot $m_{\rm F814W}$ against $m_{\rm F656N}-m_{\rm F814W}$ for stars in the cluster field (black points) and in the reference field (aqua crosses). Candidate H-$\alpha$ emitters in the cluster field and in the reference field, have been selected in the right-panel CMDs and are represented with red circles and brown crosses, respectively, in all the diagrams of this figure.  The clusters are sorted in order of increasing age.} 
 \label{fig:cmd1} 
\end{figure*} 
\end{centering}

Figures~\ref{fig:cmd1}-\ref{fig:cmd4} provide a collection of CMDs for stars in the cluster field (black points) and in the reference field (aqua crosses). Specifically, the left and the right panels show the $m_{\rm F814W}$ vs.\,$m_{\rm F336W}-m_{\rm F814W}$ and $m_{\rm F814W}$ vs.\,$m_{\rm F656N}-m_{\rm F814W}$ diagrams, respectively, while middle panels are zooms of the CMDs plotted on the left. 
In the cases of NGC\,1755 and NGC\,1866, for which there are no F656N data available, we plot the $m_{\rm F814W}$ vs.\,$m_{\rm F336W}-m_{\rm F814W}$ CMD only. The clusters are sorted by age from youngest ($\sim$40\,Myr, NGC\,330) to oldest ($\sim$1\,Gyr, NGC\,1868).

The left and middle panels show that all the clusters exhibit the eMSTO. Moreover, the upper MS, above the kink at $m_{\rm F814W} \lesssim 20.5$-$21.5$, is typically split into a blue and red sequence.
The single exception is the oldest cluster NGC 1868 (Fig.~\ref{fig:cmd4}), which does not show which does not show a split MS.  In this cluster, the faintest part of the eMSTO is only slightly brighter than the magnitude of the MS kink.

All the clusters exhibit a well-populated sequence of binary stars that is clearly visible on the red side of the red MS, and will be the subject of a forthcoming paper of this series. In this context, NGC\,330 is an intriguing case where a populous sequence of binaries seems to cross the red MS around $m_{\rm F814W} \sim 18.5$.

We plot in the right panels of Figs.~\ref{fig:cmd1}-\ref{fig:cmd4} the $m_{\rm F814W}$ vs.\,$m_{\rm F656N}-m_{\rm F814W}$ CMDs for stars in the cluster field, and notice that both the MS and the MSTO define a nearly-vertical sequence. The CMDs of the five youngest clusters, namely NGC\,330, NGC\,1818, NGC\,1805, NGC\,1850, and NGC\,2164, also display a cloud of bright stars that are significantly bluer than the bulk of MS stars and which span a wide range of $m_{\rm F656N}-m_{\rm F814W}$ colors. These stars, which populate the upper MS only and disappear below $m_{\rm F814W} \sim 18.5$, are marked with red crosses in Figs.~\ref{fig:cmd1}-\ref{fig:cmd4}. 
This reveals that the stars with an excess of F656N luminosity mostly populate the eMSTO and the red MS in the $m_{\rm F814W}$ vs.\,$m_{\rm F336W}-m_{\rm F814W}$ CMD.
Their $m_{\rm F656N}-m_{\rm F814W}$ and $m_{\rm F336W}-m_{\rm F814W}$ color distributions dramatically change from one cluster to another.

Our results are consistent with previous findings by Keller et al.\,(2000) based on Wide Field Planetary Camera 2 F160BW, F555W, and F656N imaging of NGC\,330 in the SMC, and three young LMC clusters: NGC\,1818, NGC\,2004, and NGC\,2100. 
 Keller and collaborators found that these clusters host a large fraction of Be stars and discovered that Be stars have redder F160BW$-$F555W colors than the majority of MS stars.

\begin{centering} 
\begin{figure*} 
  \includegraphics[width=12.1cm]{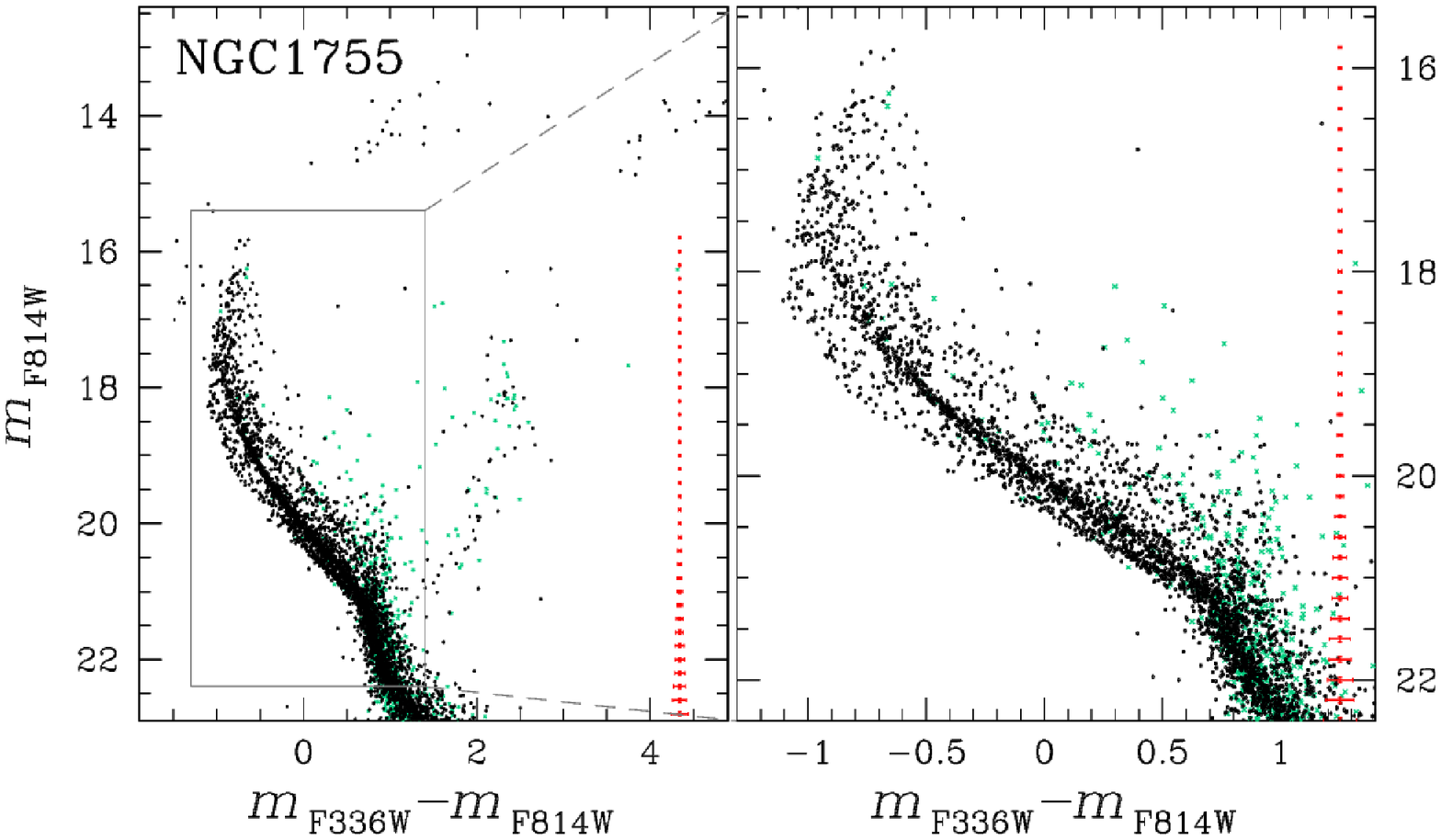}
 \includegraphics[width=5.418cm]{vuoto.ps}
  \includegraphics[width=12.1cm]{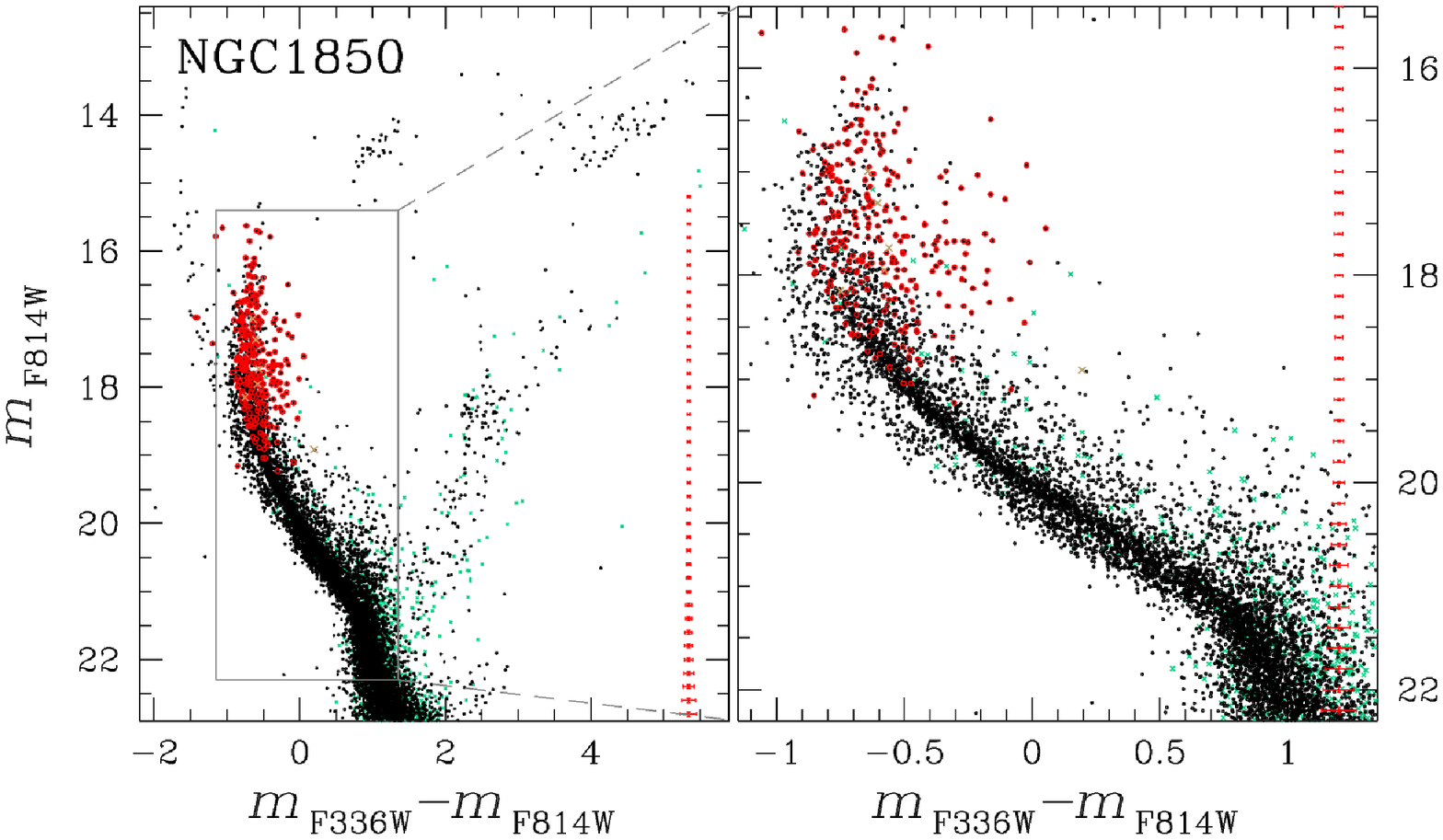}
 \includegraphics[width=5.418cm]{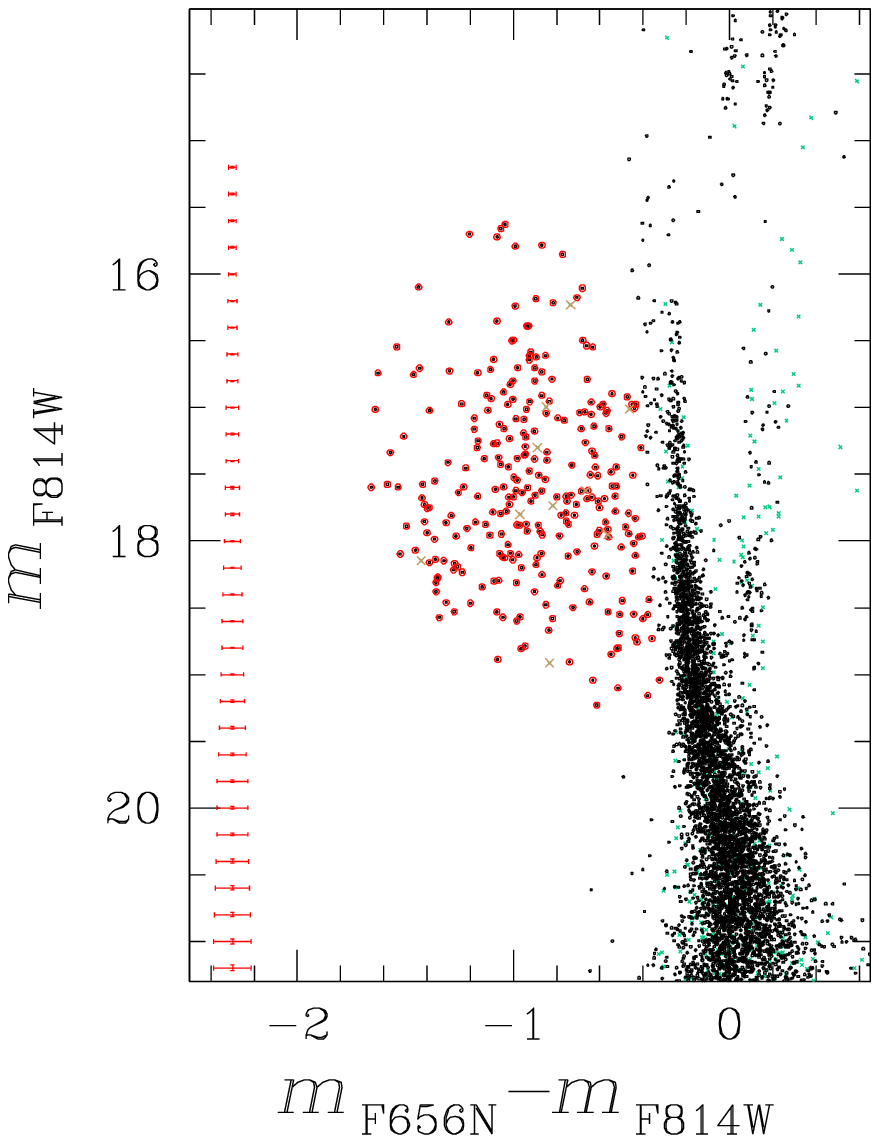}
  \includegraphics[width=12.1cm]{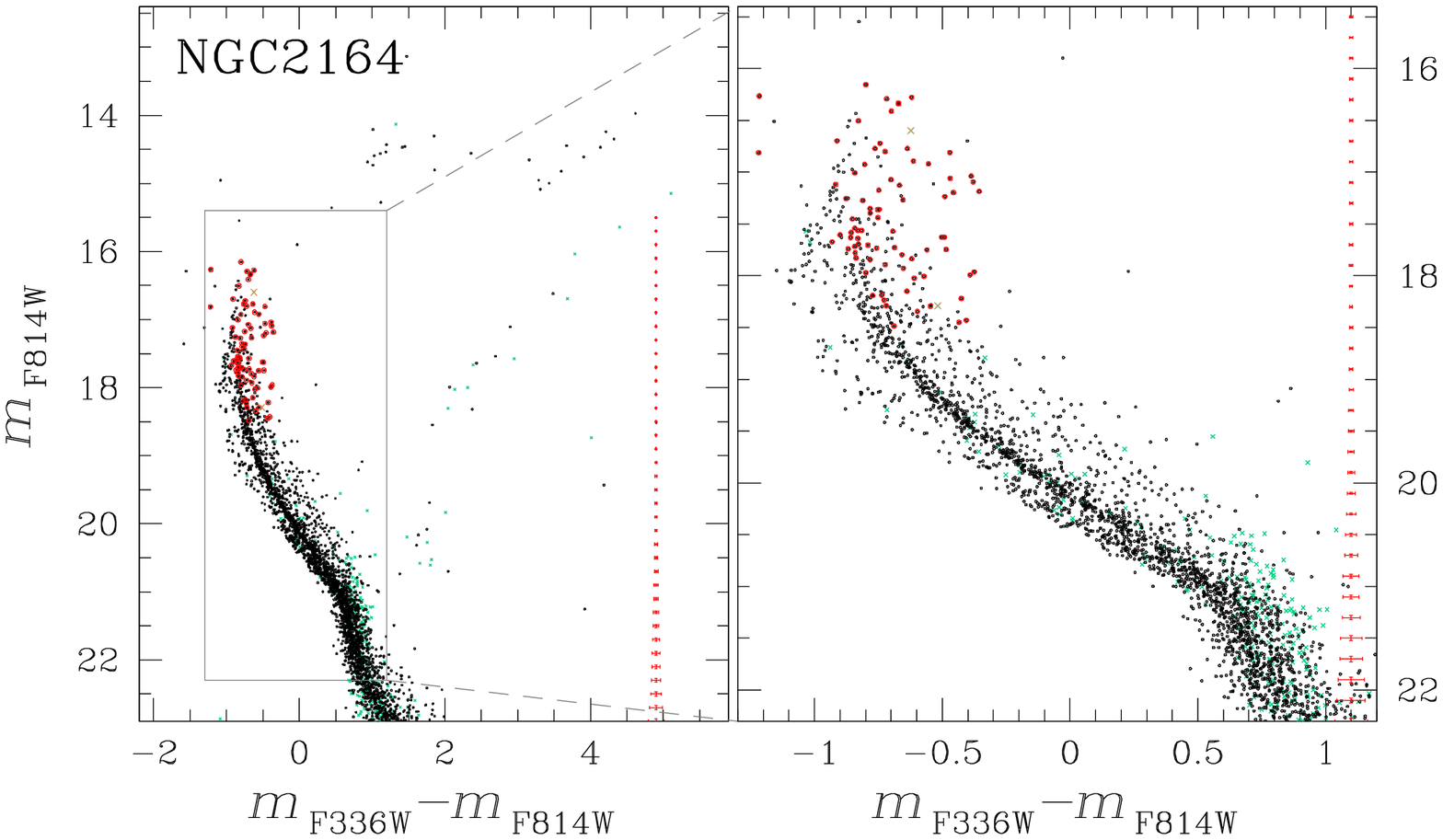}
 \includegraphics[width=5.418cm]{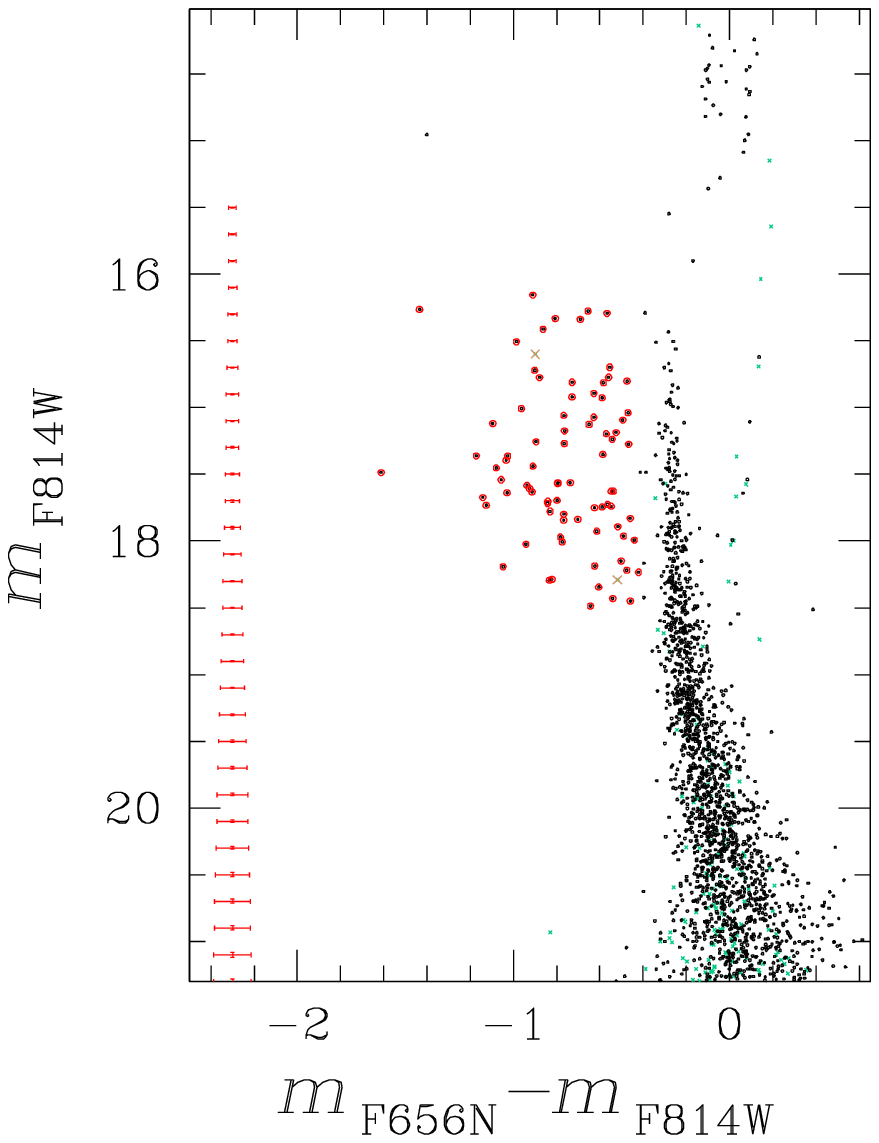}
 \caption{As in Fig.~\ref{fig:cmd1} but for NGC\,1755, NGC\,1850, and NGC\,2164. Photometry in the F656N band is not available for NGC\,1755.} 
 \label{fig:cmd2} 
\end{figure*} 
\end{centering} 

\begin{centering} 
\begin{figure*} 
  \includegraphics[width=12.1cm]{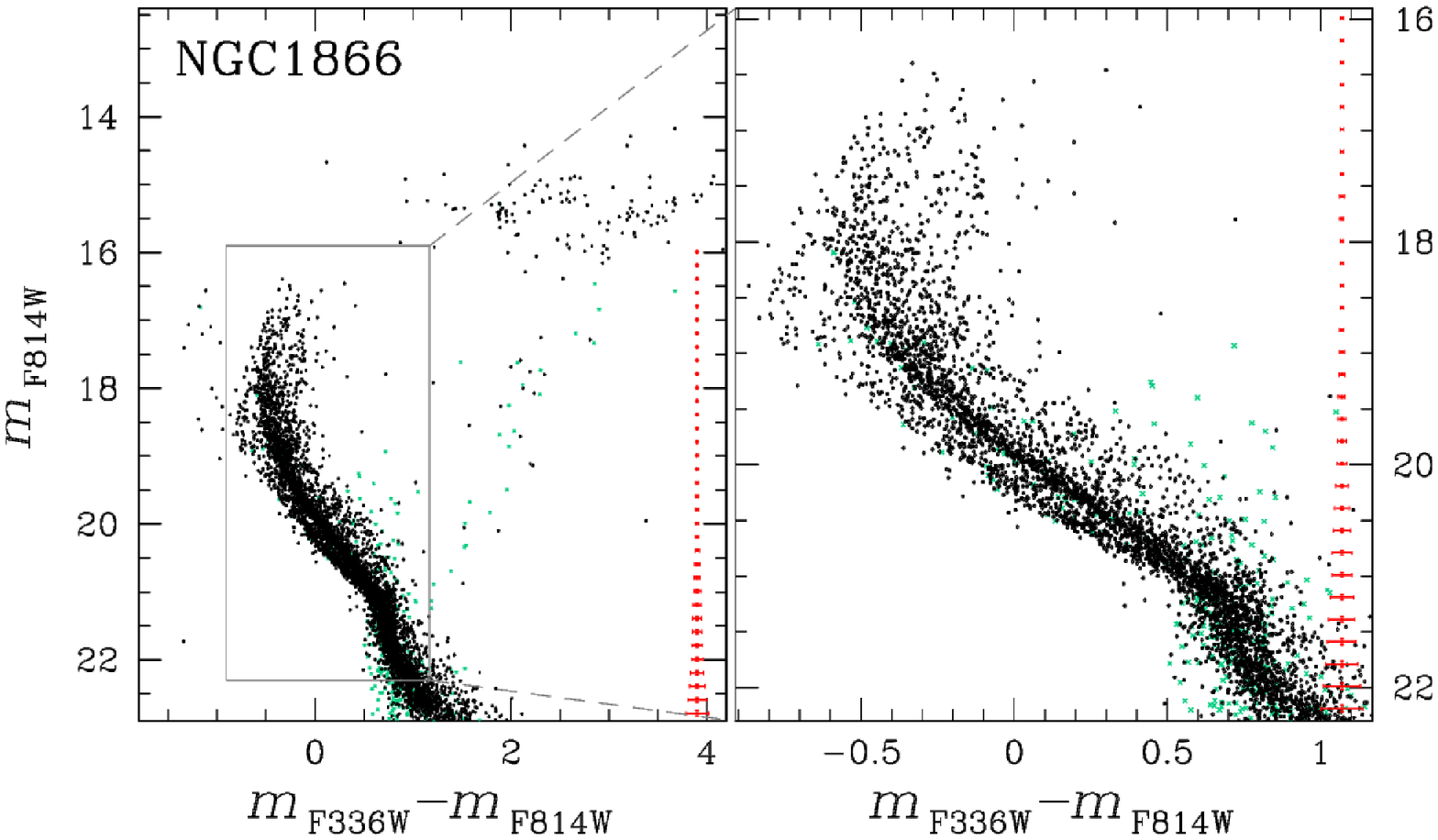}
  \includegraphics[width=5.418cm]{vuoto.ps}
  \includegraphics[width=12.1cm]{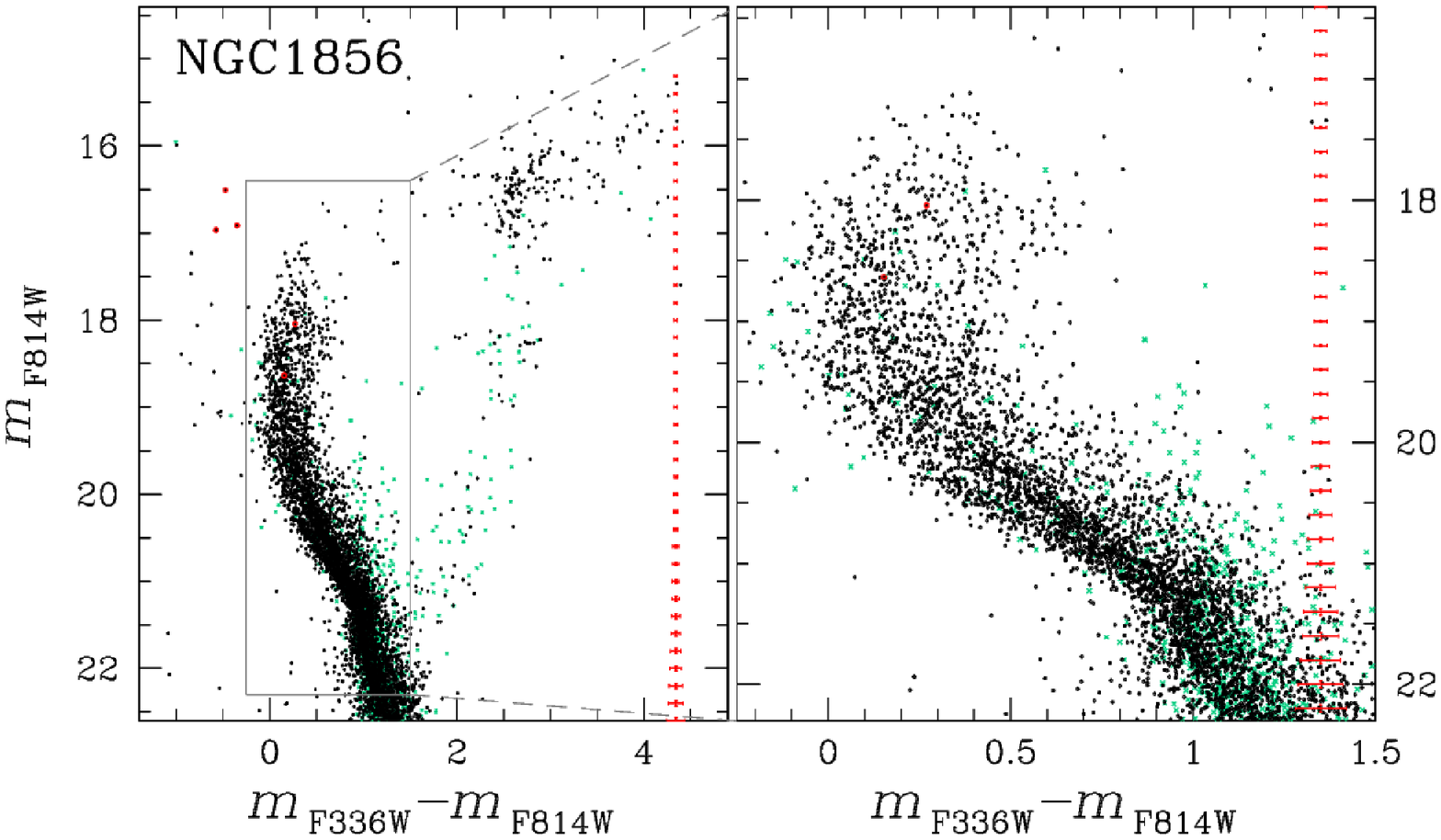}
 \includegraphics[width=5.418cm]{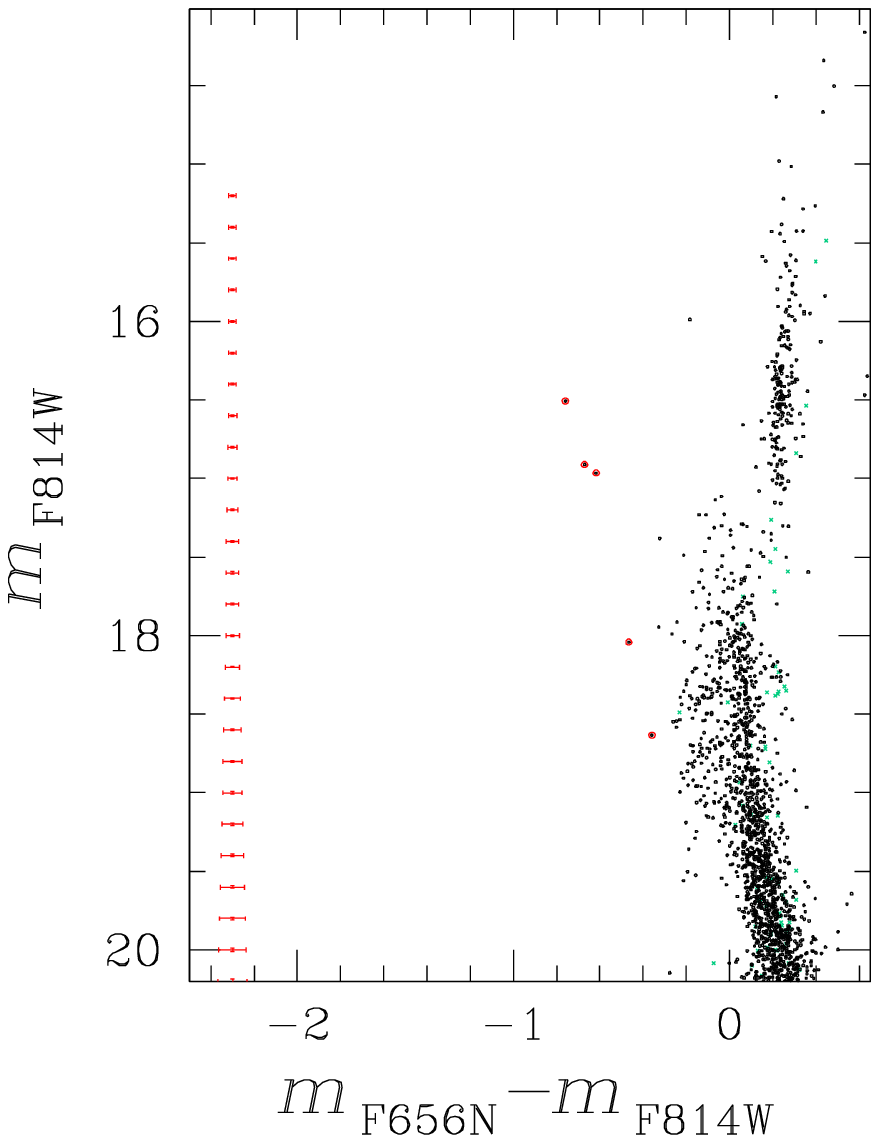}
  \includegraphics[width=12.1cm]{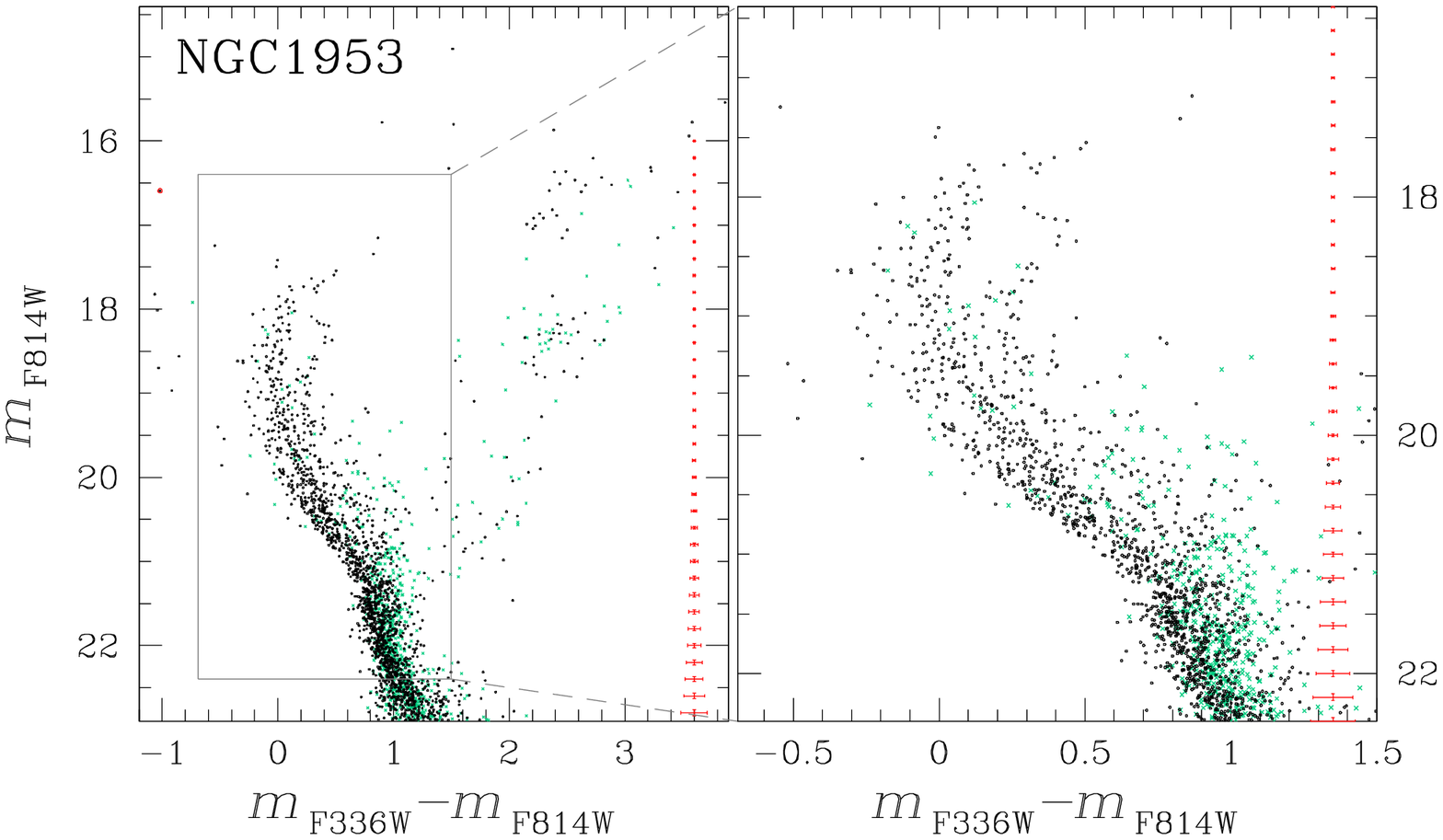}
 \includegraphics[width=5.418cm]{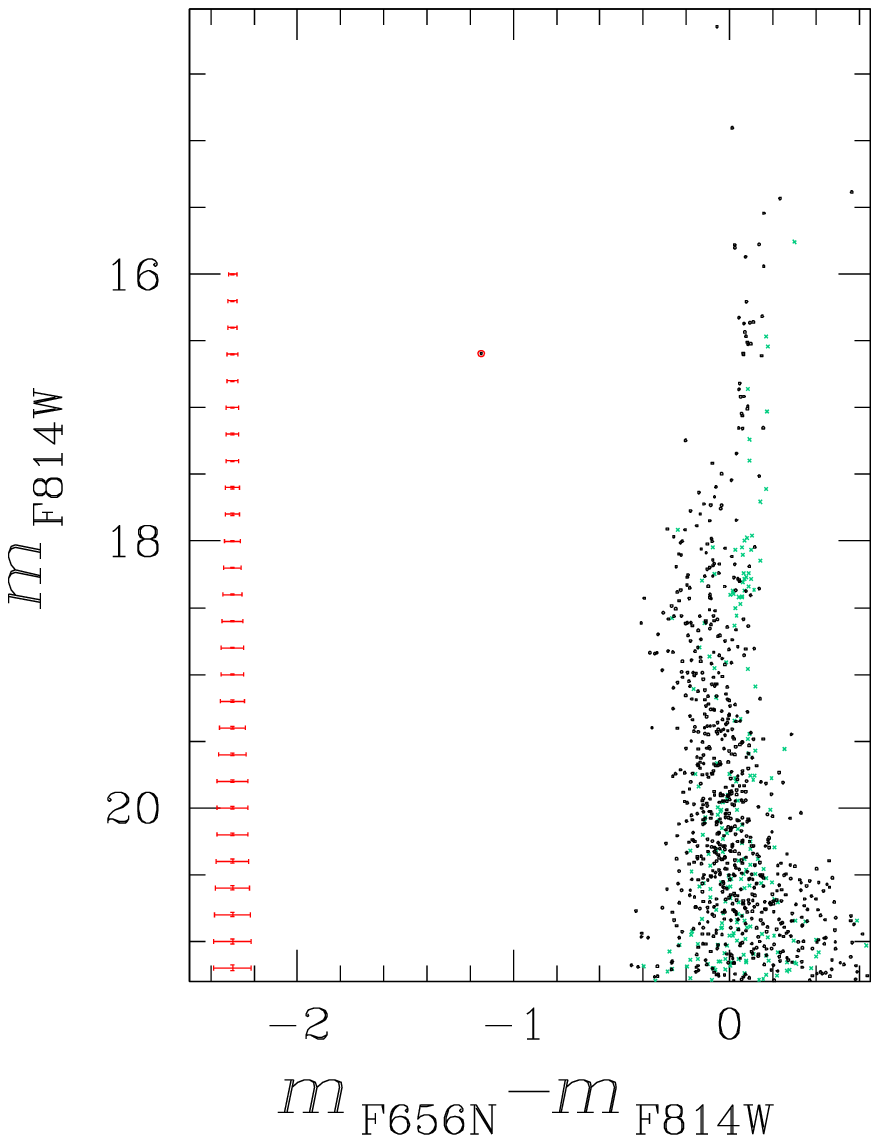}
 \caption{As in Fig.~\ref{fig:cmd1} but for NGC\,1866, NGC\,1856, and NGC\,1953. Photometry in the F656N band is not available for NGC\,1866.} 
 \label{fig:cmd3} 
\end{figure*} 
\end{centering} 

\begin{centering} 
\begin{figure*} 
 \includegraphics[width=12.1cm]{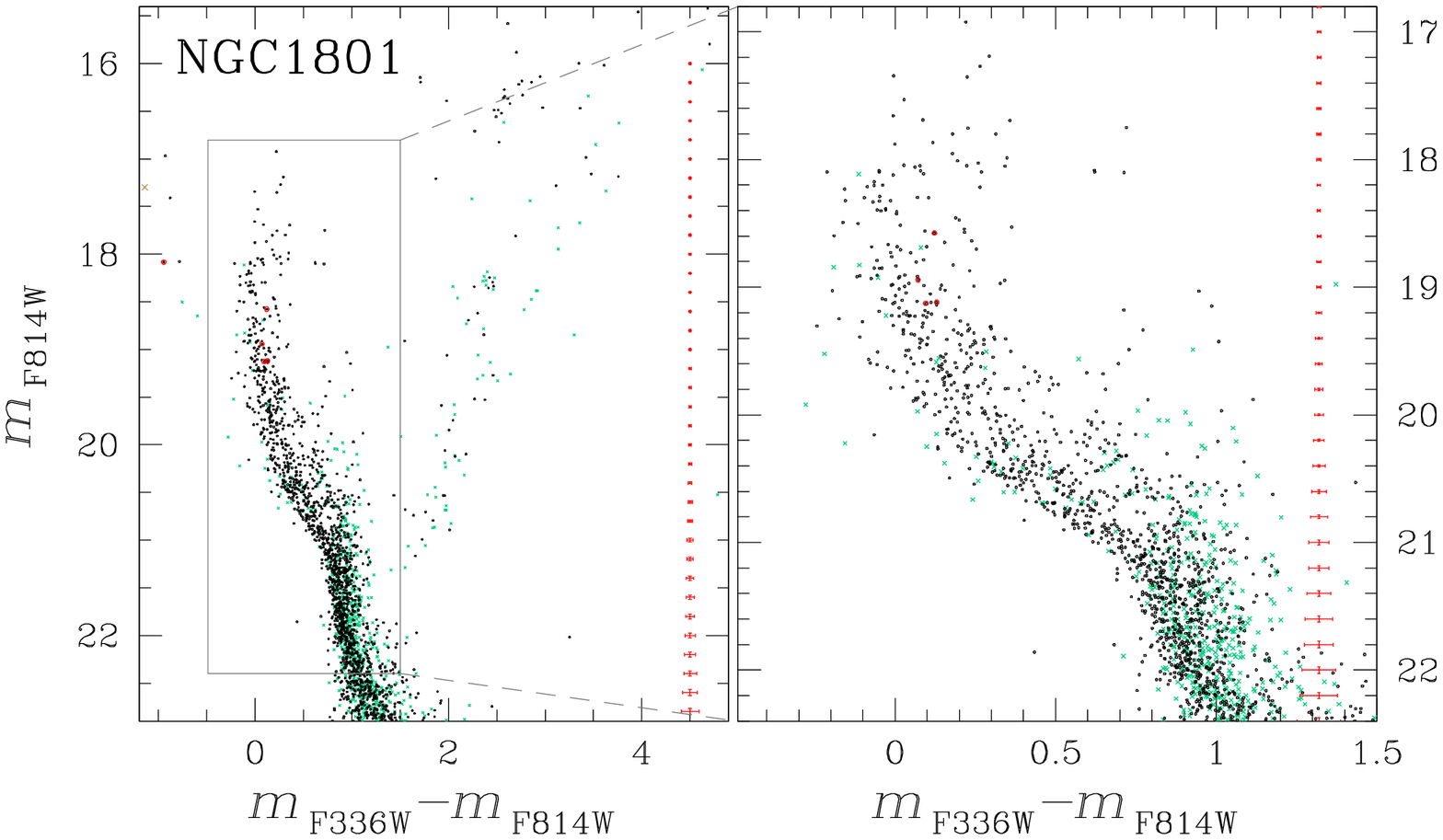}
 \includegraphics[width=5.418cm]{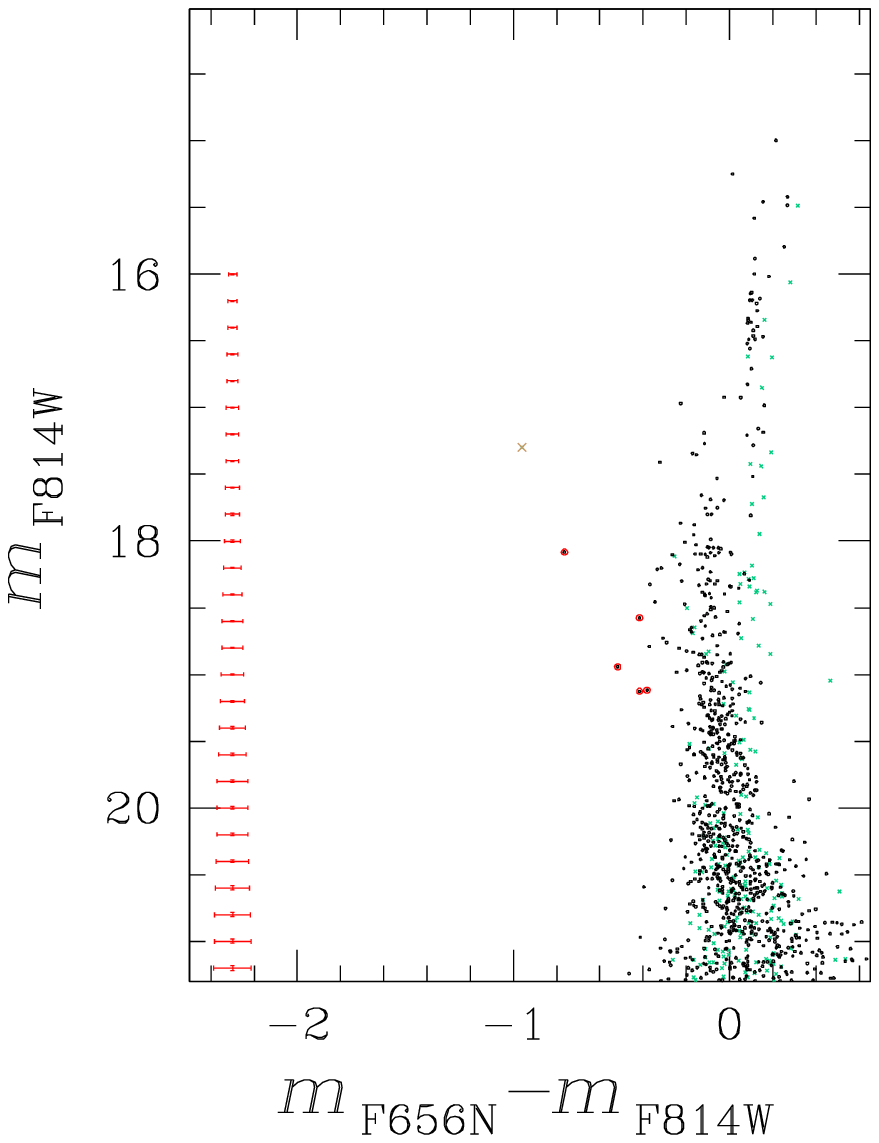}
  \includegraphics[width=12.1cm]{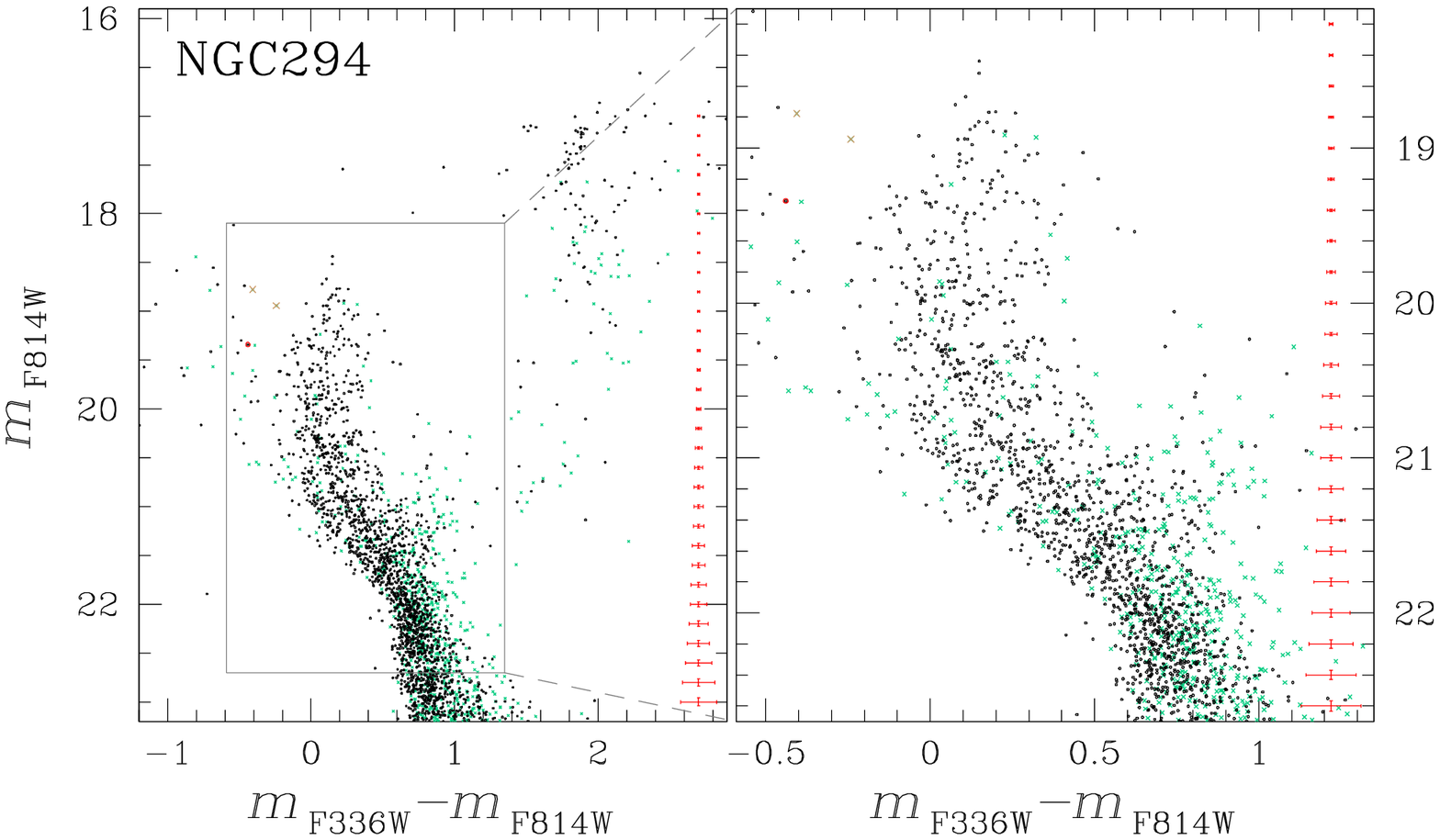}
 \includegraphics[width=5.418cm]{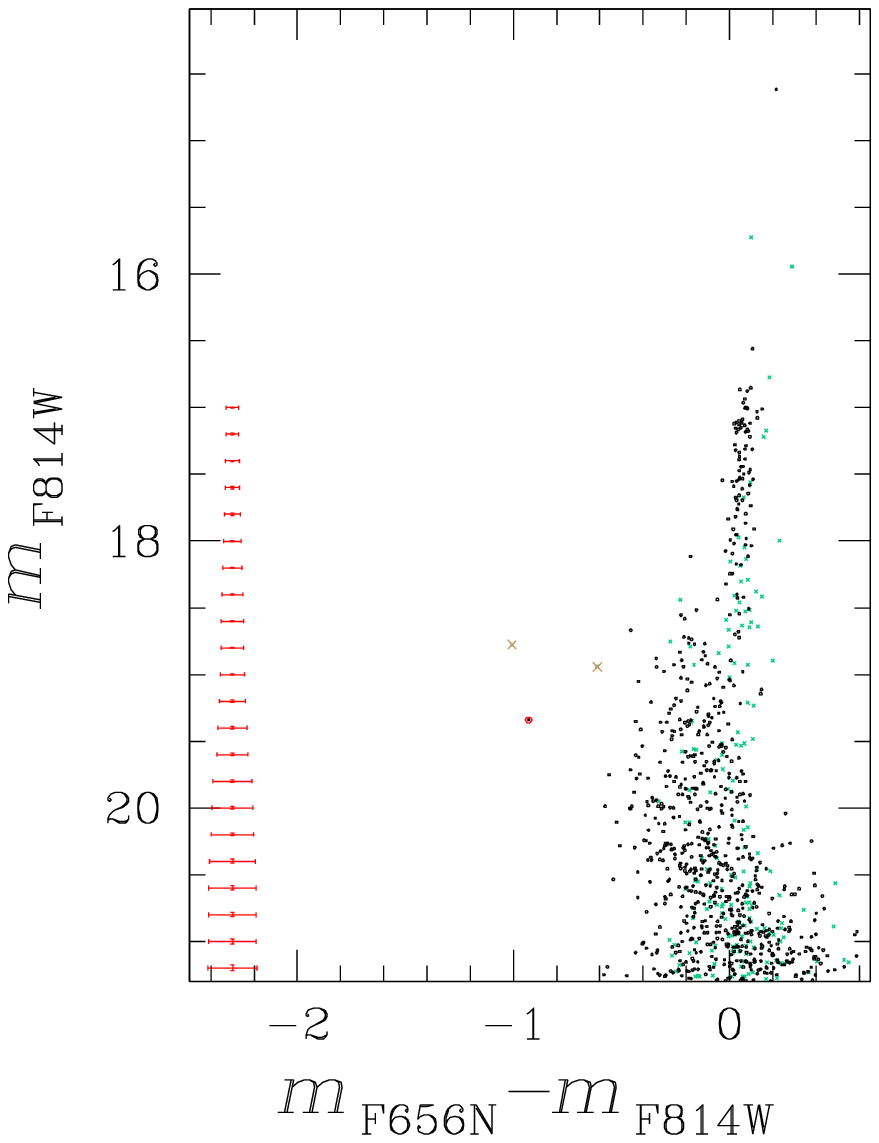}
  \includegraphics[width=12.1cm]{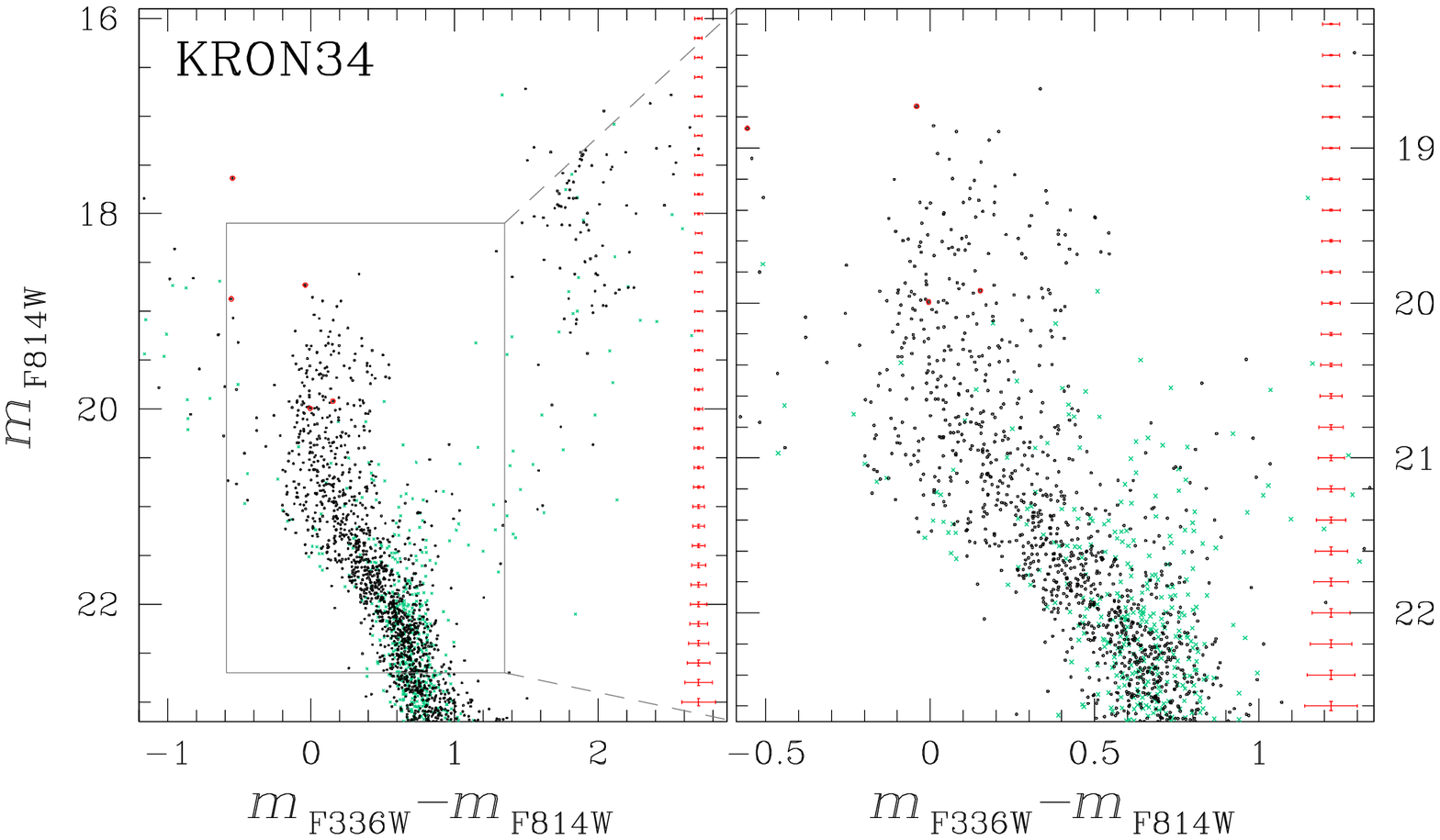}
 \includegraphics[width=5.418cm]{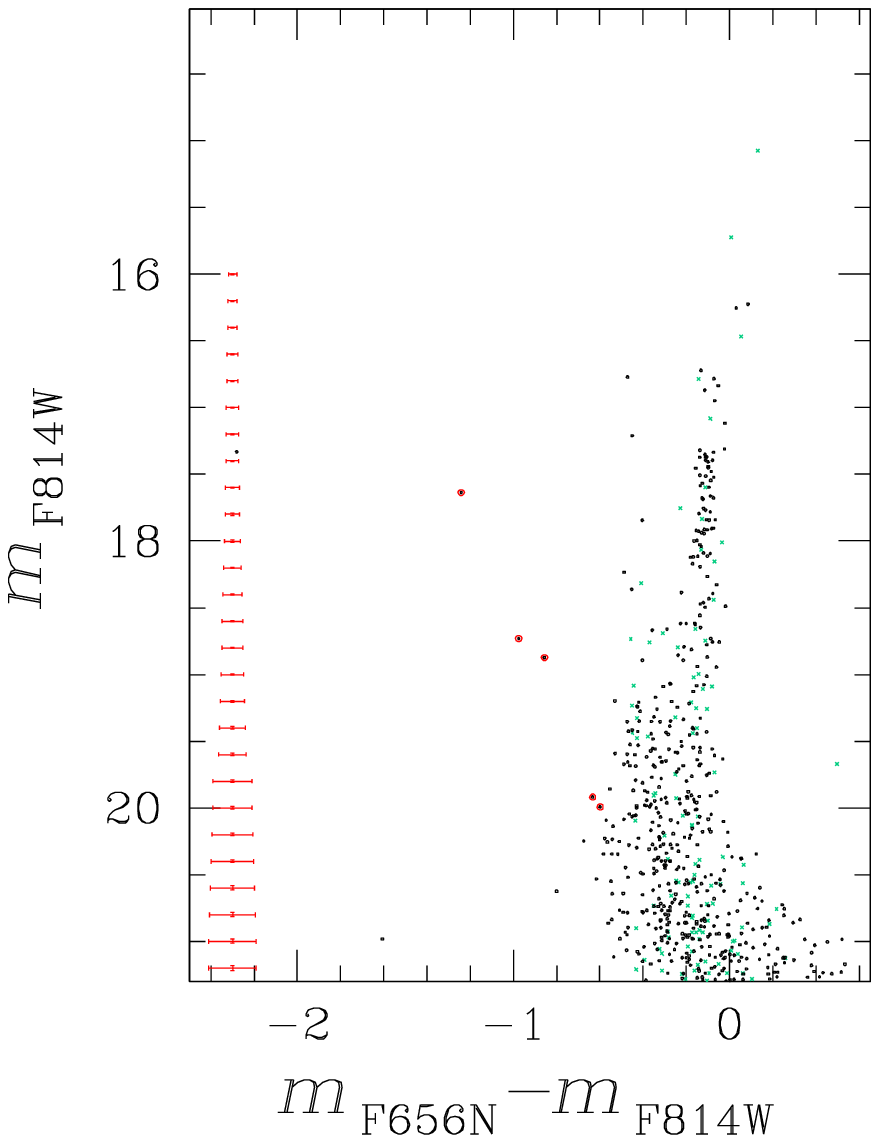}
 \caption{As in Fig.~\ref{fig:cmd1} but for NGC\,1801, NGC\,294, and KRON\,34.} 
 \label{fig:cmd4} 
\end{figure*} 
\end{centering} 

\begin{centering} 
\begin{figure*} 
  \includegraphics[width=12.1cm]{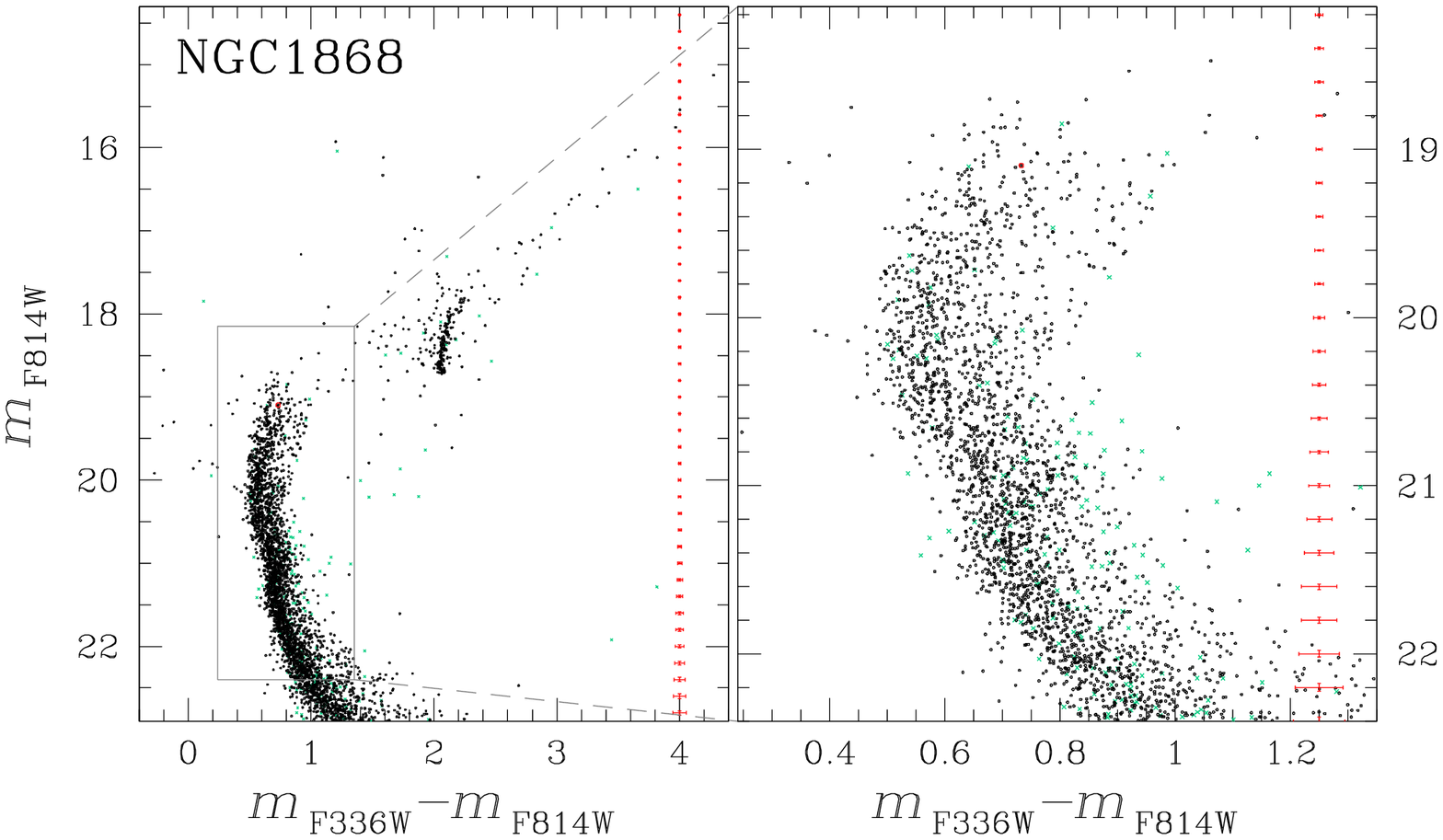}
 \includegraphics[width=5.418cm]{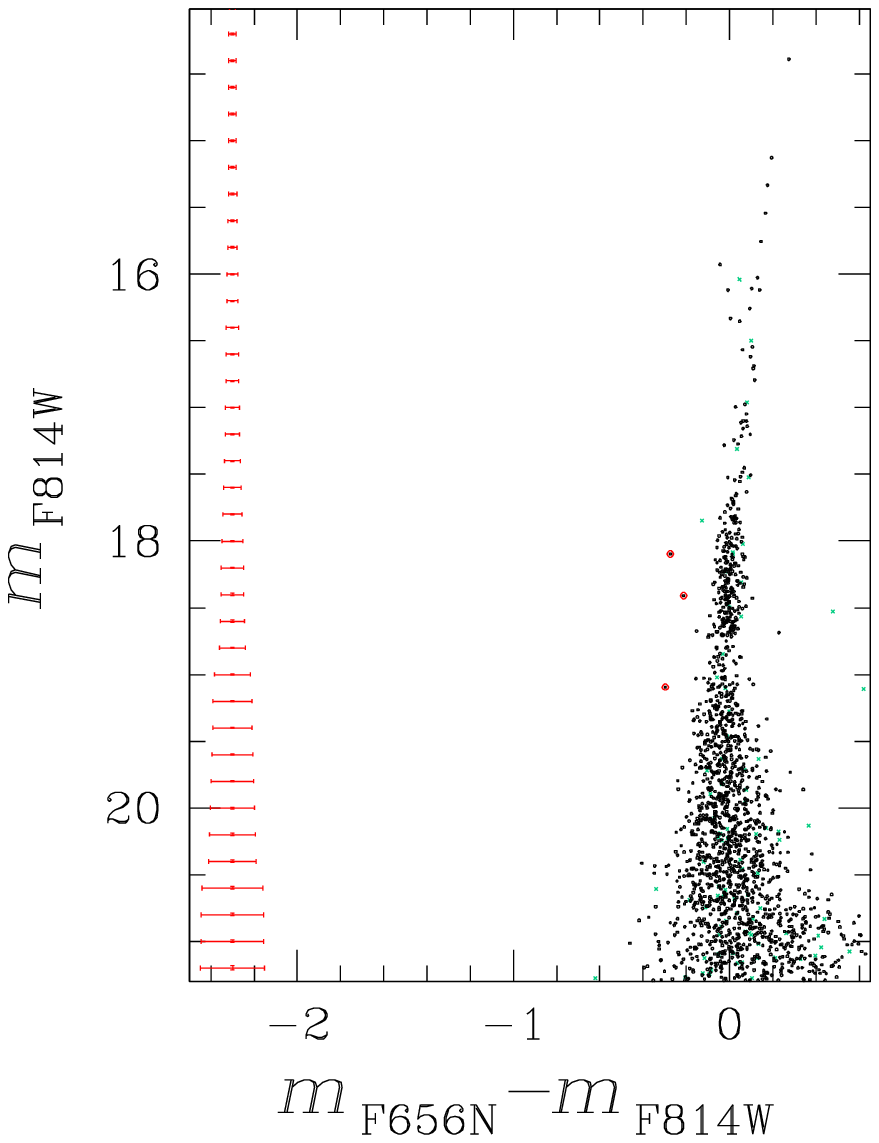}
 \caption{As in Fig.~\ref{fig:cmd1} but for NGC\,1868.} 
 \label{fig:cmd4} 
\end{figure*} 
\end{centering} 

\begin{centering} 
\begin{figure*} 
  \includegraphics[width=8.25cm]{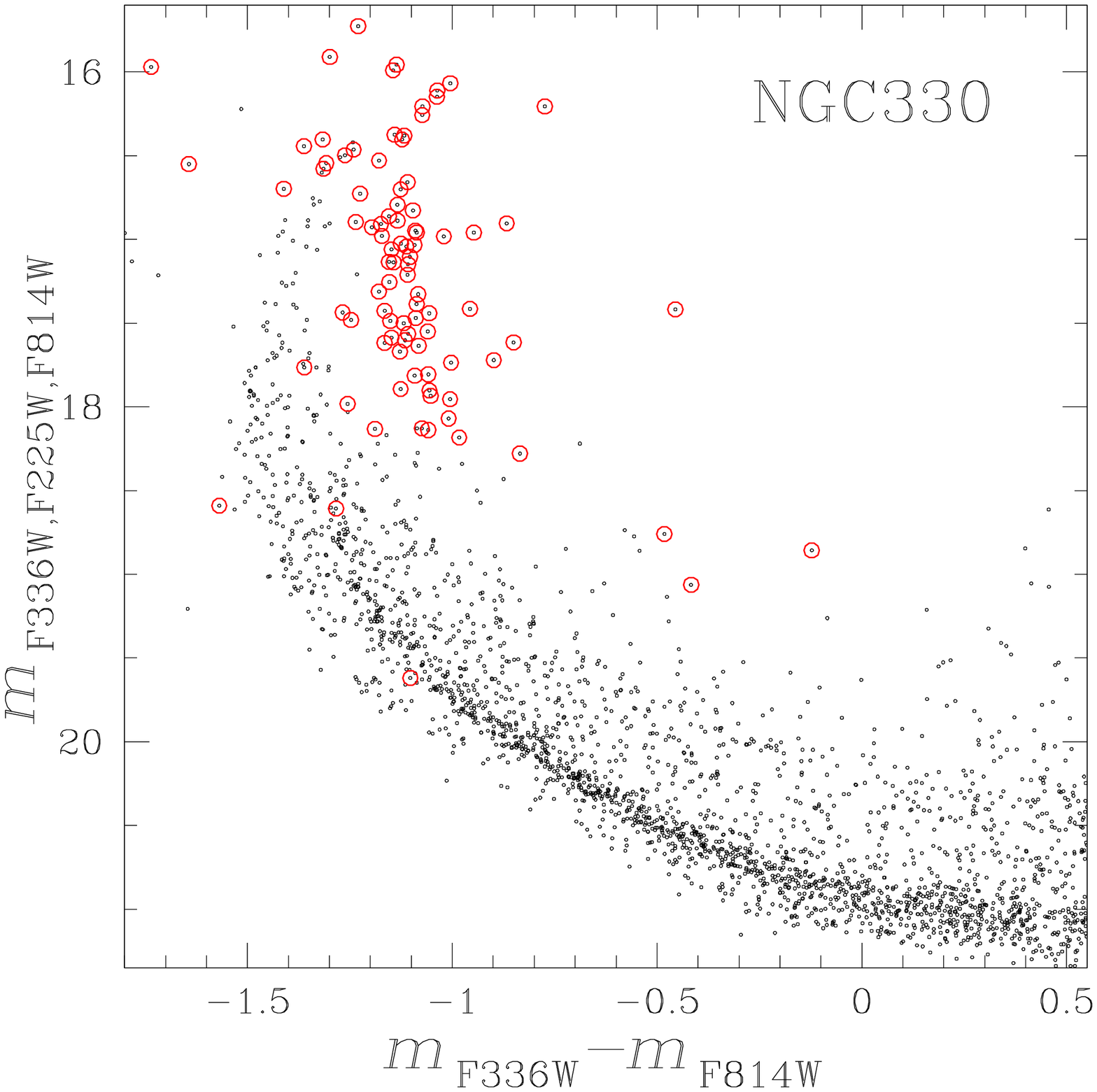}
  \includegraphics[width=8.25cm]{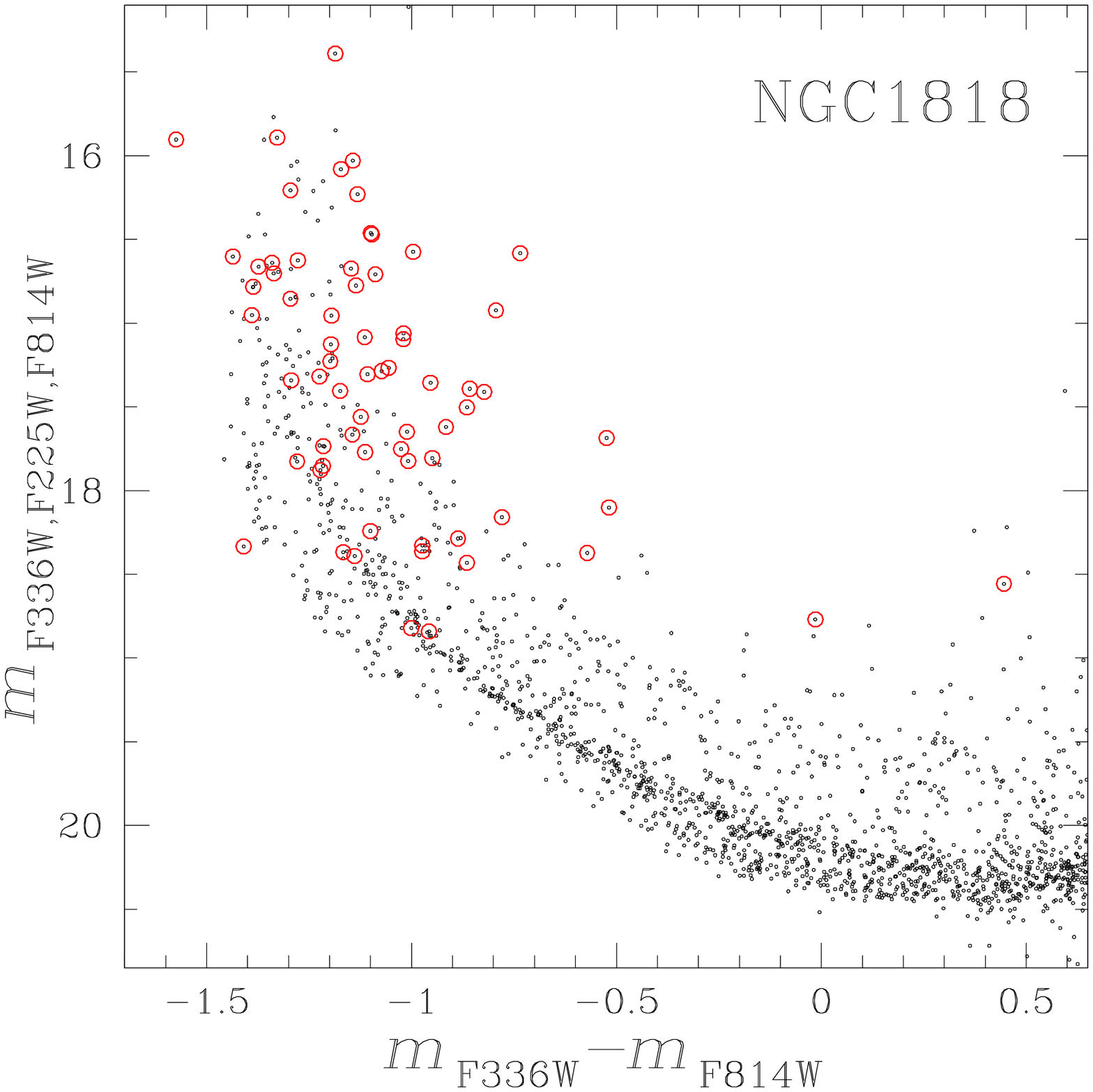}
  \includegraphics[width=8.25cm]{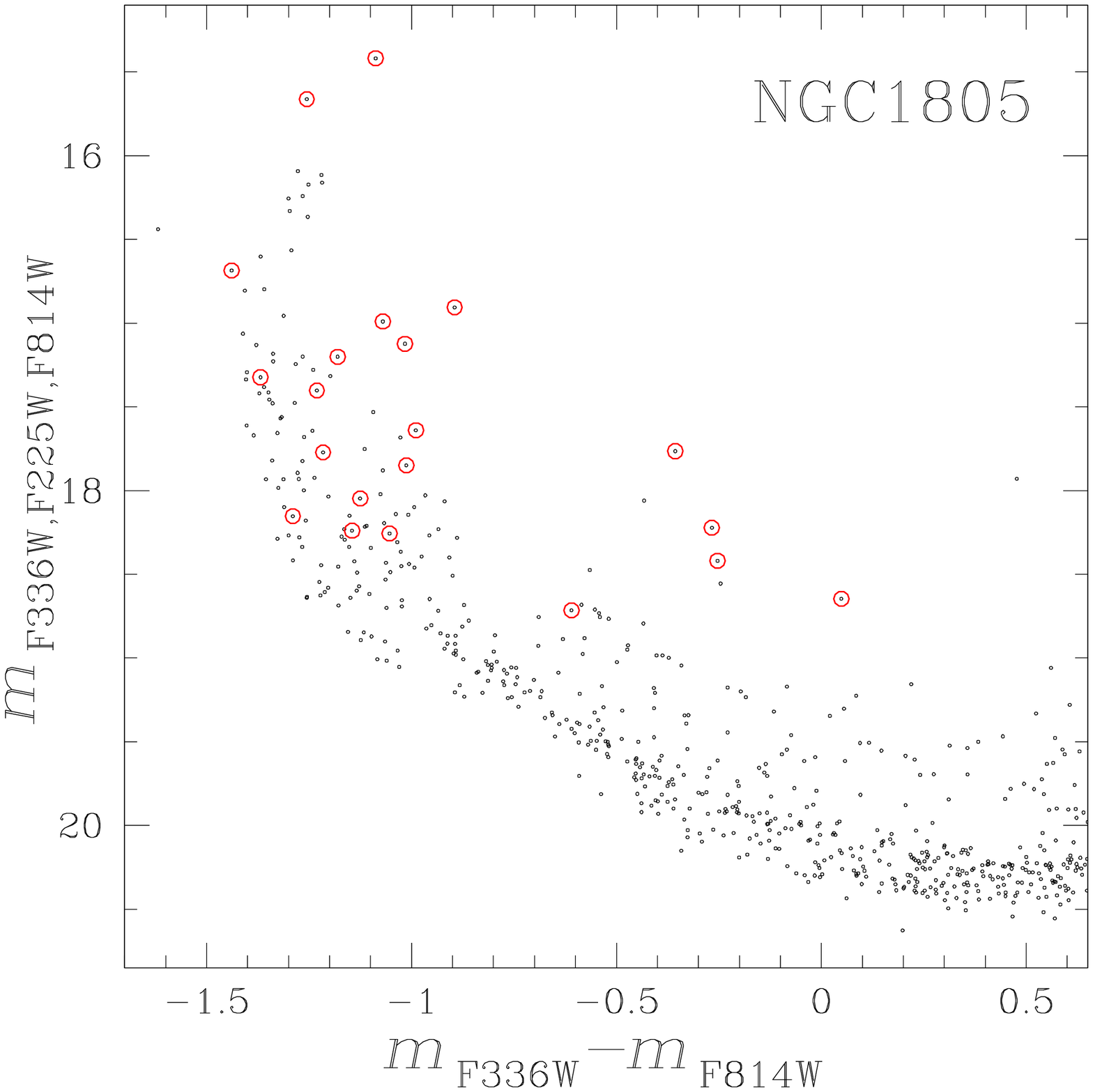}
  \includegraphics[width=8.25cm]{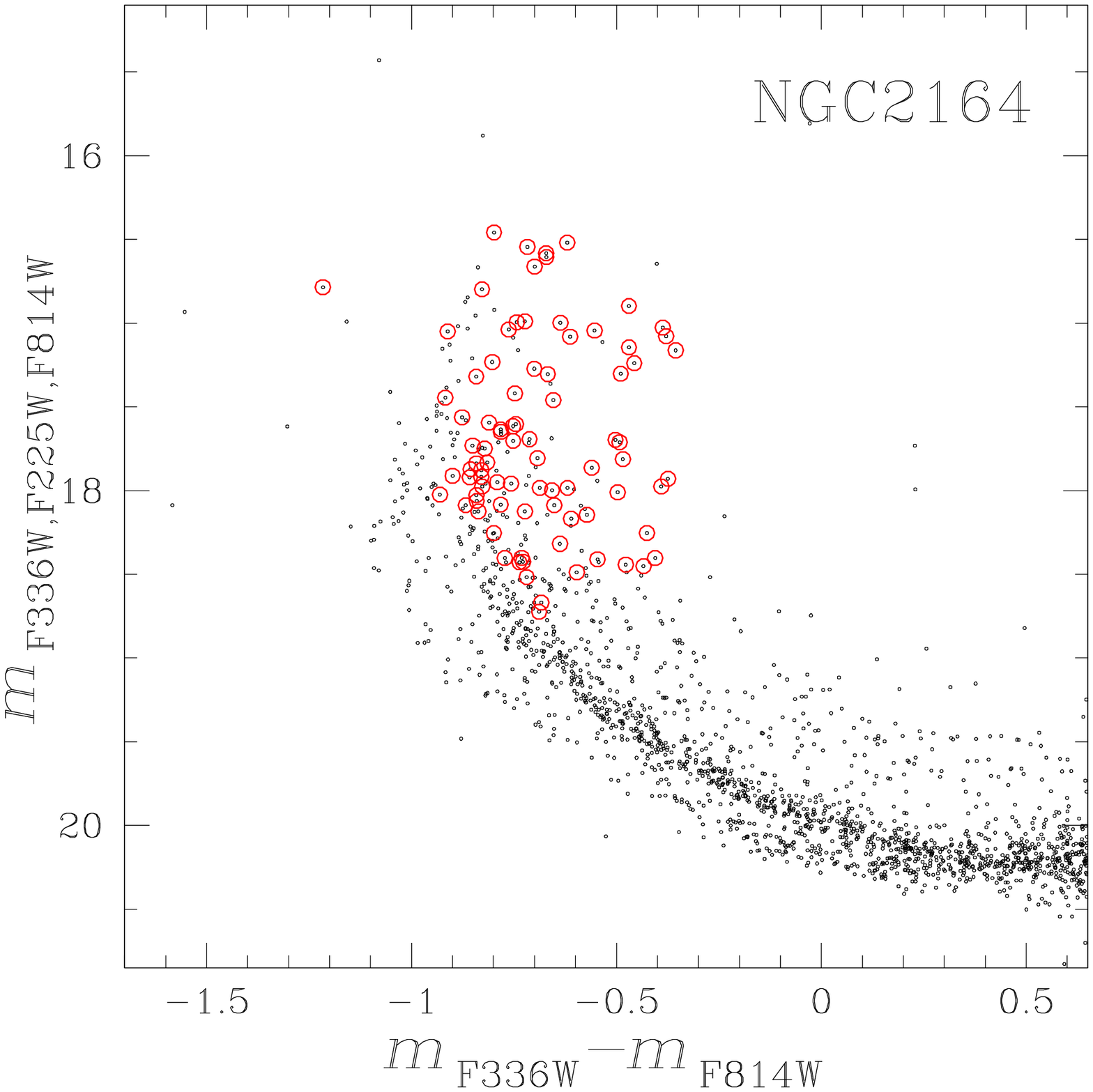}
  \caption{$m_{\rm F336W,F225W,F814W}$=$m_{\rm F336W}-m_{\rm F225W}+m_{\rm F814W}$
  vs.\,$m_{\rm F336W}-m_{\rm F814W}$ diagrams for NGC\,330, NGC\,1818, NGC\,1805, and NGC\,2164. Red circles mark candidate stars with H-$\alpha$ emission.}
 \label{fig:cmd3mag} 
\end{figure*} 
\end{centering}

\begin{centering} 
\begin{figure} 
  \includegraphics[width=8.25cm]{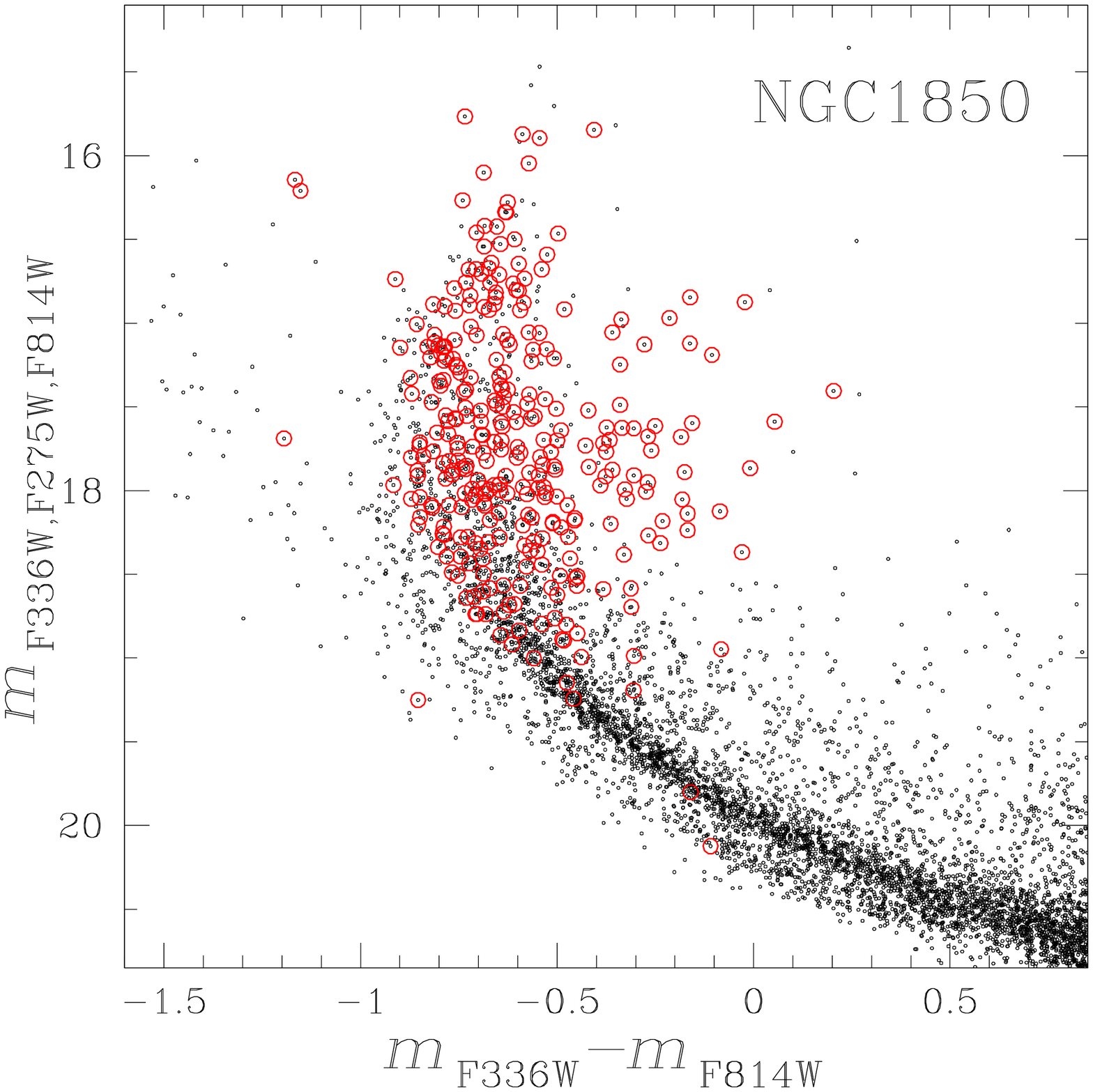}
  \caption{$m_{\rm F336W,F275W,F814W}$=$m_{\rm F336W}-m_{\rm F275W}+m_{\rm F814W}$
  vs.\,$m_{\rm F336W}-m_{\rm F814W}$ diagram for NGC\,1850. Red circles mark candidate stars with H-$\alpha$ emission.}
 \label{fig:cmd3mag1850} 
\end{figure} 
\end{centering}

In addition to F336W, F656N and F814W data, the {\it HST} archive includes UVIS/WFC3 images in F225W of NGC\,330, NGC\,1805, NGC\,1818, and NGC\,2164 and F275W images of NGC\,1850. 
We have defined the pseudo magnitudes $m_{\rm F336W,F225W,F814W}$=$m_{\rm F336W}-m_{\rm F225W}+m_{\rm F814W}$ and $m_{\rm F336W,F275W,F814W}$=$m_{\rm F336W}-m_{\rm F275W}+m_{\rm F814W}$  and plotted these pseudo-magnitudes against $m_{\rm F336W}-m_{\rm F814W}$.  The CMDs are shown in Fig.~\ref{fig:cmd3mag} for stars in the fields of NGC\,330, NGC\,1805, NGC\,1818 and NGC\,2164, and in Fig.~\ref{fig:cmd3mag1850} for NGC 1850. 
These diagrams confirm the split MSs and eMSTO and provide a better separations between blue-MS and red-MS stars than the $m_{\rm F814W}$ vs.\,$m_{\rm F336W}-m_{\rm F814W}$ CMDs.

\section{The split MS and the extended MSTO}\label{sec:MS}
\begin{centering} 
\begin{figure*} 
  \includegraphics[width=12.5cm]{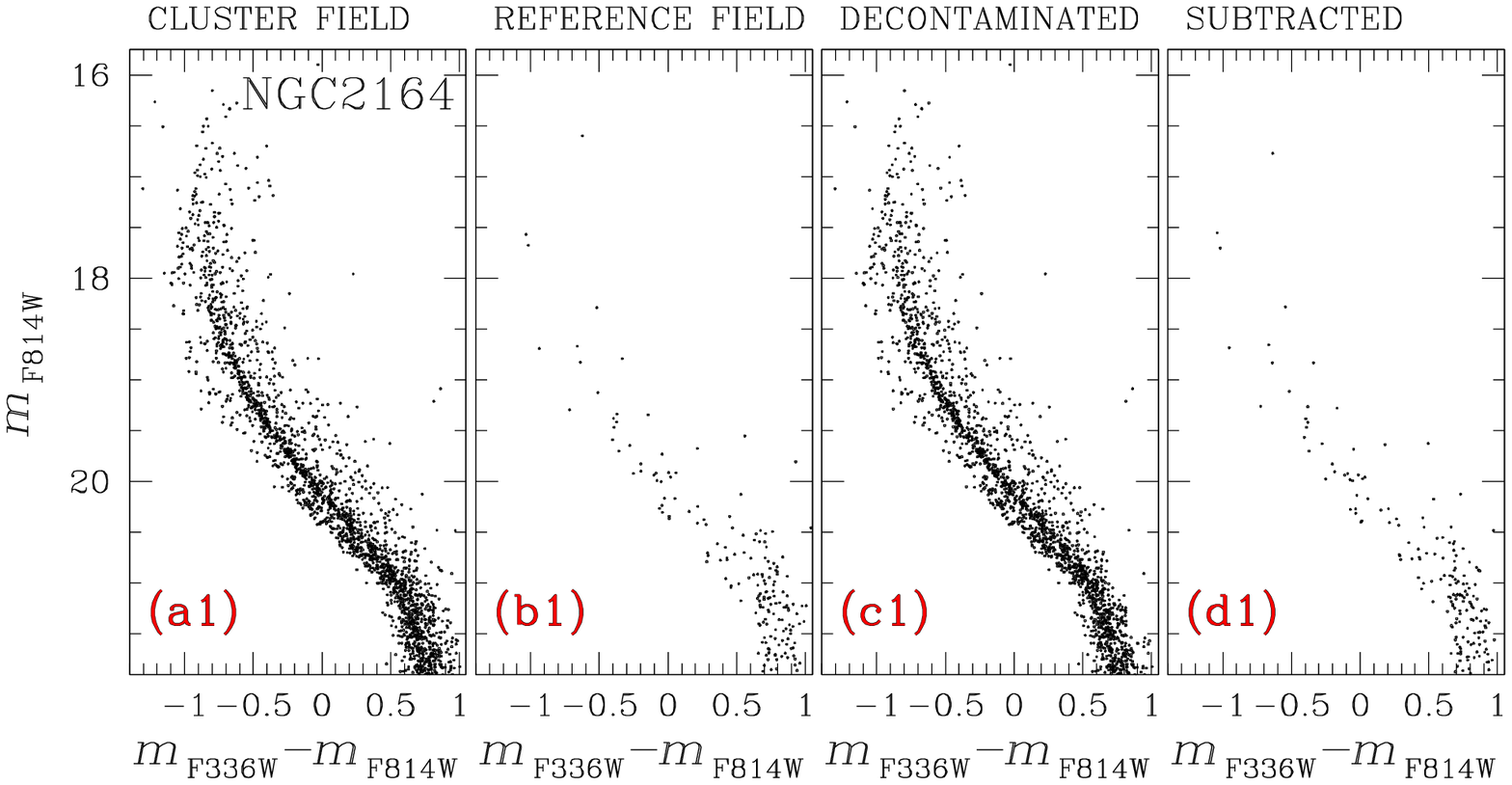}
  \includegraphics[width=12.5cm]{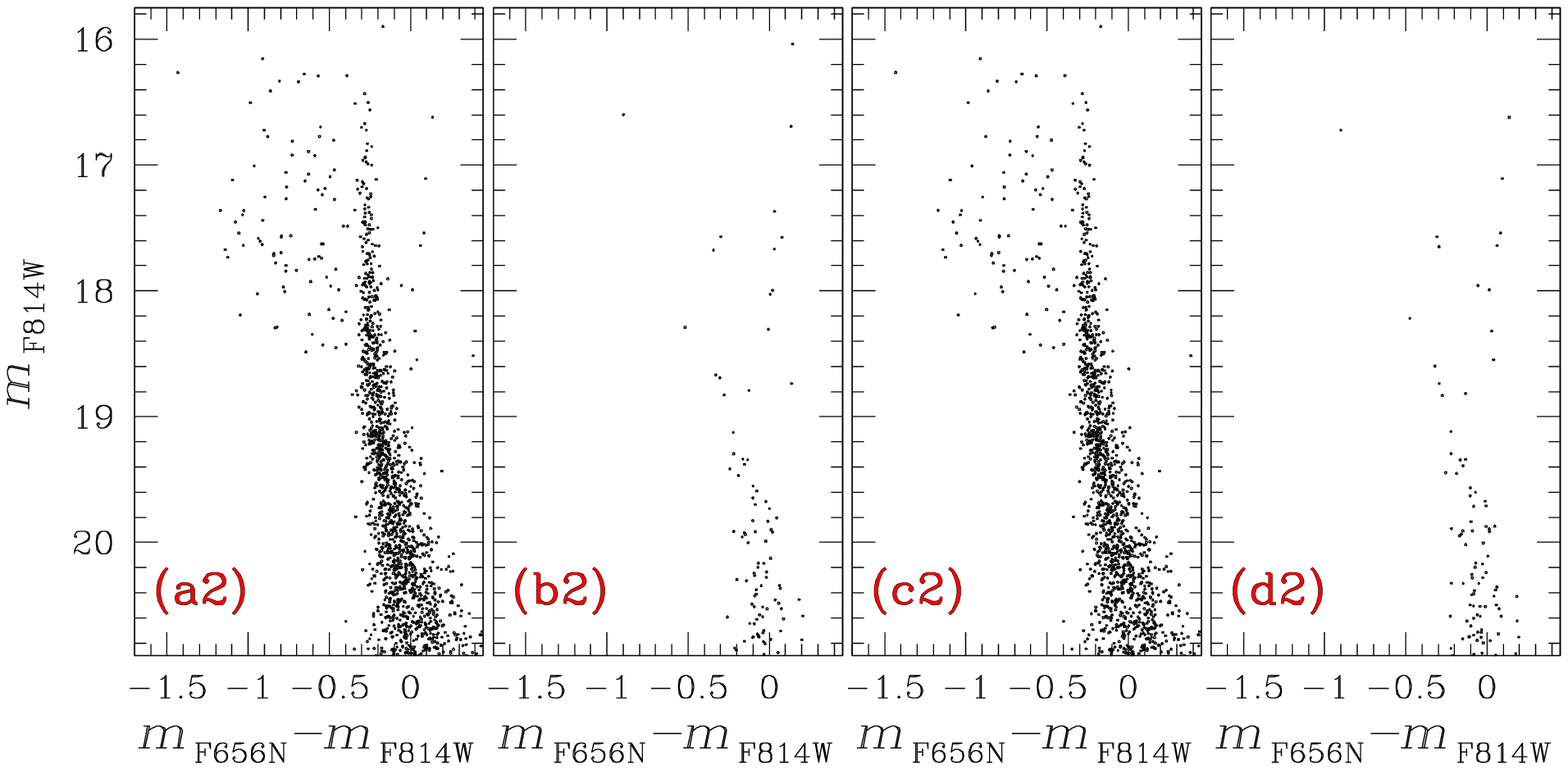}
  \caption{This figure illustrates the procedure to statistically subtract field stars from the cluster-field CMDs of NGC\,2164. The upper and lower panels refer to the $m_{\rm F814W}$ vs.\,$m_{\rm F336W}-m_{\rm F814W}$ and the $m_{\rm F814W}$ vs.\,$m_{\rm F656N}-m_{\rm F814W}$ CMDs, respectively. Panels (a1) and (a2) show the CMDs of stars in the cluster field, while the CMDs of stars in the reference field are plotted in panels (b1) and (b2). The decontaminated CMDs are shown in panels (c1) and (c2) while the subtracted stars are plotted panels (d1) and (d2). See text for details.}
 \label{fig:sub} 
\end{figure*} 
\end{centering}

In this section we demonstrate that the split MSs and the eMSTOs are not due to contamination by foreground and background stars.
To do this, we applied to each cluster, the procedure illustrated in Fig.~\ref{fig:sub} for NGC\,2164. In the panels (a1) and (b1) we compared the $m_{\rm F814W}$ vs.\,$m_{\rm F336W}-m_{\rm F814W}$ CMD of stars in the cluster field  zoomed around the upper MS and the corresponding CMD for a stars in 
 the reference field, which has the same area as the cluster field. Due to the small field of view of UVIS/WFC3, we can assume that the field stars have nearly the same distribution in the CMD of the cluster field and the reference field. 
 In this case, the fraction of field stars in the cluster-field CMD and in the range of color and magnitude populated by MS stars with $m_{\rm F814W}<22.0$ and $(m_{\rm F336W}-m_{\rm F814W})<1.2$ is typically smaller than 0.1 and ranges from 0.05 in NGC\,1866 to 0.26 in NGC\,1801.
 
 To quantify the contamination of non-cluster members in the cluster-field CMD, we statistically subtracted the reference-field CMD from the cluster-field CMD in close analogy with what we have done in the previous papers of this series. Briefly, for each star, i, in the cluster-field CMD we defined a distance in the CMD
 { \scriptsize $d_{\rm i}=\sqrt{k((m_{\rm F336W, cf}-m_{\rm F814W, cf})-(m^{\rm i}_{\rm F336W, rf}-m^{\rm i}_{\rm F814W, rf}))^{2}+(m_{\rm F814W, cf}-m^{\rm i}_{\rm F814W, rf})^{2}}   $}\\
 where
 $m_{\rm F336W (F814W), cf}$ and $m_{\rm F336W (F814W), rf}$ are the F336W (F814W) magnitudes in the cluster- and in the reference-field, respectively.
 The value of the constant $k=4.1$, which has been introduced by Gallart et al.\,(2003) to account for the fact that the color of a star is better constrained than its magnitude, has been calculated as in Marino et al.\,(2014, see their Section~3.1).
 
 For each star in the reference field, we considered the star in the cluster-field CMD  with the smallest distance to each star of the reference field as candidate to be subtracted.
 The star, i, in the reference field  and the closest star in the cluster field have different completeness levels ($c^{\rm i}_{\rm rf}$ and $c^{\rm i}_{\rm cf}$). 
 To account for this fact, we generated a random number $r_{\rm i}$ between 0 and 1 and subtracted from the cluster-field CMD all the candidates with $r_{\rm i}<c^{\rm i}_{\rm rf}/c^{\rm i}_{\rm cf}$. 

 Results are illustrated in panels (c1) and (d1) of Fig.~\ref{fig:sub} where we show the decontaminated CMD and the CMD of the subtracted stars. The fact that both the split MSs and the eMSTOs are visible in the decontaminated CMD demonstrate that they are intrinsic features of the cluster CMD.

 We applied the same method to statistically subtract field stars from the $m_{\rm F814W}$ vs.\,$m_{\rm F656N}-m_{\rm F814W}$ cluster-field CMDs. The adopted procedure, which is illustrated in panels (a2)--(d2) of Fig.~\ref{fig:sub} for NGC\,2164, has been applied to all the clusters and demonstrates that the vast majority of stars with H-$\alpha$ excess are cluster members.
 
\section{Comparison with theory}\label{sec:iso}
A number of papers in the literature have shown that the red MS of young Magellanic- Cloud clusters can be explained with a population of rapidly rotating stars with a rotation rate close to the breakup value, $\omega_{\rm c}$, while the blue MS could be composed of slow-rotating or non-rotating stars (D'Antona et al.\,2015; Milone et al.\,2016, 2017a; Li et al.\,2017a; Bastian et al.\,2017; Correnti et al.\,2017). The interpretation of the eMSTO is still debated and it is not clear whether an age variation, together with a range of stellar rotation, is needed to reproduce the observations (e.g.\,Goudfrooij et al.\,2017).

 To further investigate the physical reasons that are responsible for the eMSTO and the split MS, we used the isochrones from the Geneva and Padova databases (Mowlavi et al.\,2012; Bressan et al.\,2012; Ekstr{\"o}m et al.\,2013; Georgy et al.\,2014; Marigo et al.\,2017).
 Specifically, we compared the observed CMDs with isochrones with different age and different rotation rates in close analogy with what we have done in previous papers of this series. 
 We assumed that the stars have the same helium abundance. Indeed we have already demonstrated that the observed split MS and the eMSTO are inconsistent with multiple populations with different helium abundance (Milone et al.\,2016). Moreover, we assumed that the stars are chemically homogeneous as suggested by spectroscopic and photometric studies of NGC\,419, NGC\,1806, NGC\,1846, and NGC\,1866 (Mucciarelli et al.\,2011, 2014; Mackey et al.\,in preparation; Milone 2017; Martocchia et al.\,2017).
 
The adopted values for distance modulus ($(m-M)_{0}$), reddening (E($B-V$)), and ages which provide the best fit between the observations and the isochrones are provided in Table~\ref{tab:parameters}. We assumed metallicities of Z=0.002 and Z=0.006, for SMC and LMC clusters, respectively. The adopted reddening has been converted into absorption in the F336W and F814W bands by using the coefficients derived in Sect.~\ref{subsec:reddening}. 

\begin{centering} 
\begin{figure*} 
  \includegraphics[width=13.25cm]{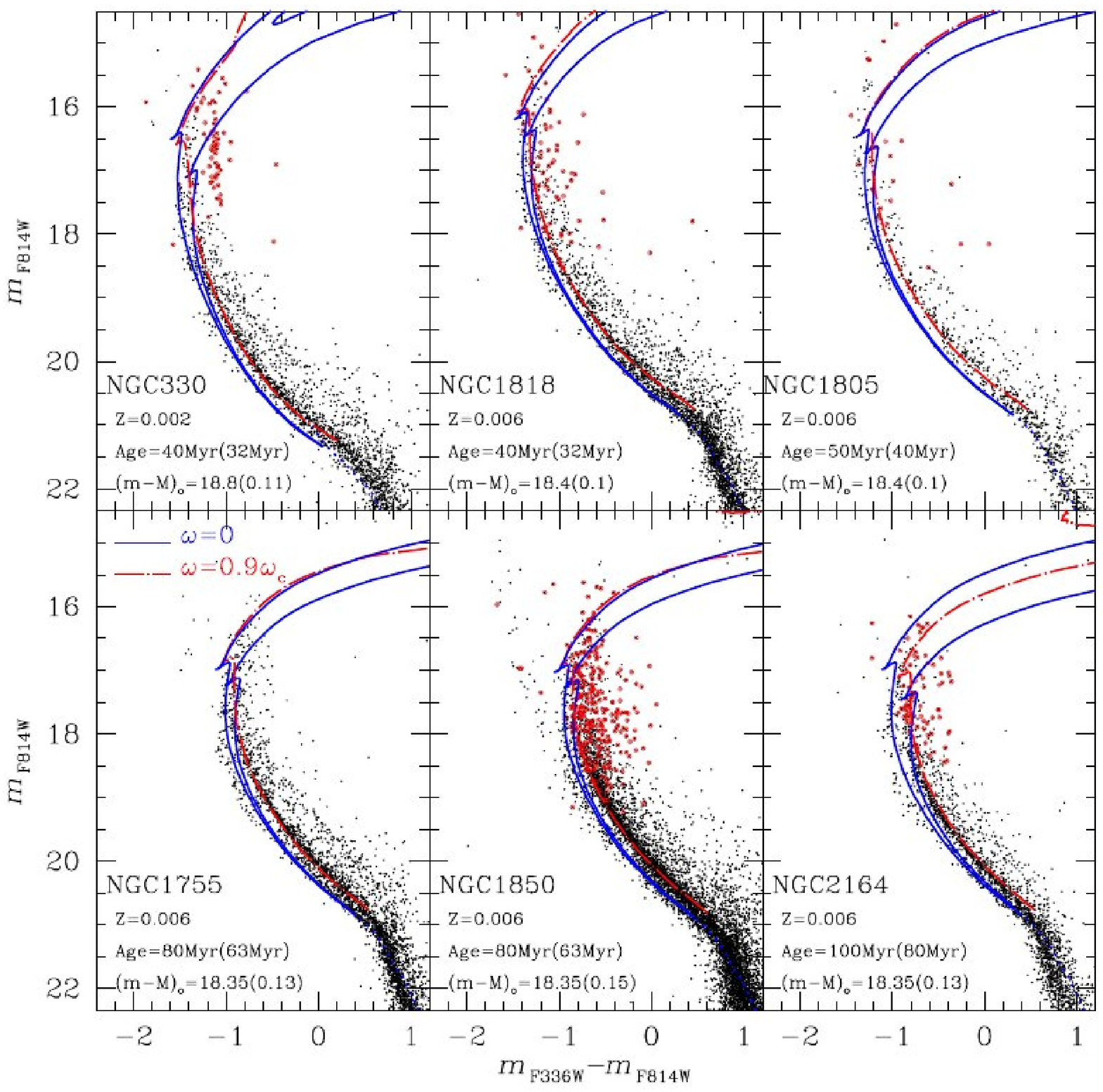}
  \caption{Comparison between isochrones (Ekstr{\"o}m et al.\,2013; Georgy et al.\,2014; Bertelli et al.\,2012; Marigo et al.\,2017) and the observed CMDs of NGC\,330, NGC\,1818, NGC\,1805, NGC\,1755, NGC\,1850, and NGC\,2164. Blue and red lines represent non-rotating isochrones and isochrones with $\omega=0.9 \omega_{\rm c}$ (Ekstr{\"o}m et al.\,2013; Georgy et al.\,2014). In each panel we quoted the adopted values for age of the red isochrones, the  intrinsic distance modulus, the reddening (in brackets), and the metallicity. The blue isochrones with the faintest MSTO have the same age as the red isochrone, while 
   the ages of the youngest blue isochrones are indicated in each panel (in brackets). 
     The dotted-blue lines are the non-rotating isochrones from Mowlavi et al.\,(2012) and Bertelli et al.\,(2012) that we used to reproduce the lower MS.}
 \label{fig:iso} 
\end{figure*} 
\end{centering}

\begin{centering} 
\begin{figure*} 
  \includegraphics[width=13.25cm]{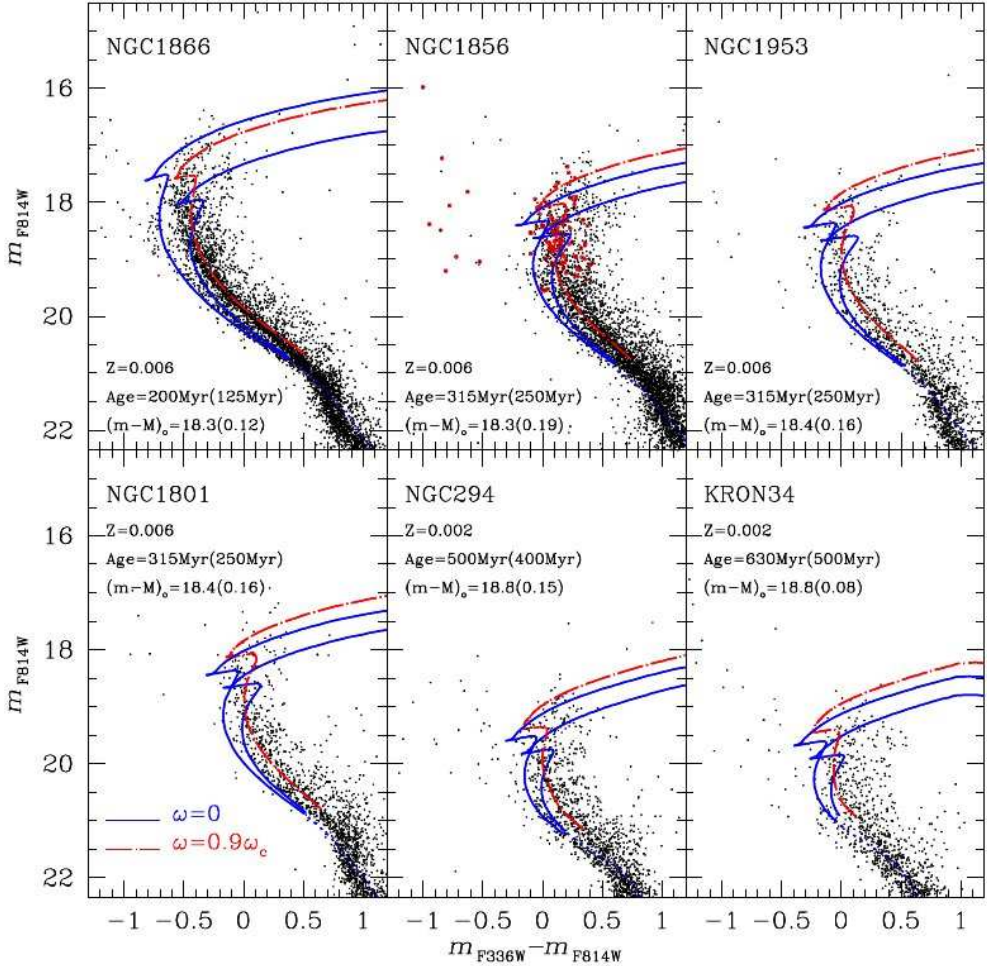}
  \caption{As in Fig.~\ref{fig:iso} but for NGC\,1866, NGC\,1856, NGC\,1953, NGC\,1801, NGC\,294, and KRON\,34.}
 \label{fig:iso2} 
\end{figure*} 
\end{centering}

\begin{centering} 
\begin{figure} 
  \includegraphics[width=8.25cm]{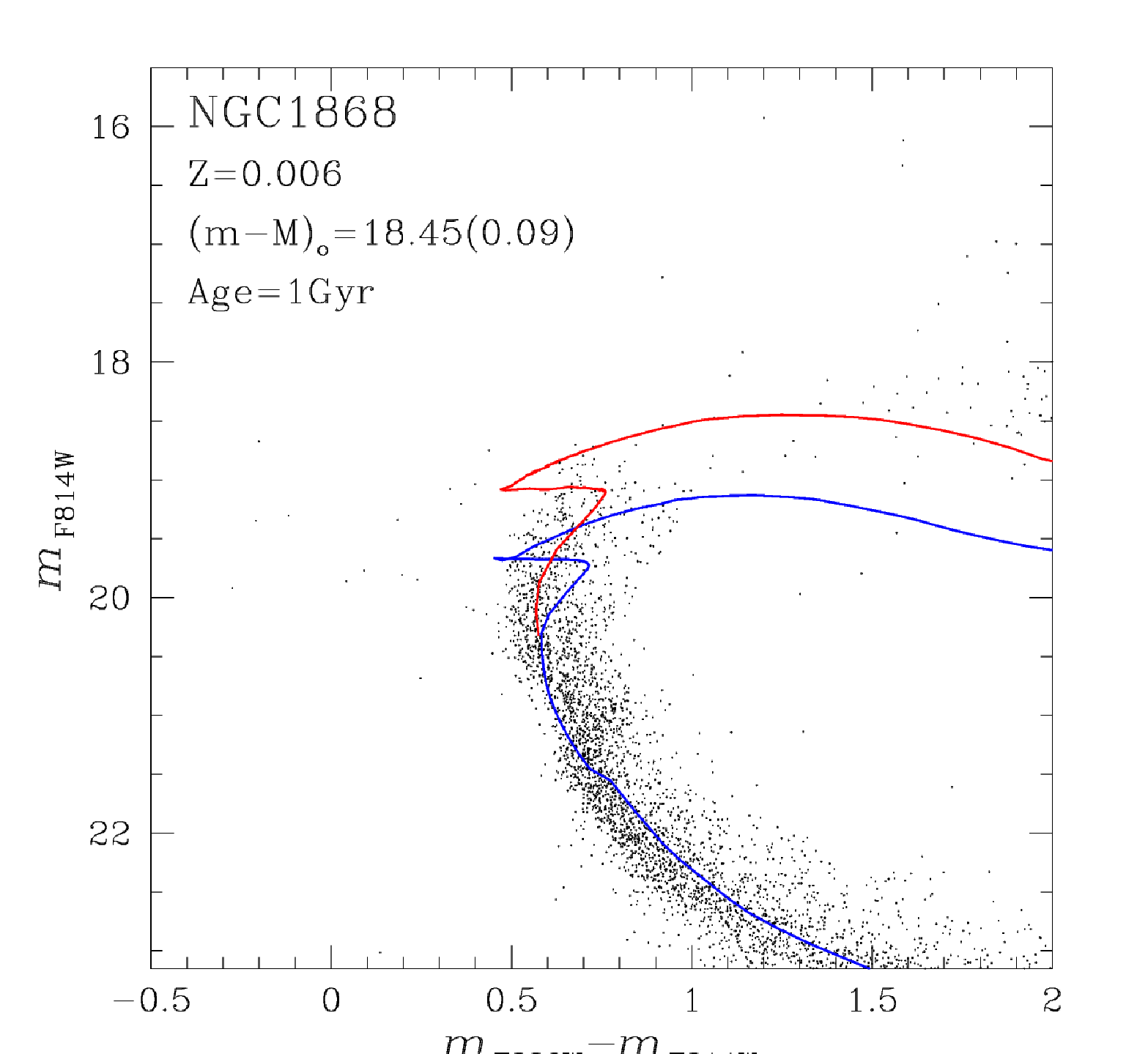}
  \caption{ Comparison between isochrones from the Geneva database and the observed CMD of NGC\,1868. The non-rotating isochrone and the isochrone with $\omega=0.9 \omega_{\rm c}$ are colored blue and red, respectively. The adopted values for the age, distance modulus, reddening, and metallicity are quoted in the figure.}
 \label{fig:iso3} 
\end{figure} 
\end{centering}

 Results are illustrated in Figs.~\ref{fig:iso}-\ref{fig:iso3}, where we superimposed on each CMD the corresponding best-fit isochrones. We find that the blue and red MS of all clusters are consistent with two coeval populations of non-rotating or slowly-rotating stars and of stars with a high rotation rate ($\omega=0.9 \omega_{\rm c}$), that we colored blue and red, respectively.
 
 All the clusters younger than $\sim 650$ Myrs (12/13) host a significant number of bright stars on the blue side of the upper MS. As noticed in previous work (e.g.\,D'Antona et al.\,2017; Milone et al.\,2017a), these stars are consistent with slow rotators and are reproduced by the bluest isochrones that are younger than the other isochrones. The age difference between the two blue isochrones is quoted in each figure.

 As previously noticed for NGC\,1856, NGC\,1755, and NGC\,1866 (D'Antona et al.\,2015; Milone et al.\,2016, 2017a), the comparison between the observed CMDs and the isochrones are not fully satisfactory, as the faint eMSTO is poorly reproduced by the models. In principle, we could improve the fit by assuming that the rotating population spans a range of ages.
 However, we note that the poor quality of the fit could be due to uncertainties in the isochrones. As noticed by D'Antona et al.\,(2015), the models of the core-hydrogen-burning phase may be affected by several second-order parameters including the effect of the inclination angle and the calculation of stellar luminosity and effective temperature.

 Since models from Ekstr{\"o}m et al.\,(2013) and Georgy et al.\,(2014) are available for bright stars only, we used non-rotating isochrones from other works (blue-dotted lines in Figs.~\ref{fig:iso}-\ref{fig:iso3}) to reproduce the lower MS of the analyzed clusters. For LMC clusters, we used isochrones from the Geneva group which are available for metallicity of Z=0.006 only (Mowlavi et al.\,2012),  while for the SMC clusters we used isochrones from the Padova database (Bressan et al.\,2012; Marigo et al.\,2017).  

 We expect that the non-rotating isochrones would overlap the fast-rotating  ones at luminosity of the MS kink, which is a feature previously observed in the CMDs of Galactic open clusters, that indicates the onset of envelope convection due to the opacity peak of partial hydrogen ionization (e.g.\,D'Antona et al.\,2002).

 Stars above the MS kink, which are present only in young clusters and belong to the spectral classes A and B, are hotter than $\sim 7,500$ K and are characterized by radiative envelopes. Rapidly-rotating A and B MS stars are $\lesssim$300\,K colder than slowly-rotating stars with the same luminosity and distribute along the blue and red MS.
In contrast, the envelopes of stars fainter than the MS kink are convective. Even though their cores might still be rapidly rotating, the effective-temperature differences between rapid rotators and non-rotators with the same luminosity become much smaller and so the  sequences effectively merge in the CMD.

The fact that the fast-rotating and the non-rotating MSs merge together around the kink is predicted by the MIST isochrones (Dotter 2016; Choi et al.\,2016), and is consistent with what we observe in the CMDs where the blue and the red MSs cross and get very close  to each other at the same luminosity.

\section{Population ratio}\label{sec:pratio}
 To determine the fraction of blue-MS and red-MS stars with respect to the total number of MS stars we 
 used a procedure which is based on the analysis of the $m_{\rm F814W}$ vs.\,$m_{\rm F336W}-m_{\rm F814W}$ CMD and has been adopted in the previous papers of this series.
 This procedure has been applied to all the clusters but NGC\,1868, which is the oldest object in our sample and shows no evidence for a split MS. 
 
\begin{centering} 
\begin{figure*} 
  \includegraphics[width=14.25cm]{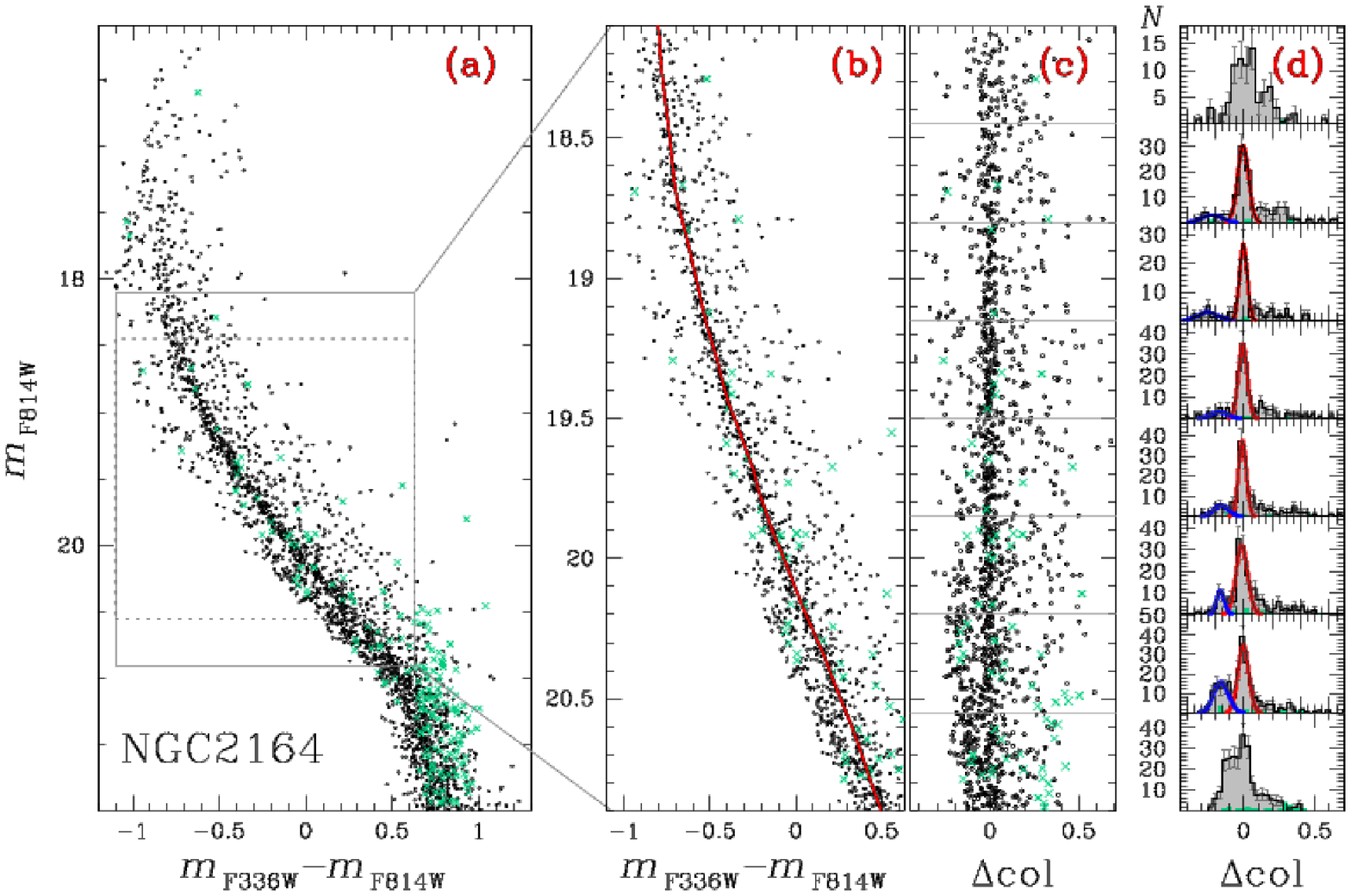}
  \caption{This figure illustrates the procedure that we used to estimate the fraction of blue- and red-MS stars with respect to the total number of MS stars in NGC\,2164. Panel (a) reproduces the $m_{\rm F814W}$ vs.\,$m_{\rm F336W}-m_{\rm F814W}$ CMD of Fig.~\ref{fig:cmd2}, while panel (b) is a zoom of the CMD region where the MS split is more-prominently visible. The red line overplotted on the CMD is the fiducial line of the red MS. Panel (c) shows the verticalized $m_{\rm F814W}$ vs.\,$\Delta$col diagram for the stars plotted in panel (b). Finally, panels (d) show the histogram distribution of $\Delta$col for stars in eight intervals of F814W magnitude. The red and blue lines overimposed on the six central histograms are the components of the best-fit bi-Gaussian function of each histogram. We did not fit with a function the upper-most and the lower-most histogram that correspond to MS stars above and below the two dotted lines in the left panel. See text for details.}
 \label{fig:pratioI} 
\end{figure*} 
\end{centering}

 The method is illustrated in Fig.~\ref{fig:pratioI} for NGC\,2164. We analyzed only the MS region within the gray rectangle shown in the CMD plotted in panel (a), where the split MS is more-clearly visible.
 We derived the fiducial line of the red MS plotted on the CMD shown in the panel (b). To derive the fiducial we divided a sample of red-MS stars selected by eye into bins of 0.2 mag, which are defined over a grid of points separated by steps of 0.05 mag (Silverman 1986). For each bin we calculated the median of the colors and magnitudes of all the stars in that bin. These median points have been smoothed with the boxcar averaging method, which replaces each point with the average of the three adjacent points.

 The verticalized $m_{\rm F336W}$ vs.\,$\Delta$col diagram plotted in the panel (c) of Fig.~\ref{fig:pratioI} has been derived by subtracting from the $m_{\rm F336W}-m_{\rm F814W}$ color of each star the color of the red fiducial line corresponding to the same magnitude in the F814W band. We defined eight intervals of 0.25 mag each and the $\Delta$col histogram distributions of the stars in the various bins are plotted in the panels (d) of Fig.~\ref{fig:pratioI}.

 Most histograms exhibit a bimodal $\Delta$col distribution that we fitted with a bi-Gaussian function. The two components of the best-fit least-squares Gaussians are represented with the blue and red lines overplotted on the histograms of Fig.~\ref{fig:pratioI}.  To minimize the effect of binary stars, we excluded from the fit all the stars with $\Delta$col larger than 0.08. 
 The areas under the blue and red histograms are used to infer the ratio between blue- and red-MS stars in each magnitude bin. 
 
 The two MSs of NGC\,2164 merge together at $m_{\rm F814W} \sim$20.8. As a consequence, we did not use the bi-Gaussian function to fit the histogram distribution plotted in the lower panel. Above $m_{\rm F814W} \sim$18.5 some MS stars evolve to the red side of the MS. For this reason, we did not estimate the population ratio of the stellar populations in the upper bin.

  In addition to the procedure described above, Milone et al.\,(2017a) used a more-sophisticated method to infer the relative numbers of blue-MS and red-MS stars in NGC\,1866 that accounts for the presence of binaries among red-MS and blue-MS stars. Since Milone and collaborators show that the two methods provide nearly-identical results, for simplicity, we defer the analysis of binaries to forthcoming work.
 
\begin{centering} 
\begin{figure} 
  \includegraphics[width=8.5cm]{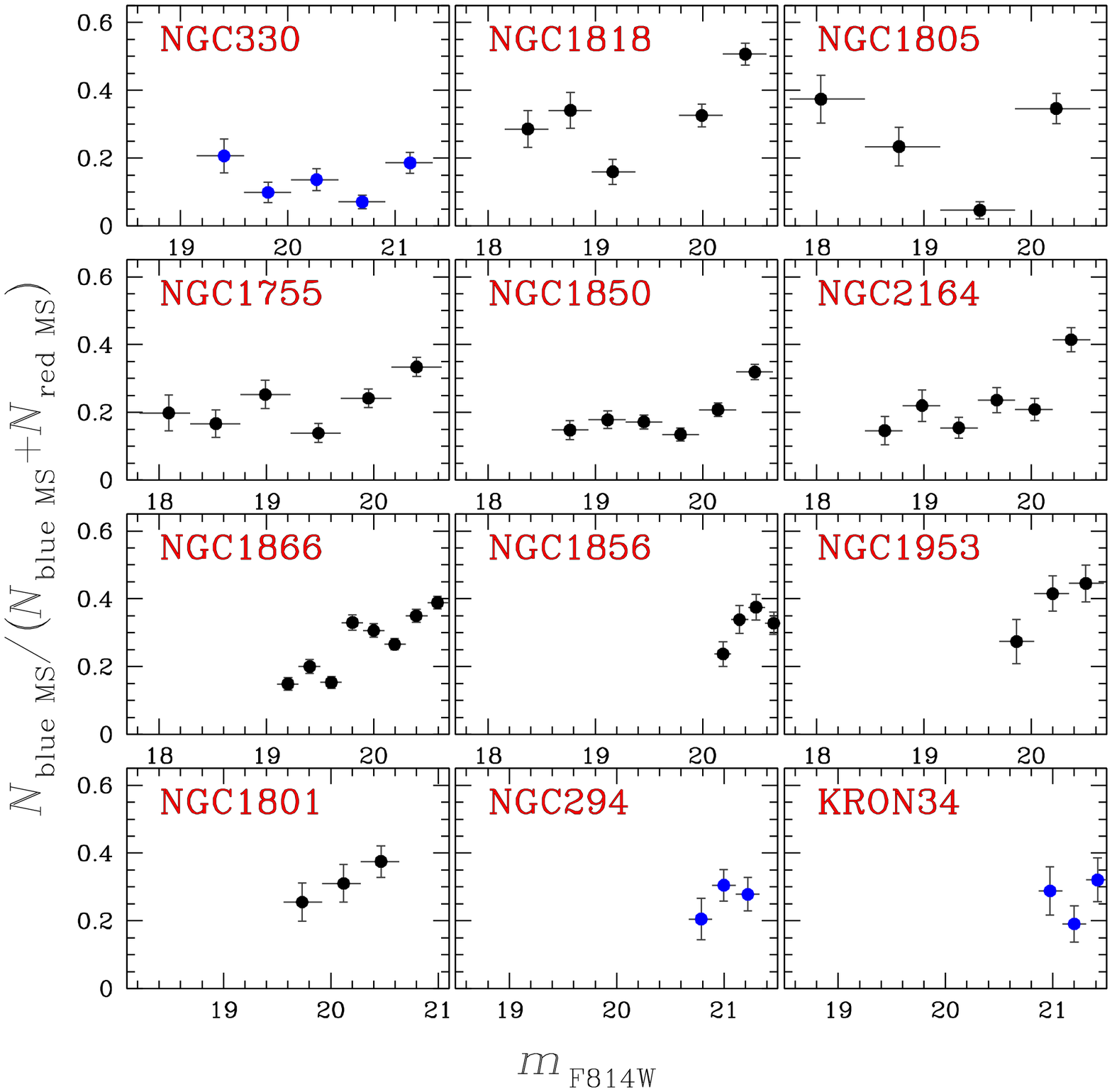}
  \caption{Fraction of blue-MS stars with respect to the overall number of MS stars as a function of the observed F814W magnitude.}
 \label{fig:lf} 
\end{figure} 
\end{centering}

\begin{centering} 
\begin{figure} 
  \includegraphics[width=8.cm]{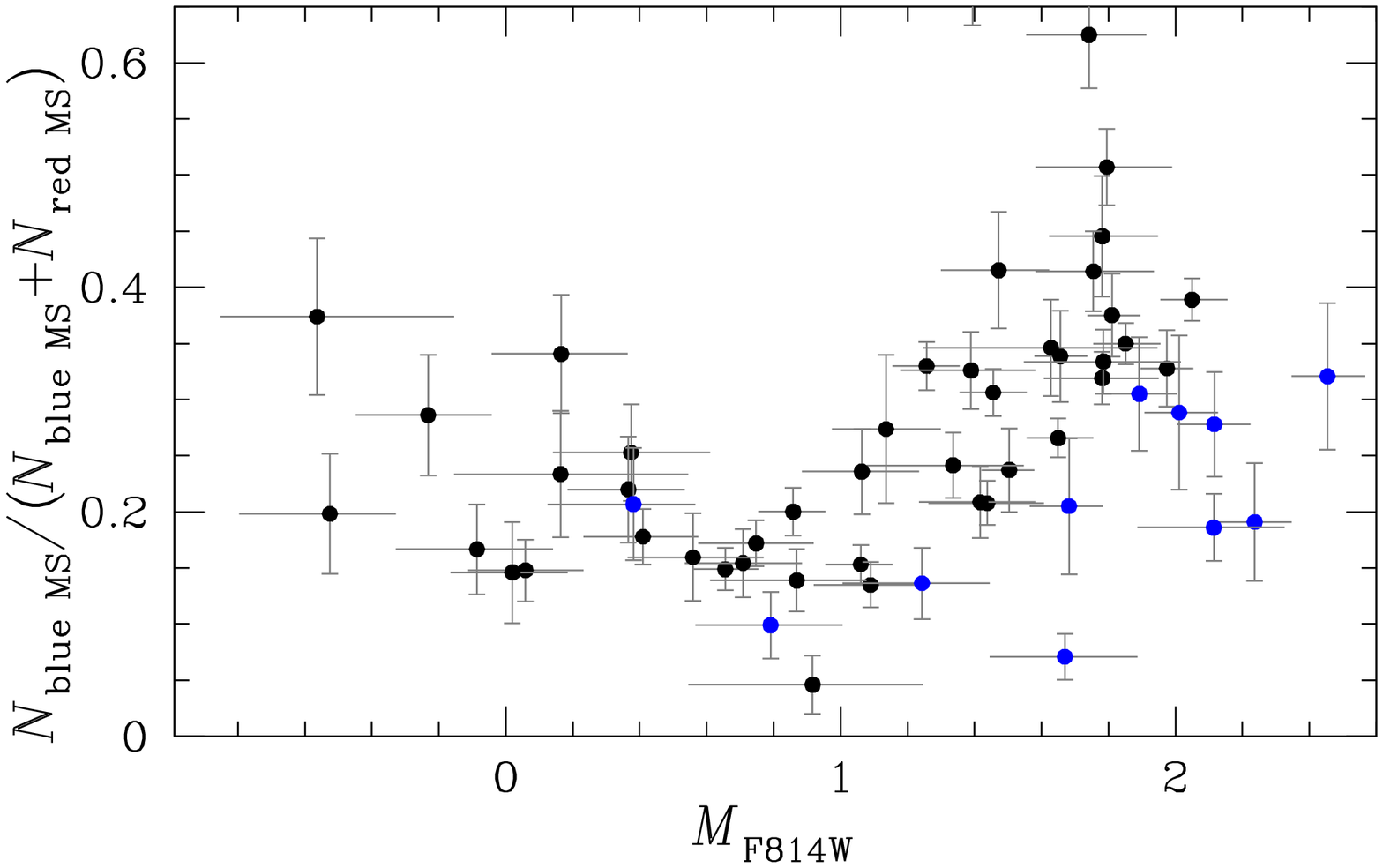}
  \includegraphics[width=8.cm]{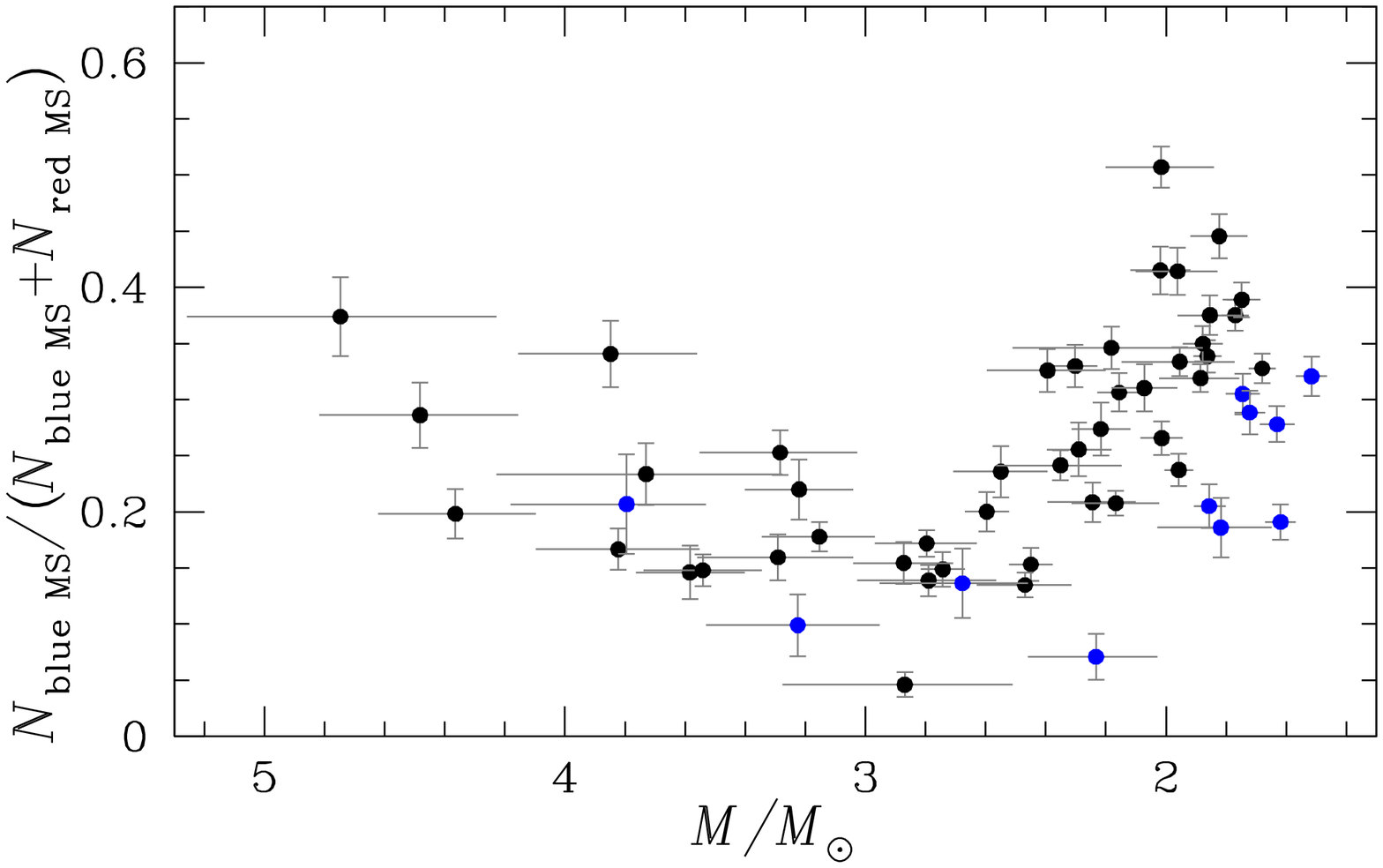}
  \caption{Fraction of blue-MS stars with respect to the overall number of MS stars as a function of the absolute F814W magnitude (upper panel) and of the stellar mass (lower panel). Data for the three SMC clusters are represented with blue dots.}
 \label{fig:lf2} 
\end{figure} 
\end{centering}

The results are illustrated in Fig.~\ref{fig:lf} where for each cluster we plot the fraction of the blue-MS stars with respect to the total number of MS stars derived in  distinct F814W magnitude bins as a function of the mean magnitude of all the MS stars in that bin.
To compare results from different clusters, in Fig.~\ref{fig:lf2} we plot the fraction of blue-MS stars as a function of the absolute F814W magnitude (upper panel) and of the stellar mass (lower panel).

We find that all the LMC clusters follow a similar trend.
 The fraction of blue-MS star with respect to the total number of MS stars reaches its maximum value of $\sim 0.4$ among faint stars with masses of $\sim$1.7 $\mathcal{M_{\odot}}$ and drops to $\sim$0.15 around $\sim$2.8 $\mathcal{M_{\odot}}$.
 In the young clusters (age $\lesssim 100 Myr$), which host massive MS stars, namely NGC\,1818, NGC\,1805, NGC\,1755, NGC\,1850, and NGC\,2164 the percentage of blue-MS stars is constant or even seems to increase towards higher masses.

 The three SMC clusters are represented with blue dots in Fig.~\ref{fig:lf2} and share a similar trend as LMC clusters but seem to host a lower fraction of blue-MS stars. This fact is particularly evident in NGC\,330, which is the youngest analyzed SMC cluster, where the fraction of blue-MS stars is almost constant along the entire analyzed interval of mass.

\section{Stars with H-$\alpha$ emission}\label{sec:Be}
As shown in Figs.~\ref{fig:cmd1}--\ref{fig:cmd4}, the five youngest analyzed clusters for which photometry in the F656N band is available, namely NGC\,330, NGC\,1805, NGC\,1818, NGC\,1850, and NGC\,2164 host a large number of bright MS stars with $m_{\rm F656N}-m_{\rm F814W}$ colors bluer than the bulk of MS stars by more than one magnitude.
 Such low $m_{\rm F656N}-m_{\rm F814W}$ colors, which involve stars around the MSTO only, are likely due to strong H-$\alpha$ emission of Be stars.

In contrast, a visual inspection at the $m_{\rm F814W}$ vs.\,$m_{\rm F656N}-m_{\rm F814W}$ CMD of NGC\,1856 plotted in Fig.~\ref{fig:cmd3} reveals that this clusters does not host MS stars with strong $m_{\rm F656N}-m_{\rm F814W}$ color difference from the majority of MS stars (see also Correnti et al.\,2015). 
 Nevertheless, the color and magnitude distribution of stars around the MSTO is broadened with a tail of stars with bluer $m_{\rm F656N}-m_{\rm F814W}$ values than the bulk of MS stars with the same luminosity. A similar conclusion comes from the analysis of the $m_{\rm F814W}$ vs.\,$m_{\rm F625W}-m_{\rm F656N}$ diagram by Bastian et al.\,(2017).

 In the other clusters older than $\sim$100 Myrs, namely NGC\,1953, NGC\,1801, NGC\,294, and KRON\,34, the upper MS exhibits a wider $m_{\rm F656N}-m_{\rm F814W}$ color than what we expect from observational errors alone. This fact suggests that weak H-$\alpha$-emitters are likely present in all clusters younger than $\sim 600$ Myr. In the following we analyze the $m_{\rm F656N}-m_{\rm F814W}$ color and the $m_{\rm F814W}$ magnitude distribution of the H-$\alpha$ emitters.

\begin{centering} 
\begin{figure*} 
  \includegraphics[width=14.25cm]{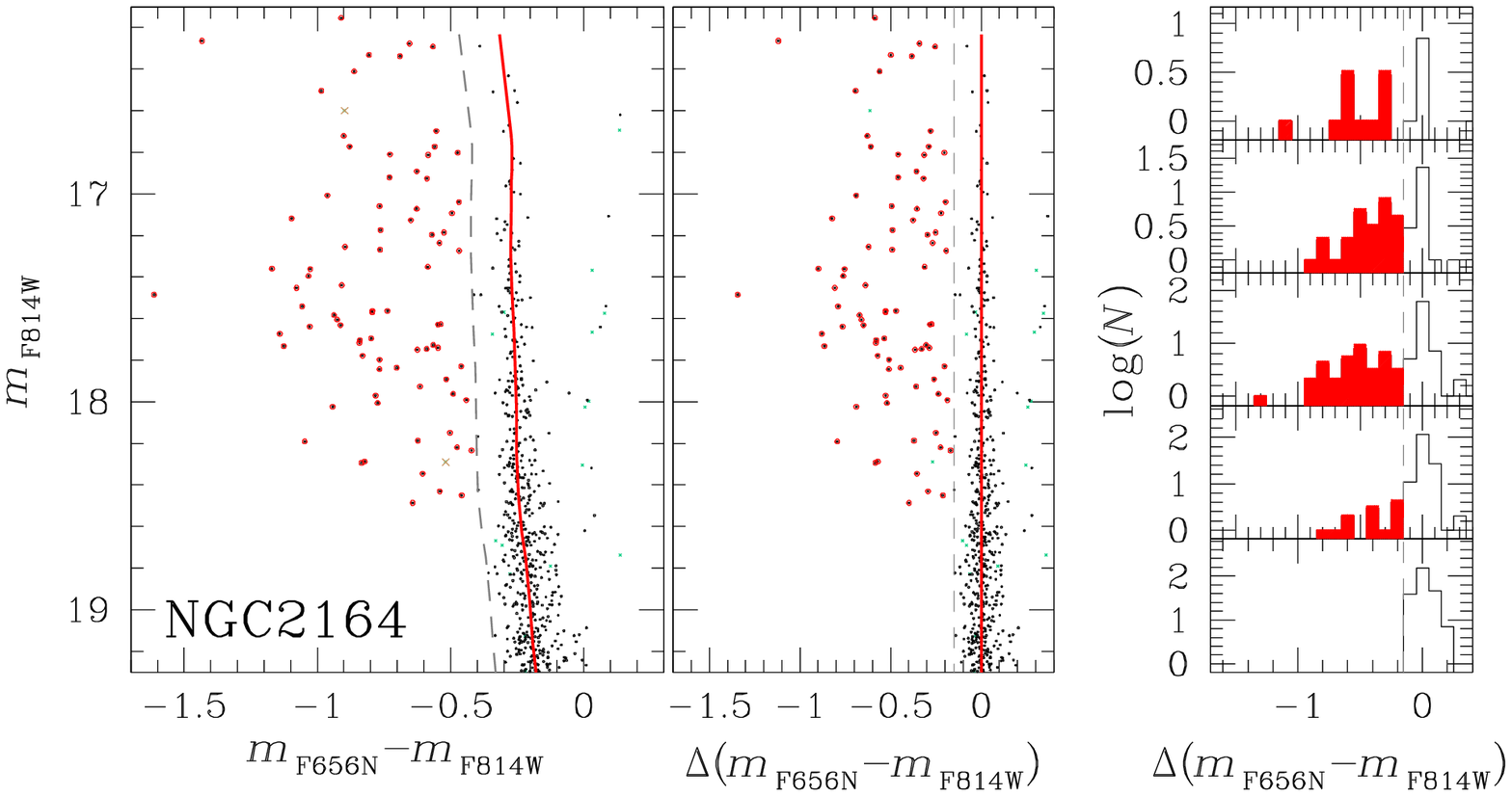}
  \caption{Method to estimate the fraction of stars with H-$\alpha$ emission in NGC\,2164. Left panel reproduces the $m_{\rm F814W}$ vs.\,$m_{\rm F656N}-m_{\rm F814W}$ CMD of Fig.~\ref{fig:cmd2} with the fiducial line of non-emitters stars represented in red. The left panel shows the verticalized $m_{\rm F814W}$ vs.\,$\Delta (m_{\rm F656N}-m_{\rm F814W})$ diagram, while right panels show the $\Delta (m_{\rm F656N}-m_{\rm F814W})$ histogram distribution for stars in five magnitude intervals.
     The vertical dashed lines plotted in the left and middle panels are used to separate candidate Be stars from the remaining MS stars.
     Candidate H-$\alpha$ emitters are marked with red open dots, while the areas of the corresponding histogram distribution are colored in red.}
 \label{fig:selBe} 
\end{figure*} 
\end{centering}

 To investigate how the fraction of Be stares with respect to the total number of MS stars changes as a function of the F814W magnitude we used the procedure illustrated in Fig.~\ref{fig:selBe} for NGC\,2164. We selected by eye a sample of MS stars in the cluster-field CMD with no evidence for H-$\alpha$ emission and derived the red fiducial line as described in the previous section and shown in the left panel of Fig.~\ref{fig:selBe}. This fiducial has been used to generate the verticalized $m_{\rm F814W}$ vs.\,$\Delta (m_{\rm F656N}-m_{\rm F814W})$ diagram plotted in the middle panel of Fig.~\ref{fig:selBe}. All the stars in both the cluster and the reference field with $\Delta (m_{\rm F656N}-m_{\rm F814W}) < 0.15$ mag are considered as candidate stars with strong emission in H-$\alpha$.

 This procedure has been applied to all the clusters where the observational error in the $m_{\rm F656N}-m_{\rm F814W}$ color for stars with possible H-$\alpha$ emission is smaller than 0.037 mag, including the five clusters younger than $\sim$100 Myrs with available F656N photometry and NGC\,1856. This choice guarantees that the selected candidate Be stars are unlikely non-emitting MS stars with large photometric uncertainties, as confirmed by AS experiments. We excluded from the analysis the remaining clusters with larger photometric errors.

 We derived the $\Delta (m_{\rm F656N}-m_{\rm F814W})$ histogram distributions for stars in five magnitude intervals in both the cluster and the reference field by accounting for the completeness level of each star. The histograms plotted in the right panels of Fig.~\ref{fig:selBe} are calculated as the difference between the histograms of stars in the cluster and in the reference field. The shaded-red histograms represent the candidate emitters.

\begin{centering} 
\begin{figure} 
  \includegraphics[width=8.cm]{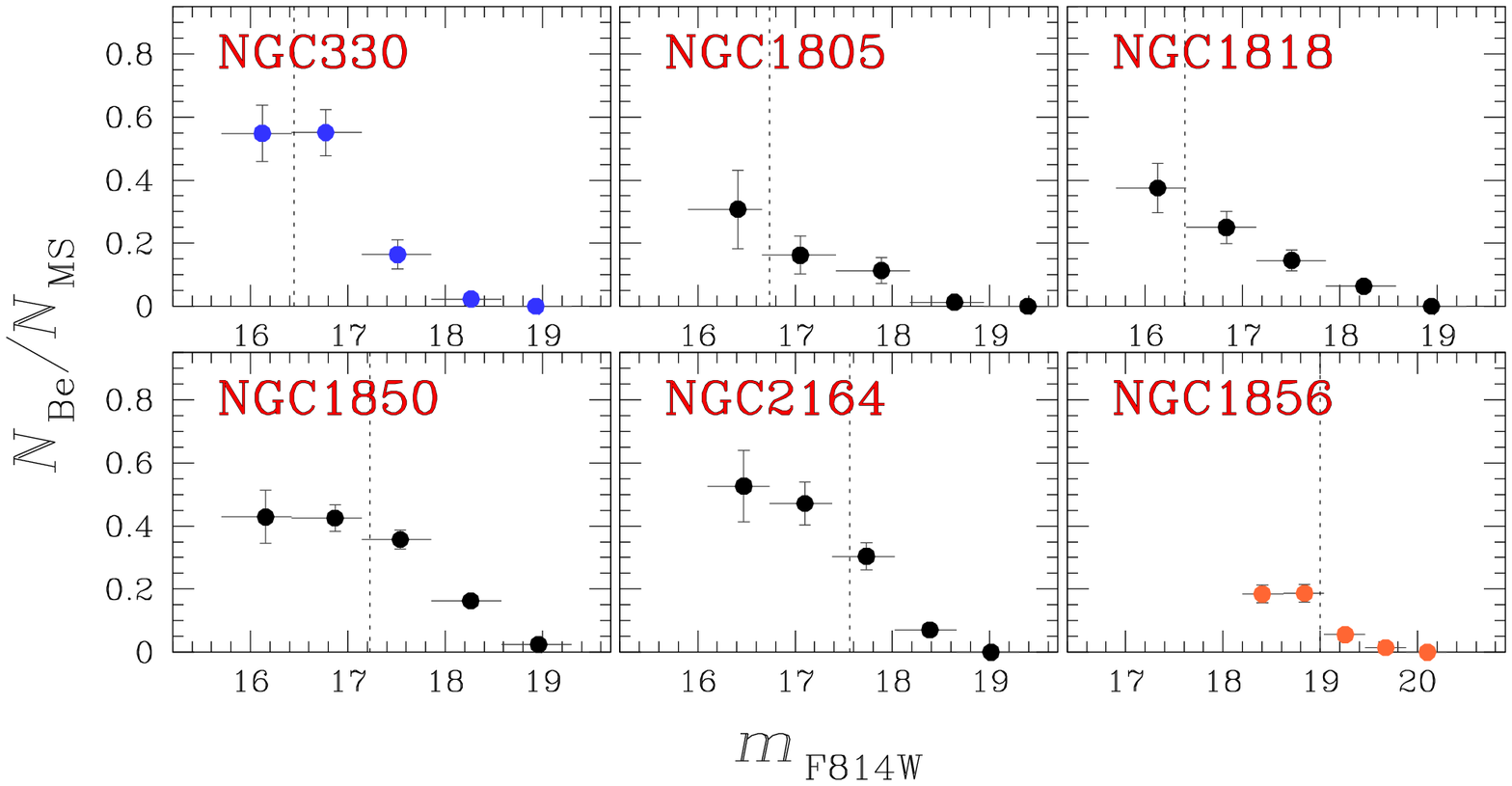}
  \includegraphics[width=8.cm]{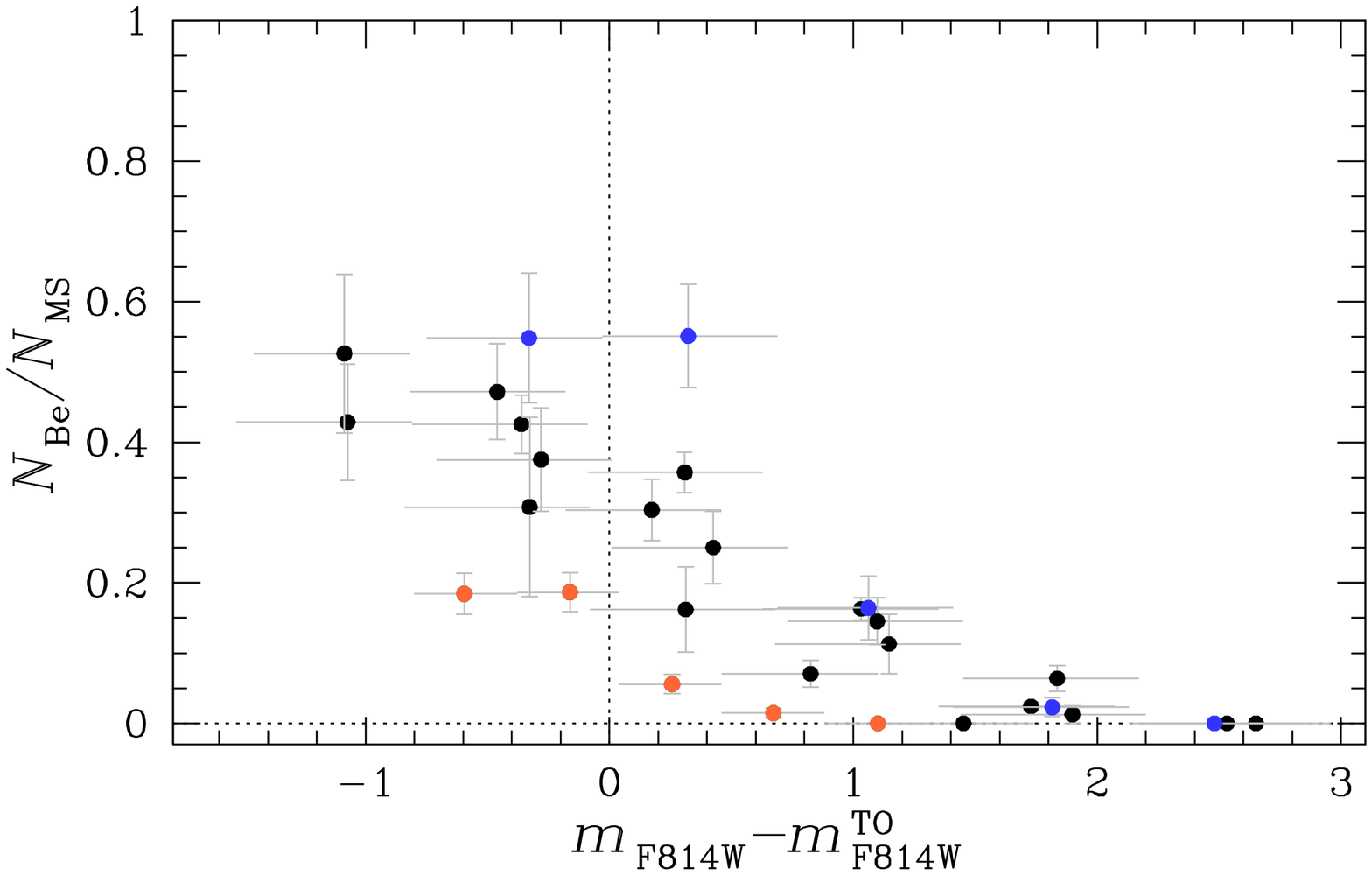}
  \caption{Fraction of stars with H-$\alpha$ emission with respect to the total  number of MS stars as a function of the F814W magnitude (upper panels) and as a function of the relative magnitude to the turn off of the red MS (lower panel). The blue points refer to the SMC cluster NGC\,330. Orange points indicate the $\sim 300$-Myr old LMC cluster NGC\,1856. The magnitude of the red-MS turn off is marked with vertical dotted lines. }
 \label{fig:Be} 
\end{figure} 
\end{centering}

The results are illustrated in the upper panels of Fig.~\ref{fig:Be} where we show for each cluster the fraction of H-$\alpha$ emitters, calculated in five distinct magnitude bins, as a function of the average F814W magnitude of all the stars in the corresponding bin.
We find that the fraction of Be stars in each cluster reached its maximum around the MSTO, which is indicated with the dotted vertical lines. This conclusion corroborates similar findings by Keller et al.\,(2000) who
 derived the luminosity function of Be stars in NGC\,330 and in the LMC GCs NGC\,1818, NGC\,2004, and NGC\,2100 and concluded that their fraction of Be stars peaks towards the MS turn off (see also Iqbal \& Keller\,2013).

 To better compare results from different clusters, we calculated the difference between the average magnitude of stars in the various bins and the magnitude of the red-MSTO, $m^{\rm TO}_{\rm F814W}$ and plotted in the lower panel of Fig.~\ref{fig:Be} the fraction of  H-$\alpha$ emitters as a function of $m_{\rm F814W}-m^{\rm TO}_{\rm F814W}$.
 
 We find that the ratio between the number of strong H-$\alpha$ emitters and the number of MS stars follow similar trends in the analyzed clusters. It is maximal above the turn off, and monotonically decreasing to zero around $m_{\rm F814W}-m^{\rm TO}_{\rm F814W} \sim 2.0$. There is no evidence for H-$\alpha$ emitters with fainter luminosities. In NGC\,330, NGC\,1818, NGC\,2164 and NGC\,1850 the emitters comprise up to 40-55\% of MS stars, while in the less-massive cluster, NGC\,1805, the fraction of Be stars is always smaller than $\sim$0.3.
In this context, it should be noticed that in clusters with very high radial velocities the H-$\alpha$ line is redshifted to the edge or even outside the passband of the F656N filter. For this reason, we can not exlude that the small fraction  of Be stars with excess in the F656N band in NGC\,1805 is due to its radial velocity of $\sim + 390$ km~s$^{-1}$ (Fehrenbach\,1974).
  The oldest cluster analyzed in this section, NGC\,1856, seems to host a smaller fraction of emitting stars than the other GCs.

The comparison of results in the various clusters reveals that stars with the same absolute magnitude can be H-$\alpha$ emitters or not depending on their relative luminosity with respect to the MSTO. Indeed, while faint H-$\alpha$ emitters in NGC\,1856 have absolute magnitude $M_{\rm F814W} \sim 0.7$, there are no Be stars fainter than $M_{\rm F814W} \sim -0.5$ in NGC\,330 and no emitting stars with $M_{\rm F814W} \sim 0.5$ in the other analyzed clusters. As a consequence, the Be-phenomenon seems connected with the evolution of a star along the MS.

\begin{centering} 
\begin{figure} 
  \includegraphics[width=8.5cm]{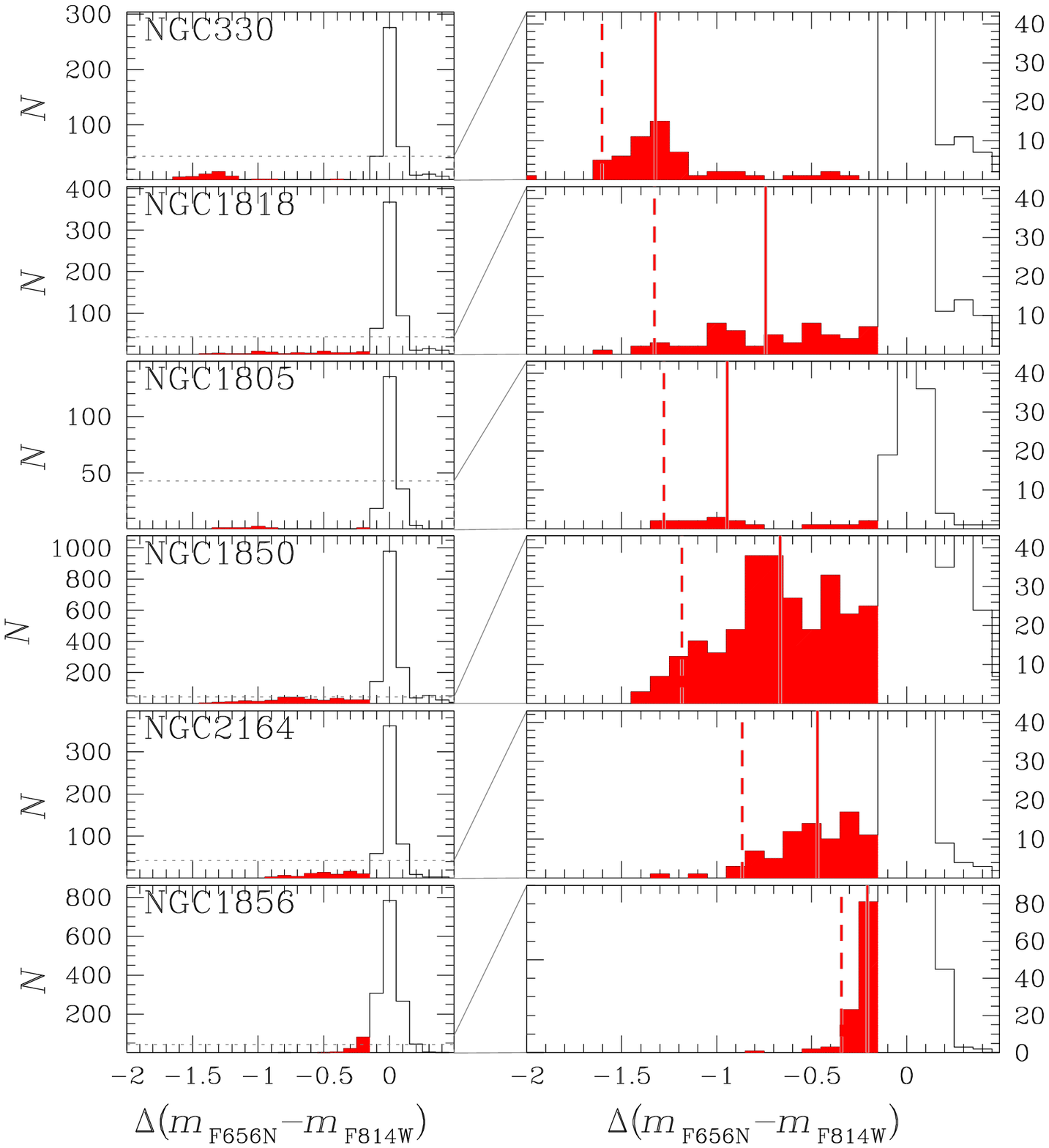}
  \includegraphics[width=8.5cm]{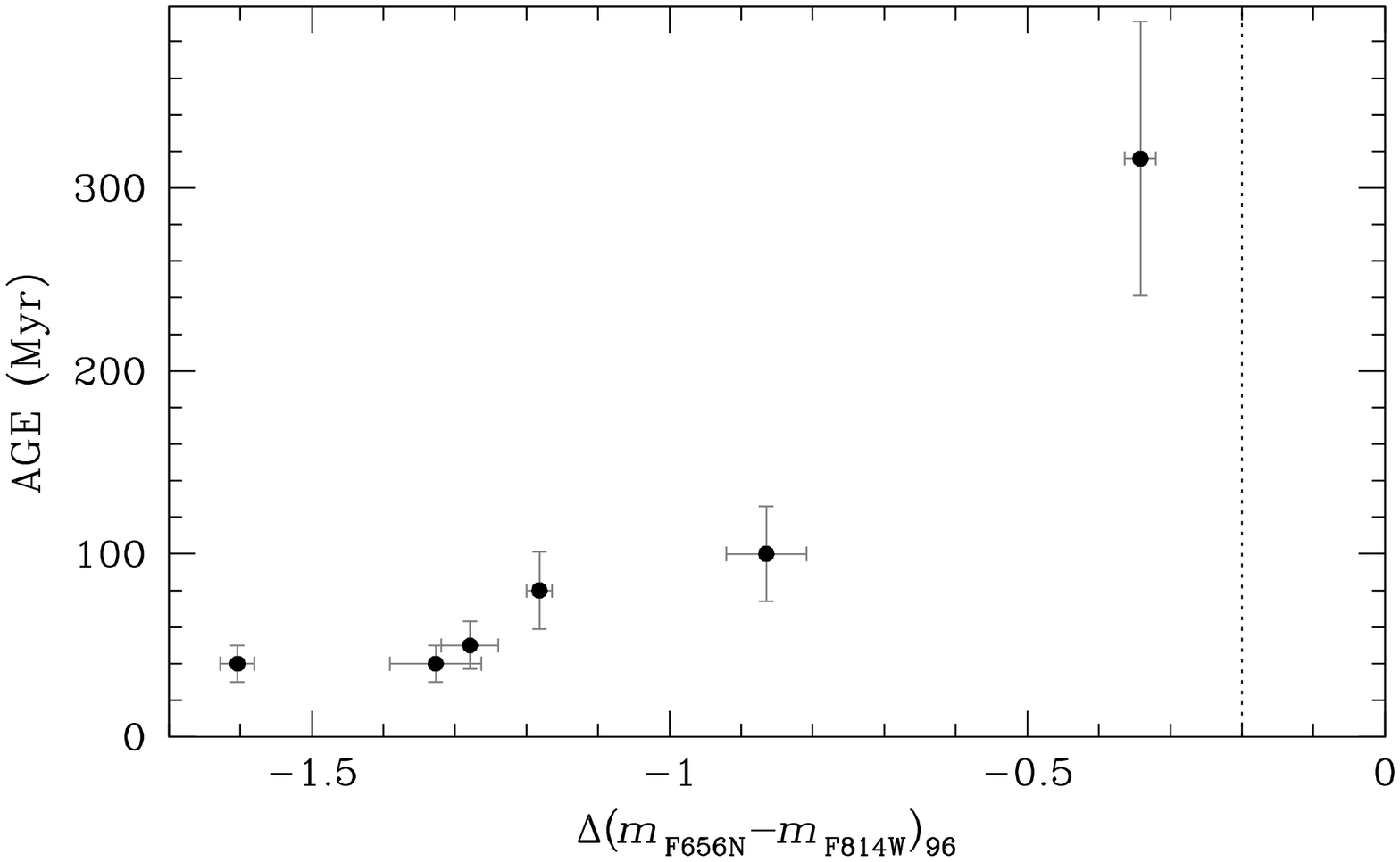}
  \caption{\textit{Upper Panels.} Histograms of the $\Delta (m_{\rm F656N}-m_{\rm F814W})$ distribution of stars brighter than $m_{\rm F814W}=19.3$ (left panels). The right panels are a zoom of the left panels. The red-shaded areas indicate the histograms of candidate H$_{\rm \alpha}$ emitters, while the continuous and the dashed vertical line mark the values corresponding to the median  and the 96$^{\rm th}$ percentile of the $\Delta (m_{\rm F656N}-m_{\rm F814W})$ distribution, respectively. \textit{Lower Panel.} Cluster age as a function of the 96$^{\rm th}$ percentile of the $\Delta (m_{\rm F656N}-m_{\rm F814W})$ distribution. The vertical dotted line indicates the adopted separation between candidate stars with strong H-$\alpha$ emission and the other MS stars.}
 \label{fig:histo} 
\end{figure} 
\end{centering}

 The $m_{\rm F656N}-m_{\rm F814W}$ color of MS stars is analyzed in the upper panels Fig.~\ref{fig:histo} where we show that the $\Delta (m_{\rm F656N}-m_{\rm F814W})$ histogram distribution for all the MS stars brighter than $m^{\rm TO}_{\rm F814W}=1.7$ changes from one cluster to another.
In NGC\,330, the H-$\alpha$ emitters are peaked around $\Delta (m_{\rm F656N}-m_{\rm F814W}) \sim -1.3$, with few stars only having $\Delta (m_{\rm F656N}-m_{\rm F814W}) \gtrsim -1.0$. In contrast, the H-$\alpha$ emitters of NGC\,1818 exhibit a rather flat $\Delta (m_{\rm F656N}-m_{\rm F814W})$ distribution.

In NGC\,1850, which is the cluster with the largest number of stars with H-$\alpha$ excess, we identify some hints of three peaks of emitting stars around $\Delta (m_{\rm F656N}-m_{\rm F814W}) \sim -1.1, -0.8$, and $-0.4$. In NGC\,2164 the number of H-$\alpha$ emitters increases when moving towards higher values of $\Delta (m_{\rm F656N}-m_{\rm F814W})$.
The average value of $\Delta (m_{\rm F656N}-m_{\rm F814W})$ increases when moving from young towards older clusters. The $96^{\rm th}$ percentile of  the $\Delta (m_{\rm F656N}-m_{\rm F814W})$ distribution,  $\Delta (m_{\rm F656N}-m_{\rm F814W})_{96}$, which is indicative of the maximum color distance of Be stars from the MS fiducial ranges from $\sim$1.6 in NGC\,330 to $\sim$0.25 in NGC\,1856 and correlates with the cluster age as illustrated in the lower panel of Fig.~\ref{fig:histo}.
These results provide empirical evidence that the H-$\alpha$ emission is stronger in the young clusters, where the effective temperature of the Be stars around the MSTO is hotter than that of older clusters. 

\section{Summary and discussion}\label{sec:conclusion}
 We derived high-precision photometry from UVIS/WFC3 images of thirteen young clusters in the LMC and SMC to investigate their multiple stellar populations. 
 The analyzed clusters span a wide range of masses from $\sim10^{3.6}$ to $\sim10^{5.1}$ $\mathcal{M_{\odot}}$ and ages from $\sim$40 Myrs to $\sim$1\,Gyr.
 
We confirm previous evidence that the MS and the MSTO of NGC\,330, NGC\,1755, NGC\,1805, NGC\,1818, NGC\,1850, NGC\,1856,  NGC\,1866 and NGC\,2164 are not compatible with a simple stellar population (Milone et al.\,2015, 2016, 2017a, Correnti et al.\,2015, 2017; Bastian et al.\,2017; Li et al.\,2017a).  
Moreover, we find that all the analyzed clusters exhibit the eMSTO and that all the clusters except for the oldest, NGC\,1868, show a split MS. 

The fraction of blue-MS stars with respect to the total number of MS stars significantly changes with the stellar luminosity and follows similar trends in the analyzed clusters.
 It reaches its maximum value at $M_{\rm F814W} \sim 1.8$ where the blue MS comprises about $\sim$30-40\% of the total number of MS stars and monotonically drops down to $\sim$15-25\% at $M_{\rm F814W} \sim 0.7$.
 The trend seems reversed and the ratio between the number of blue-MS and the total number of MS stars seems increased towards higher lumonisity. The red and the blue MS merge brighter than $M_{\rm F814W} \sim 2.0$, thus indicating that there is no evidence for double MS among stars less-massive than $\sim$1.6$\mathcal{M}_{\odot}$  (see also Milone et al.\,2017a; D'Antona et al.\,2017).
NGC\,1868, which is the only cluster with no evidence for multiple populations along the MS, is the oldest analyzed cluster and its brightest unevolved MS stars  are fainter than $M_{\rm F814W} \sim 1.8$.
 The fact that all the young clusters follow similar trends suggests that the physical mechanism that is responsible for the split MS is not dependent on the cluster mass.

 A possible exception to this notion is NGC\,330 in the SMC, which is the youngest analyzed cluster, and shows a lower fraction of blue-MS stars than the other clusters in the entire range of magnitudes $M_{\rm F814W} \lesssim 2$. This fact could suggest that all stars form as red-MS stars and only stars older than a certain age move to the blue-MS.
 However we note that also the other two analyzed SMC clusters, Kron\,34 and NGC\,294, seem to exhibit a slightly smaller fraction of blue MS stars than the LMC clusters. Hence it would be tempting to speculate that either the environment or the cluster metallicity would play a major role in determining the relative numbers of blue-MS and red-MS stars.

 Several authors have concluded that a low-metallicity environment such us the SMC hosts a larger fraction of Be stars that higher-metallicity environments such us the Milky Way and the LMC do (e.g.\,Maeder et al.\,1999; McSwain \& Gies 2005; Igbal \& Keller\,2013).  A larger sample of clusters is needed to properly investigate these possibilities.

The evidence of multiple populations in all the analyzed clusters suggests that the split MS is a common feature in the CMDs of all the star clusters younger than $\sim$1 Gyr and that its presence is closely connected with the evolution of stars brighter than $M_{\rm F814W} \sim 2$. This luminosity corresponds to a stellar mass of $\mathcal{M} \sim 1.6 \mathcal{M_{\rm \odot}}$ and corresponds to the MS kink occurring at $T_{\rm eff} \sim$7500\,K.
 Such a feature, that occurs also in Galactic open clusters, indicates the onset of envelope convection due to the opacity peak of partial hydrogen ionization (e.g.\,D’Antona et al.\,2002).

 Recent work on old Galactic GCs clearly shows that the incidence and the complexity of the multiple-population phenomenon both increase with the cluster mass.  In particular, the fraction of second-generation stars with respect to the total number of stars correlates with the mass (Milone et al.\,2017b). Our  analysis of young clusters reveals that the properties of their multiple sequences  analyzed in this paper, including the fraction of blue-MS stars with respect to the total number of stars and the stellar rotation rates needed to reproduce the MSs, do not depend on the mass of the host cluster.
   These facts provide another major difference between the multiple-population phenomenon in old and young clusters.

The comparison between the observed CMDs and isochrones from the Geneva database suggests that the red MS is composed of stars with high rotation rates ($\omega =0.9 \omega_{\rm c}$), while the blue MS is made of non-rotating or slowly-rotating stars in close analogy with what has been suggested for NGC\,1856, NGC\,1755, NGC\,1850, and NGC\,1866 (D'Antona et al.\,2015;  Milone et al.\,2016, 2017a; Correnti et al.\,2017). 

Red-MS stars and the majority of blue-MS stars are consistent with coeval stellar populations.
Moreover, the CMDs of all the clusters younger than $\sim$1\,Gyr host stars that are fitted by the corresponding slowly-rotating isochrones but are younger than the majority of blue-MS stars. These stars have been interpreted by D'Antona et al.\,(2017) to the braking of the rapidly-rotating stars, which end up in having a younger nuclear evolution stage than that of stars slowly rotating from the beginning of their life.
 D'Antona and collaborators have suggested that the braking could be due to some mechanism of angular-momentum loss. 

Several papers have shown that young clusters host a large population of Be stars (e.g.\,Grebel et al.\,1992, 1997; Mazzali et al.\,1996; Keller et al.\,1999; Correnti et al.\,2017; Bastian et al.\,2017).
The clusters younger than $\sim$600 Myrs for which F656N photometry is available host a large populations of  H-$\alpha$ emitters which are distributed along the upper part of the red MS and the eMSTO. The fact that the H-$\alpha$ emission is associated with the Be phenomenon supports the idea that the red MS is made of fast rotators. 
The fact that the SMC cluster NGC\,330 hosts both a slightly-higher fraction of Be stars and a higher fraction of red-MS stars than the studied LMC clusters could suggest that both phenomena depend on the cluster metallicity. 

In each of the analyzed young clusters, the fraction of H-$\alpha$ emitters stars with respect to the total number of MS stars dramatically changes as a function of magnitude offset from the turnoff. We derived the fraction of Be stars in distinct magnitude intervals and plotted this quantity against the difference between the average F814W magnitude of all the stars in each bin and the magnitude of the TO of the red MS.  In this plane all the analyzed clusters follow similar trends, regardless their mass.  The fraction of Be stars is maximal above the MSTO, where the stars with flux excess in the F656N include about one half of the MS stars, and drops to zero about one F814W magnitude below the MSTO.
These results are consistent with previous evidence that the Be star fraction in a cluster peak at the MSTO luminosity and decreases with decreasing luminosity  (Keller et al.\,2000; Iqbal \& Keller 2013).

   Moreover, the evidence that H-alpha emitters are present among the upper MS in clusters with different age, where MSTO stars have different masses, suggests that the stars are not emitters at the formation. The mechanism responsible for the Be phenomenon seems closely connected to the evolution of MS stars instead.

\section*{Appendix. Discovery of a new star cluster}
\begin{centering} 
\begin{figure*} 
  \includegraphics[height=13.25cm]{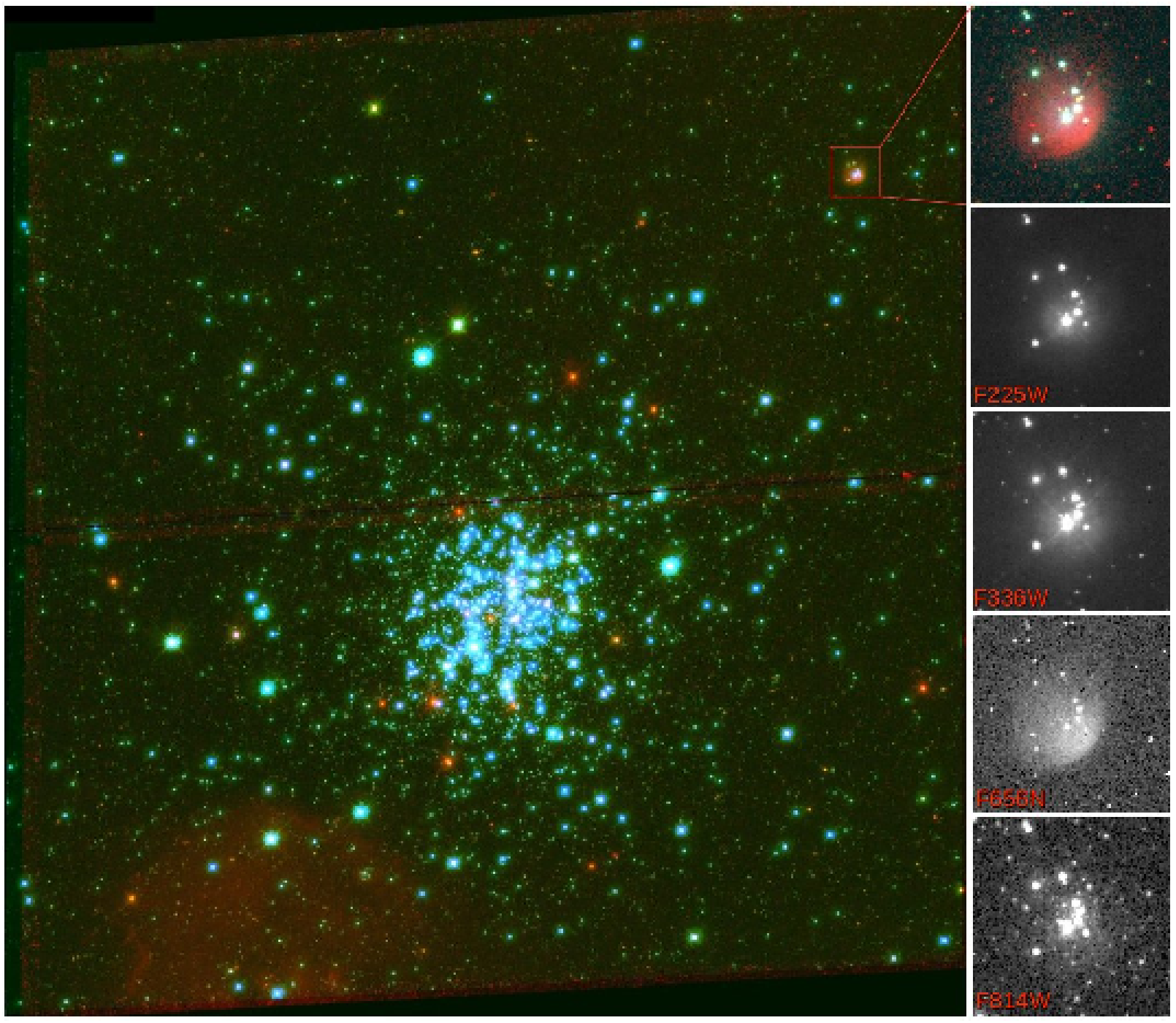}
  \caption{Three-color image of NGC\,330 (left panel). Right panels are zoom of the region around the cluster, GALFOR\,1, that we have identified in this work. From the top to the bottom we show the three-color image and images in the F225W, F336W, F656N, and F814W bands. South is up and east is on the right.}
 \label{fig:NGC330} 
\end{figure*} 
\end{centering}

We report the serendipitous discovery of a small star cluster in the outskirts of NGC\,330. In the following we will name this object as GALFOR\,1, after the project financed by the European Research Council that supports this paper.
The three-colour image of the field of view of NGC\,330 is provided in the left panel of Fig.~\ref{fig:NGC330} and is derived from the  combination of the F656N (R channel), F336W (G channel), and F225W (B channel) master frames. The right panels are zooms of the region around GALFOR\,1 in the left-panel color image and in the monochromatic F225W, F336W, F656N, and F814W bands. 

\begin{centering} 
\begin{figure*} 
  \includegraphics[width=13.25cm]{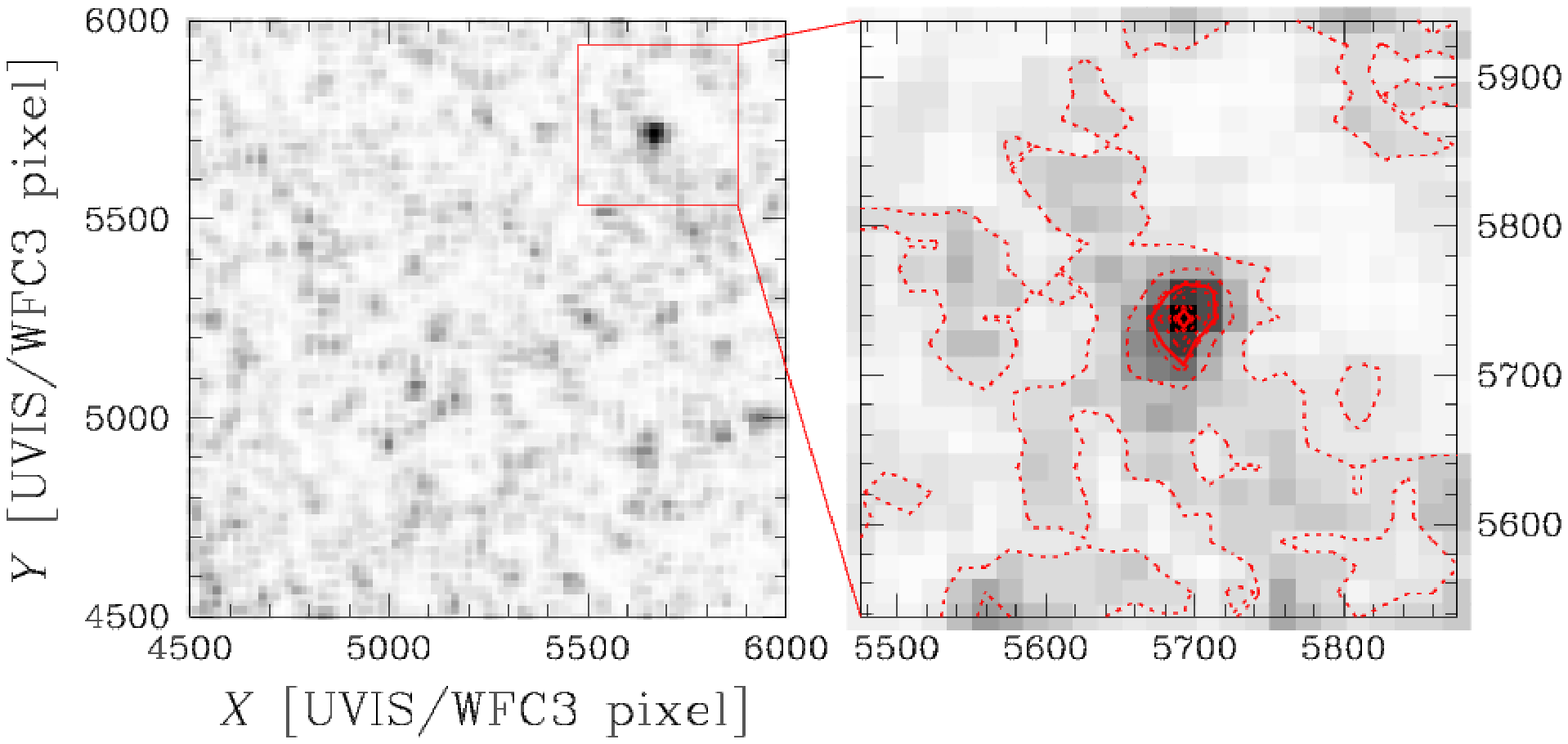}
  \includegraphics[width=13.25cm]{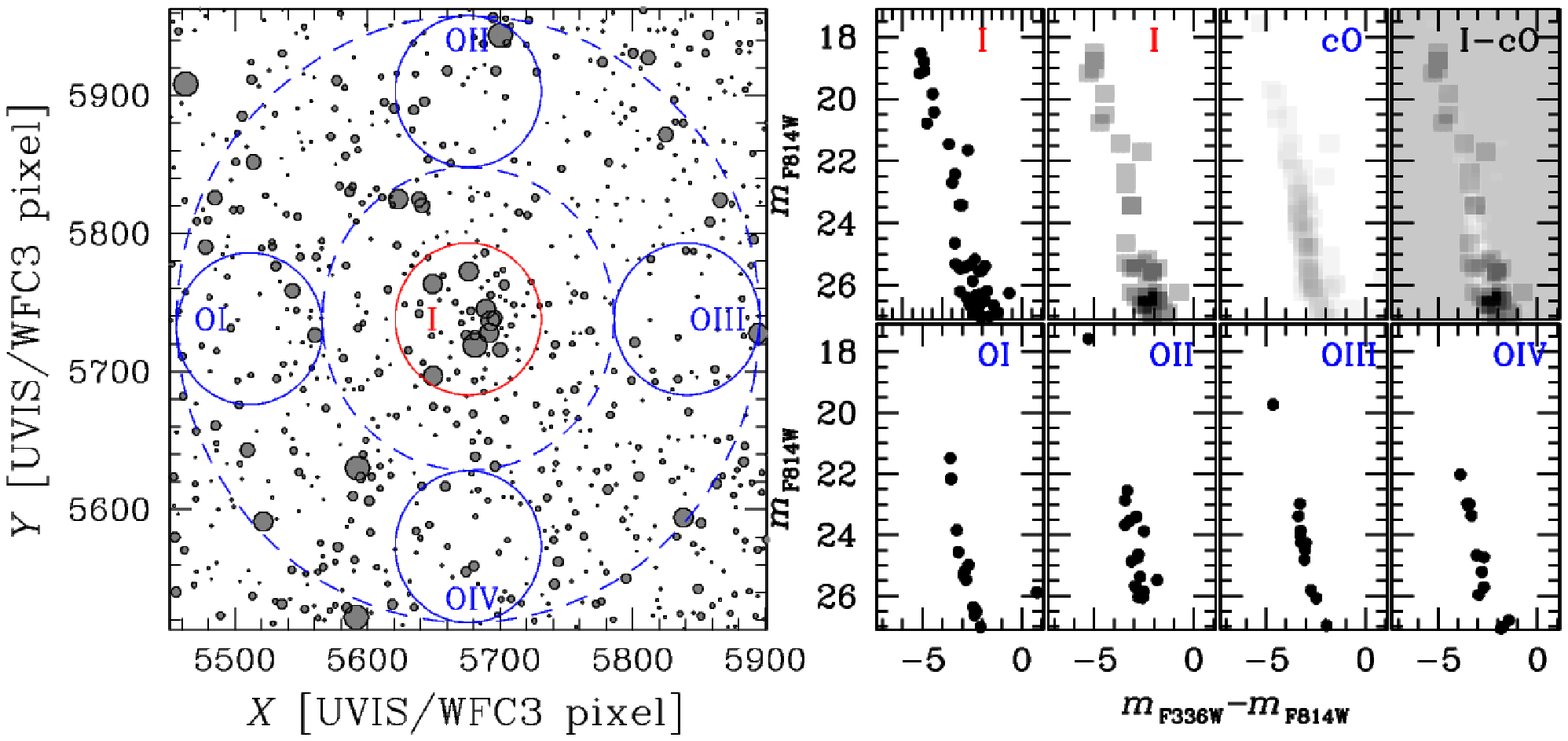}
  \caption{Map of stellar density for stars in a  field of 1500$\times$1500 UVIS/WFC3 pixels (1.0$\times$1.0 arcmin) south-east to NGC\,330 (upper-left panel). Upper-right panel shows a zoom of the upper-left-panel plot in a region centered on the newly-discovered star cluster. The contours of stellar density are represented with red lines.
    Lower-left panel show a map of stars around the newly-discovered star cluster. The red circle marks the inner field, I, while the region between the two dashed circles is indicated as outer field. The four sub-regions of the outer field (OI, OII, OIII, and OIV) are marked with blue-continuous fields. The $m_{\rm F814W}$ vs.\,$m_{\rm F336W}-m_{\rm F814W}$ CMDs of stars in the fields I, OI--OIV are shown in the lower-right panels where we also plotted the Hess diagram of stars in the fields I and O and the Hess diagram of the corresponding difference. See text for details.   
   }
 \label{fig:newcl} 
\end{figure*} 
\end{centering}

The cluster is clearly visible on the F814W image and the most prominent feature of the F656N image is the nebula that surrounds it.
The presence of an overdensity of stars corresponding to the position of the nebula is confirmed by the analysis illustrated in the upper-left panel of Fig.~\ref{fig:newcl} where we show the density map of stars brighter than $m_{\rm F814W}=25.5$ in a region south-east to NGC\,330. In the upper-right panel we overplotted  the contours on the density map and infer from the contours the position of the cluster center: RA=00:56:38.6 DEC=$-$72:28:23.7, J2000.

The lower-left panel of Fig.~\ref{fig:newcl} shows a map of all the stars brighter than $m_{\rm F814W}=25.5$ around GALFOR\,1. We defined a circle with radius of 2 arcsec centered on GALFOR\,1 that includes most of the cluster members. Moreover, we derived a concentric circular annulus, with inner and outer radius of 4 and 8 arcsec, respectively, which includes field stars only. The regions within the inner circle and the outer region within the annulus are indicates as I and O regions, respectively. We also delimit with blue circles four sub-regions, namely OI--OIV, within the field O with the same area as the field I.

The $m_{\rm F814W}$ vs.\,$m_{\rm F336W}-m_{\rm F814W}$ CMDs of stars in the fields I and OI--OIV are plotted in the lower-right panels and confirms that the field I host a larger number of stars with respect to the surrounding regions.
We also show the Hess diagrams of stars in the fields I and O. To properly compare their stellar density we multiplied the density of stars in the field O by a constant, c=1/12, which corresponds to the ratio between the areas of the fields O and I. We find that the difference between the stellar density in the field I and the stellar density in the field O multiplied by the constant c is significantly larger than zero as demonstrated by the corresponding Hess diagram plotted in Fig.~\ref{fig:newcl}. These facts demonstrate that the overdensity of stars in the inner field is real and is due to the presence of a small young star cluster. 

In their catalogue of H-$\alpha$ emission-line stars and small nebulae in the SMC, Meyssonnier \& Azzopardi (1993, MA93) marked at the cluster position an emission-line star, MA93-952, in close analogy with Murphy \& Bessell (2000\, see their Table\,1, line 150). The F656N photometry analyzed in our work shows no evidence of emitting stars in this cluster. As a matter of fact, the {\it HST} multi-band images reveal that GALFOR\,1 is immersed in a diffuse nebula with both a reflection and an emission component. 

The reflection component, made of scattered star-light by dust, is prominent in the F225W and F336W filters, where it appears of virtually spherical shape.
That a significant amount of interstellar dust is indeed present in the cluster is further supported by the corresponding WISE images, that reveal at that position a (clearly not resolved) point source of magnitudes m$_{\rm W1}$=14.454$\pm$0.028, m$_{\rm W2}$=14.383$\pm$0.039, m$_{\rm W3}$=9.615$\pm$0.032, m$_{\rm W4}$=6.625$\pm$0.057 (Cutri et al. 2013). The derived infrared excess would thus indicate the presence of moderately cold dust grains at temperature $T_{\rm dust}\lesssim$ 150 K.

The H$-{\alpha}$ emission component, as from the F656N filter, appears with a characteristic fan-like shape whose sharp edge 
 is reminiscent of a stellar wind bow-shock (Brownsberger \& Romani 2014). The star responsible for the ionization of the interstellar gas is likely the brightest star of the cluster: if our assumption about the presence of a bow shock is correct, then it would imply a supersonic velocity between the star itself and the surrounding ambient medium. In that case, the bow-shock orientation may even be taken as the direction of the star proper motion.\\
Nevertheless, without further data and a dedicated analysis, we are not in the position of distinguishing whether the bow-shock orientation and morphology are the result of stellar motion of the local gas dynamics or of density gradients (Kiminki et al.\,2017); nor we can confidently exclude the possibility that what we are seeing is actually an interstellar bubble within a HII region that surrounds slowly moving stars (Mackey et al.\,2015).

A forthcoming paper will be dedicated to address this and other issues, and to provide a better characterization of this newly-discovered SMC star cluster.

\section*{acknowledgments} 
\small 
 We thank the anonymous referee for several suggestions that have improved the quality of this manuscript.
This work has been supported by the European Research Council through the ERC-StG 2016 project 716082 `GALFOR'. 
APM, AFM and HJ acknowledge support by the Australian Research Council through Discovery Early Career Researcher Awards DE150101816 and DE160100851 and Discovery project DP150100862.
This work  is based on observations with the NASA/ESA Hubble Space Telescope,
    obtained at the Space Telescope Science Institute, which is
    operated by AURA, Inc., under NASA contract NAS 5-26555, under
    GO-13397, GO-14204 and GO-14710.
    This publication makes use of data products from the Wide-field Infrared Survey Explorer, which is a joint project of the University of California, Los Angeles, and the Jet Propulsion Laboratory/California Institute of Technology, funded by the National Aeronautics and Space Administration.

\bibliographystyle{aa}

\end{document}